\documentclass[useAMS,usenatbib]{mnras}

\usepackage{amssymb}
\usepackage{xcolor}
\usepackage{graphicx}
\usepackage{amsmath}
\usepackage{multirow}
\usepackage{color}
\usepackage{morefloats}
\usepackage{booktabs}
\usepackage{threeparttable}
\usepackage{caption}
\usepackage{hyperref}

\newcommand{\espadons}{{\it ESPaDOns}}
\newcommand{\narval}{{\it NARVAL}}

\newcommand{\teff}{$T_{eff}$}
\newcommand{\xspec}{{\sc xspec}}
\newcommand{\cmfflux}{{\sc cmf\_flux}}
\newcommand{\Ro}{R$_{{\rm 0}}$}
\newcommand{\Tx}{T$_{{\rm X}}$}
\newcommand{\apec}{{\sc apec}}
\newcommand{\xcmfgen}{{\sc xcmfgen}}
\newcommand{\cmfgen}{{\sc cmfgen}}

\newcommand{\fastwind}{{\sc fastwind}}

\newcommand{\hipparcos}{{\it HIPPARCOS}}
\newcommand{\epsori}{$\epsilon$~O\lowercase{ri}}

\newcommand{\ghrs}{{\it GHRS}}
\newcommand{\copernicus}{{\it COPERNICUS}}
\newcommand{\chandra}{\emph{Chandra}}
\newcommand{\xmm}{\emph{XMM-Newton}}
\newcommand{\xmms}{\emph{XMM}} 

\newcommand{\ionn}[2]{#1\,{\sc #2}}

\newcommand{\ionnl}[3]{\ionn{#1}{#2} $\lambda$#3}
\newcommand{\ionnll}[3]{\ionn{#1}{#2} $\lambda\lambda$#3}

\newcommand{\scie}[2]{#1$\times$10$^{#2}$}
\newcommand{\rsun}{R$_\odot$}
\newcommand{\sunyr}{\msun ~yr$^{-1}$}

\newcommand{\massrate}[2]{#1$\times$10$^{#2}$ \sunyr}

\newcommand{\ebv}{{\it E(B-V)}}
\newcommand{\iue}{{\it IUE}}
\newcommand{\fuse}{{\it FUSE}}
\newcommand{\hst}{{\it HST}}

\newcommand{\msun}{\mbox{\thinspace M$_{\odot}$}}

\newcommand{\kms}{\mbox{\thinspace km\thinspace s$^{-1}$}}

\def\Mdot{\hbox{$\dot{\rm M}$}}

\title[X-ray, UV and optical analysis of supergiants: \boldmath$\epsilon$~Ori]
{X-ray, UV and optical analysis of supergiants: \boldmath$\epsilon$~Ori}
\author[Raul E. Puebla et al.]{Raul E. Puebla,$^{1}$\thanks{E-mail:
rep54@pitt.edu}  D. John Hillier,$^1$ Janos Zsarg\'{o},$^2$ David H. Cohen$^3$
\newauthor and
 Maurice A. Leutenegger$^{4,5}$\thanks{Based on observations obtained with 
XMM-Newton, an ESA science mission with instruments and contributions directly
funded by ESA Member States and NASA.}\thanks{The scientific results reported in this article 
are based on data obtained from the Chandra Data Archive.}\thanks{Some of the data 
presented in this paper were obtained from the Mikulski Archive for Space Telescopes 
(MAST). STScI is operated by the Association of Universities for Research in Astronomy,
Inc., under NASA contract NAS5-26555. Support for MAST for non-HST data is provided by
the NASA Office of Space Science via grant NNX09AF08G and by other grants and contracts}
\thanks{Optical data for this work were obtained from POLARBASE \citep{petit14}}\\
$^{1}$Department of Physics and Astronomy \& Pittsburgh Particle Physics, Astrophysics, 
and Cosmology Center (PITT PACC), \\
University of Pittsburgh, 3941 O'Hara Street, Pittsburgh, PA 15260, USA.\\
$^{2}$Escuela Superior de F\'{i}sica y Matem\'{a}tica. Instituo Polit\'{e}cnico Nacional.
Av. Instituto Polit\'{e}cnico Nacional, Edificio 9,\\ C.P. 07738, M\'{e}xico, DF. M\'{e}xico.\\
$^{3}$Department of Physics and Astronomy. Swarthmore College, 500 College Ave. Swarthmore, PA 19081, USA.\\
$^{4}$CRESST/University of Maryland, Baltimore County, 1000 Hilltop Circle, Baltimore, MD 21250, USA.\\
$^{5}$NASA/Goddard Space Flight Center, 8800 Greenbelt Road, Greenbelt, MD 20771, USA.}

\begin{document}

\date{}

\pagerange{\pageref{firstpage}--\pageref{lastpage}} \pubyear{2015}

\maketitle

\label{firstpage}


\begin{abstract}

We present a multi-wavelength (X-ray to optical) analysis, 
based on non-local thermodynamic equilibrium photospheric+wind models,
of the B0 Ia-supergiant: $\epsilon$~Ori. The aim is to test the consistency of  
physical parameters, such as the mass-loss rate and CNO abundances,
derived  from different spectral bands.
The derived  mass-loss rate is  $\Mdot/\sqrt{f_\infty}\sim$1.6$\times$10$^{-6}$ 
M$_\odot$ yr$^{-1}$ where $f_\infty$ is the volume filling factor.  However, the \ionnll{S}{iv}{1062,1073} 
profiles are too strong in the models; to fit the observed profiles it is necessary to use
$f_\infty<$0.01.  This value is a factor of 5 to 10 lower than
inferred from other diagnostics, and implies
$\Mdot \lesssim1 \times 10^{-7}$ M$_\odot$ yr$^{-1}$.
The discrepancy could be related to porosity-vorosity effects or
a problem with the ionization of sulfur in the wind.
To fit the UV profiles of \ionn{N}{v} and \ionn{O}{vi}
it was necessary to include emission from an interclump medium with
a density contrast ($\rho_{cl}/\rho_{ICM}$)  of $\sim$100.
X-ray emission in H-He like and Fe L lines was 
modeled using four plasma components located within the wind.
We derive plasma temperatures from $1 \times 10^{6}$ to $7\times 10^{6}$ K, with lower temperatures
starting in the outer regions (R$_0\sim$3-6 R$_*$), and a hot component starting closer to the star
(R$_0\lesssim$2.9 R$_*$). From X-ray line profiles we infer $\Mdot <\, $\massrate{4.9}{-7}. 
The X-ray spectrum ($\geq$0.1 kev) yields an X-ray luminosity  
$L_{\rm X}\sim 2.0\times10^{-7} L_{\rm bol}$, consistent with the superion line profiles. 
X-ray abundances are in 
agreement with those derived from the UV and optical analysis: 
$\epsilon$~Ori is slightly enhanced in nitrogen and depleted 
in carbon and oxygen, evidence for CNO processed material.

\end{abstract}


\section{INTRODUCTION}\label{intro}

Massive stars play a fundamental role in the Universe. 
Considered the progenitors of core collapse supernovae,
they are also responsible for galactic 
\ionn{H}{ii} regions, the transfer of mass, momentum and 
energy to the interstellar medium (ISM), and for metal
enrichment in their host galaxy.

Among the stellar wind properties, the mass-loss rate (\Mdot)
is crucial for understanding the evolution of massive stars. 
\Mdot\ affects the star's lifetime on the main sequence,
the rotation rate of the star, and the star's subsequent evolution
\citep[e.g.][]{chiosi86,maeder00}. 
The first estimates for \Mdot\ were derived assuming 
spherical smooth winds. However, subsequent observations 
reveal inconsistencies. For instance, the
far UV resonance line profiles (e.g. \ionn{P}{v} and \ionn{S}{iv}) 
observed by \copernicus\ and {\it FUSE} cannot be simultaneously fit 
with H$\alpha$ \citep[e.g.][]{crowther02,hillier03}. 
Futhermore, P~Cygni profiles from highly ionized species, such as 
\ionn{N}{v} and \ionn{O}{vi}, were detected in the UV spectrum \citep{snow76}, but such
ionization states are incompatible with radiative equilibrium in a smooth wind. 
\cite{cassinelli79} suggested that these ions
could be produced by Auger ionization by X-rays (double
ionization due to ejection of an inner shell electron by a
X-Ray photon.), a suggestion confirmed 
by {\it Einstein} X-ray Observatory which found that 
massive stars are strong X-ray sources 
\citep{harnden79,seward79}. Two scenarios were proposed to 
explain the observed X-ray emission: emission from a corona just 
above the photosphere \citep{hearn75,cassinelli79} and X-ray 
emission from shock-heated plasma distributed throughout the wind
\citep{lucy80,lucy82,owocki88}. The latter is now generally
accepted as the dominant X-ray emission mechanism in single stars.     

Because of the deficiencies discussed above, the ``standard'' smooth wind
model needed to be revised. At present, it is widely
accepted that the winds of massive stars are strongly 
structured (clumped). The existence of such winds is supported by
hydrodynamical time-dependent simulations that
predict that radiation-driven winds are unstable \citep{owocki88,
feldmeier97,runacres02}. Evidence for these 
inhomogeneities has been found, for instance, by observations 
of variable discrete absorption components (DAC) in UV and optical lines   
\citep[e.g.][]{fullerton96,kaper96,morel04,prinja06},
and from observations of stochastic variability \citep[e.g.,][]{eversberg98}. 
\cite{hillier91} showed that the electron scattering wings 
of recombination lines of Wolf-Rayet (W-R) stars cannot be reproduced 
by smooth winds, necessitating clumped models and
lower mass-loss rates to fit them. 

Clumping can be treated using two approximations. The simplest
approach considers optically thin clumps at all wavelengths 
(``{\it microclumping}''). The second approach 
allows for the  optical thickness of clumps,
especially for lines  (``{\it macroclumping}''). 
Commonly it is assumed that the interclump medium is void. However,
recently it has been shown that its influence is not negligible \citep[e.g.][]{zsargo08}. 

Microclumping has been used to lower the discrepancy between
\ionn{P}{v} UV resonance lines and H$\alpha$.
Typical filling factors in O stars are 0.01 to 0.1,  although,
in some cases, it still may be necessary to reduce the P abundance
by a factor of 2 from the expected value \citep[e.g.][]
{crowther02,hillier03,bouret05,fullerton06,bouret12}.
The principal consequence of microclumping is a
reduction of mass-loss rates by factors from 3 to 10.

A 3D Monte-Carlo simulation performed by \cite{surlan13} 
showed that when macroclumping is included the  
H$\alpha$ and \ionn{P}{v} lines discrepancy is fixed
for higher values of the volume filling factor, yielding higher
mass-loss rates when compared with those based on pure microclumping
 \citep[see also][]{oskinova07,sundqvist10,surlan12}.

The problem with all these analyses is degeneracy --
there are several parameters which can be varied 
but only a few lines that can be modeled, making it
difficult to reach definitive 
conclusions \citep[see details in][]{surlan13}.

As noted by several different authors \citep[e.g.][]{cohen10},
the fitting of X-ray lines provides an 
independent method to estimate the mass-loss rates of
OB stars. \cite{cohen10} estimated 
the characteristic optical depth for X-rays ($\tau_*=
\kappa \dot{M}/4\pi R_*v_\infty$), fitting separately 
X-ray lines and then fitting the best optical depth
wavelength distribution using the mass-loss rate 
as a free parameter. The main reported problem with 
X-ray line emission is that most of the observed 
profiles are fairly symmetric. This contradicts 
models that predict lines skewed to the blue 
\citep{macfarlane91,owocki01}. Three explanations 
are currently proposed: resonance scattering, 
lower mass-loss rates and porosity (see \cite{oskinova11} for 
a summary of X-ray emission properties of OB stars). 
However, resonance scattering cannot explain all observed line profiles 
\citep{ignace02,leutenegger07}, while the needed reduction
in the mass-loss rates inferred from H$\alpha$
is large (factors of 2 to 10) 
\citep[e.g.][]{cohen14}. It has also been argued that the porosity lengths 
(a measure of mean free path of photons between clumps) 
required to get symmetric lines are unlikely to be as large as needed
\citep{owocki06}. \cite{sundqvist12} 
and \cite{leutenegger13} concluded that a porosity 
consistent with the observed X-ray line profiles
cannot affect the mass-loss rate determination
significantly. Furthermore, \cite{herve13} showed
that porosity is not important to explain the 
X-ray spectrum of $\zeta$ Pup.

With the aim of providing more rigorous constraints, and reducing
systematic errors, we present our work on a multi-wavelength
analysis of the supergiant $\epsilon$ O\lowercase{ri}. The analysis 
is based on X-ray data from \chandra\ and \xmm, UV
data from \iue, \hst, and the \copernicus\ satellite, and optical data
taken from {\it POLARBASE} archive. 
The photospheric, wind and hot-plasma parameters are 
obtained from a modified version of \cmfgen\ \citep{hillier98}, 
and the consistency of the derived parameters is examined.

The paper is constructed as follows: In the next section, 
we present results from previous analyses of \epsori\ and 
its main spectral features. Observational data are described 
in Section \ref{data} while the method and model assumptions 
are presented in Section \ref{method}. The results of optical, 
UV and X-ray analysis are given in Section \ref{results}. 
A discussion of the results and conclusions are provided in Sections 
\ref{discussion} and \ref{conclusions}.

\section{\boldmath{$\epsilon$} O\lowercase{ri} (HD 37128)}

The B0 Ia supergiant star \epsori\ (HD 37128)  is the
central star of the Orion Belt, and is also known as {\it Alnilam}
(an Arabic word that means ``{\it string of pearls}'' 
\citep{knobel09}) and belongs to the Orion OB1(b1) 
association.

The first attempts to derive the physical parameters
for \epsori\ using photospheric models in non-local thermodynamic 
equilibrium (NLTE) were made by \cite{auer72} and \cite{lamers74} who used optical line 
profiles, and found an effective temperature around 29000 K and $\log g$=3.0.  
 \cite{mcerlean98} using an NLTE model found similar parameters, while
\cite{kudritzki99}, using the original version of the 
code \fastwind\ \citep{santolaya97,puls05} 
found   T$_{eff}$=28000 K and $\log g$=3.0.

\cite{crowther06}  and \cite{searle08} used the NLTE 
transfer code \cmfgen\ \citep{hillier98}
to derive effective temperatures of 27000 K 
and 27500 K respectively. However, there exists a discrepancy
between their $\log g$ values -- \citeauthor{crowther06} reported
a value of 2.9 while \citeauthor{searle08} reported 3.1.
Both works also analyzed the CNO abundances of \epsori.
In contrast to the nitrogen deficiency reported by \cite{walborn76},
\citeauthor{crowther06} reported a slight 
enrichment of nitrogen and depletion of carbon. On the other hand, 
\citeauthor{searle08} found  a nitrogen and carbon deficiency
and a solar oxygen abundance.

Previous values for mass-loss rate from different diagnostics
encompass values from \scie{1.5}{-6} to \massrate{4.3}{-6}. Diagnostics include H$\alpha$ strength \citep{kudritzki99,crowther06,searle08,urbaneja04}, UV P~Cyg profiles
\citep{lamers99}, thermal radio fluxes \citep{blomme02,lamers93}
as well as H and He infrared lines in the H and K bands (\citealt{repolust05};
they derived an upper limit of \Mdot=\massrate{5.25}{-6}).
These values where derived using smooth winds. In a subsequent
analysis by  \cite{najarro11}, lines from the L band were used 
in conjunction with UV and optical data to obtain a mass-loss rate of 
\mbox{$\dot{\rm M}/\sqrt{f_\infty}$}=\massrate{2.65}{-6} 
and a filling factor $f_\infty$=0.03.

Mass-loss rate determinations should all be scaled to the same 
distance, since the mass-loss rate typically scales as
$d^{1.5}$ for $\rho^2$ dependent diagnostics (or $d$ for $\rho$ dependent
diagnostics such as X-ray profiles). Distance estimates for \epsori\ tend to cluster around 400 pc
\citep{lesh68,lamers74,savage77,brown94}, and are broadly consistent
with the initial  \hipparcos\ determination of  412~pc \citep{perryman97}. 
However, with the latest calibration the new estimate 
is 606~pc \citep{vanleeuwen07}. This  distance is the 
highest ever estimated, and substantially increases the luminosity 
and mass-loss rate for \epsori.

\iue\ and \copernicus\ spectra of \epsori\ show line emission 
from \ionn{N}{v} and \ionn{O}{vi} resonance transitions, 
which, given the effective temperature of  \epsori, 
provides evidence of X-ray emission in wind. \chandra\ and \xmms\ 
spectra of HD~37128 show strong emission lines from H/He-like 
atoms of C, N, O, Ne, Mg and Si as well as \ionn{Fe}{xvii}
lines.

Previous analyses of X-ray emission from \epsori\
found a wide spatial distribution 
of the hot plasma in the wind. \cite{cohen14} fitted
\chandra\ lines and did not find any correlation
between the onset radius for emitting plasma and the emitting ion.
This confirmed the results by \cite{leutenegger06}
who found no evidence for different spatial distributions for different ions.

Recently, \cite{cohen14} estimated the mass-loss rate 
for \epsori\ by fitting X-ray line profiles.
They reported two values of mass-loss rate: 
\massrate{2.1}{-7} and \massrate{6.5}{-7}. The first value
uses nine X-ray lines while the second value excludes three
lines that might be influenced by resonance scattering.

\epsori\ shows spectral variability in both the optical and UV.
The main variability detected in the UV is associated with
 DACs, especially in the blue wing of the \ionn{Si}{iv} and \ionn{N}{v} doublets.
Some other UV lines, including string photospheric lines, also show variability \citep{prinja02}. 
In the optical  H$\alpha$ shows strong variability 
with  changes in both shape and strength occurring on a time
scales of hours to tens of days \citep[e.g.][and references therein]{ebbets82,morel04,prinja04,thompson13}. One possible cause
of the variability is radial and non-radial oscillations
that produce a mass-loss rate modulation 
\citep{thompson13}. Our calculations show that variations
of $\pm30\%$ in \Mdot\ about our derived value can explain the
observed variations of H$\alpha$.

We chose $\epsilon$ Ori as a standard early B supergiant due to the availability 
of  high resolution X-Ray, UV and optical data. In practice,
spectral variability is common is B supergiants, and hence
unavoidable. Its variability will introduce uncertainties in 
the derived wind parameters, but these uncertainties can be qualified,
and will not affect our main aim of evaluating the consistency of the
main physical parameters of \epsori\ derived in this multi-wavelength analysis.

\section{THE DATA}\label{data}

For this work we collected optical, UV 
and X-ray data from different archives as  described 
below. The sources of the data is presented in Table \ref{tdata}.

\subsection{Optical Data}

The optical data were obtained from 
{\it POLARBASE}\footnote{http://polarbase.irap.omp.eu/},
the stellar spectra archive for the {\it NARVAL} and {\it ESPaDOns}
echelle spectropolarimeters. The former is installed in the 
T\'{e}lescope Bernard Lyot (TBL, Pic du Midi Observatory) 
and the latter at the Canada-French-Hawaii Telescope. 
A detailed description of these
instruments can be found in \cite{silvester12} and \cite{petit14}.
They are twin spectropolarimeters except for the aperture 
(2.8 arcsec for \narval\ and 1.6 arcsec for \espadons). 
Both of them have a spectral resolution
R=$\lambda/\Delta\lambda$ $\simeq$ 65000 and a spectral
coverage from 3690 to 10000 \AA. The wavelength calibration is 
performed using a Th-Ar spectra and it is refined using telluric 
bands as references for radial velocity. The instruments have two 
modes of operation: spectroscopic and polarimetric.
In the  polarimetric mode, the beam is split with a Wolloston prism, and the new beams
are then conducted to a spectrometer through fibers. 
The $I$ Stokes parameter is obtained by adding these two spectra; the rest of the Stokes 
parameters (V, Q and U) are extracted by combinations as described 
by \cite{bagnulo09}\citep[see also][]{petit14}.

The early B supergiant star \epsori\ has been observed by both of these 
instruments at different epochs. Because of its spectral variability
we selected for this work one set of 112 exposures taken by \narval\ on
October 19th, 2007 during 4 hours (universal time from archive) 
in polarimetric mode. Every exposure has a signal to noise ratio of $\sim$ 600.
The reduction for each of these exposures was automatically performed by 
the Libre-Esprit reduction pipeline \citep{petit14}.
We chose the ``$I$'' Stokes spectra from the archive for this work. All of
the exposures were combined and the resulting spectrum normalized using
soft spline functions between nodes chosen by visual exploration 
to avoid spurious oscillations. This task was performed through 
the \texttt{line\_norm} procedure from FUSE IDL 
tools\footnote{http://fuse.pha.jhu.edu/analysis/fuse\_idl\_tools.html}
This combined and normalized spectrum was used for the analysis. 

\begin{table*}
\begin{threeparttable}
\caption{Summary of data of \epsori: optical (NARVAL),
              ultraviolet (\iue, \hst\ and \copernicus) and
              X-ray (\chandra\ and \xmm).\label{tdata}}

\begin{tabular}[ht!]{ccccc}
\hline
\hline
ID Obs. & Date &  Julian Date & Spectral Range (\AA) & $\lambda / \Delta \lambda$\\
\hline
\multicolumn{5}{c}{\narval\ (Optical)}\\
\hline
-- & Oct-19-2007 & 2454392.55907 & 3690 - 10000 & 65000 \\
\hline
\multicolumn{4}{c}{\iue\ (UV)} \\
\hline
SWP30177 & Jan-28-1987 & 2446823.57668 & 1150-1975 & 10000 \\
SWP30196 & Jan-30-1987 & 2446825.56254 & 1150-1975 & 10000 \\
SWP30204 & Jan-31-1987 & 2446826.54533 & 1150-1975 & 10000 \\
SWP30216 & Feb-01-1987 & 2446827.57926 & 1150-1975 & 10000 \\
SWP30225 & Feb-01-1987 & 2446828.47190 & 1150-1975 & 10000 \\
SWP30242 & Feb-03-1987 & 2446829.64887 & 1150-1975 & 10000 \\
SWP30249 & Feb-03-1987 & 2446830.47746 & 1150-1975 & 10000 \\
SWP30257 & Feb-05-1987 & 2446831.64990 & 1150-1975 & 10000 \\
SWP30266 & Feb-06-1987 & 2446832.66278 & 1150-1975 & 10000 \\ 
SWP30272 & Feb-06-1987 & 2446833.47835 & 1150-1975 & 10000 \\
LWR02238 & Sep-01-1978 & 2443753.35590 & 1900-3080 & 14000 \\
LWR02239 & Sep-01-1978 & 2443753.40463 & 1900-3080 & 14000 \\
LWR02240 & Sep-01-1978 & 2443753.42454 & 1900-3080 & 14000 \\
\hline
\multicolumn{4}{c}{\ghrs\ (UV)} \\
\hline
Z1BW040TT & Nov-11-1994 & 2449302.5975 & 1180-1218 & 20000 \\ 
Z1BW040UT & Nov-11-1994 & 2449302.5000 & 1229-1268 & 20000 \\
Z1BW040VT & Nov-11-1994 & 2449302.5019 & 1273-1311 & 20000 \\ 
Z1BW040WT & Nov-11-1994 & 2449302.5389 & 1324-1363 & 20000 \\
Z1BW040XT & Nov-11-1994 & 2449302.5408 & 1385-1423 & 20000 \\
Z1BW040YT & Nov-11-1994 & 2449302.5428 & 1527-1564 & 20000 \\ 
Z1BW040ZT & Nov-11-1994 & 2449302.5634 & 1588-1625 & 20000 \\
Z1BW0400T & Nov-11-1994 & 2449302.5656 & 1649-1684 & 20000 \\ 
\hline                    
\multicolumn{4}{c}{\copernicus\ (UV)} \\
\hline
027 & Nov-30-1972 & 2441652.49738 & 1000-1420 & 5500\\
\hline
\multicolumn{4}{c}{\chandra\ (X-ray)} \\
\hline 
3753 & Dec-12-2003 & 2452986.00475 & 2-26 & 150-1100 \\
\hline
\multicolumn{4}{c}{\xmm\ (X-ray)} \\
\hline
0112400101 & Mar-06-2002 & 2452339.84921 &5-35 & 150-800 \\
\hline
\hline
\end{tabular}
\end{threeparttable}
\end{table*}

\subsection{UV Data}

\subsubsection{IUE}

The International Ultraviolet Explorer (IUE) has observed \epsori\ in
both low and high dispersion modes. For this work, we 
selected data obtained for the BSIGP  program (PI: Geraldine Peters) 
from the MAST\footnote{http://archive.stsci.edu/} archive.
These 10 exposures of \epsori\ were undertaken
from January 28th, 1987 to February 6th, 1987 in high dispersion 
and utilizing the large aperture with the short-wavelength prime camera (SWP). 
This instrumental configuration yields a spectral resolution 
of $\lambda/\Delta \lambda$ $\sim$10000 with a spectral 
coverage of $\lambda \lambda$1150-1975. When extracted, 
the spectra did not show strong variability, so we combined them 
to obtain the final mean spectrum. 

The IUE observation program HSCAD (PI: A.K. Dupree) has 33 
exposures of \epsori\ taken through the small aperture. 
When combined and scaled to BSIGP fluxes, no
significant differences were detected when compared with 
the higher signal-to-noise BSIGP observations.

For long-wavelength we averaged the three exposures 
made on January 9th, 1979 using the LWR camera (1900-3080 \AA)
at high dispersion ($\lambda/\Delta \lambda$ $\sim$14000) and
small aperture. This data were collected within the program
ID: UK022 (PI: P. Byrne).

\subsubsection{GHRS/HST}

Hubble Space Telescope (HST) observed \epsori\ only with the Goddard
High Resolution Spectrograph (GHRS) (programs: 3472, 6070
6541, 3859 and 6249). As the programs focused on interstellar 
abundances using specific diagnostic lines, none of the 
observations covered a wide spectral range. 
We used the observations from program 3472 
(PI: Lewis Hobbs), that were collected on November 11th, 1994.
We retrieved the calibrated data from the G160M first-order
grating at intermediate resolution (R$\simeq$25000) from the
MAST archive. We treated the data using the STSDAS package and 
related tasks from analysis software
IRAF\footnote{http://iraf.noao.edu/}.

\begin{table*}
\begin{threeparttable}
\caption{Species and ions included in \cmfgen\ models. F is the number
of levels for each atomic model and S denotes the super-levels
used in calculations. ``E\textbackslash I'' denotes Element\textbackslash 
Ionization.\label{tatomic}}
\begin{tabular}{cccccccccccccccccccccc}
\hline
\hline
E\textbackslash I & \multicolumn{2}{c}{I} & & \multicolumn{2}{c}{II} & 
           & \multicolumn{2}{c}{III} & & \multicolumn{2}{c}{IV} & & \multicolumn{2}{c}{V} &
           & \multicolumn{2}{c}{VI} & & \multicolumn{2}{c}{VII}\\
 & F & S && F & S && F & S && F & S && F & S && F & S && F & S \\         
\hline
H & 30 & 20 \\
He & 69 & 45  && 30 & 22 \\
C  & & && 92 & 40 && 84 & 51 && 64 & 64  \\
N  & & && 85 & 45 && 82 & 41 && 76 & 44 && 49 & 41 \\
O  & & && 123 & 54 && 170 & 88 && 78 & 38 && 56 & 32 \\
Ne & & && & && 71 & 28 && 52 & 17  && 166 & 37 \\
Si & & && 100 & 72 && 33 & 33 && 33 & 22 \\
P  & & && & && & && 90 & 30  && 62 & 16  \\
S  & & && & && 44 & 24 && 142 & 51 && 98 & 31 \\
Fe & & && & && 1433 & 104 && 520 & 74 && 220 & 50 && 433 & 44 && 153 & 29 \\
\hline
\end{tabular}
\end{threeparttable}
\end{table*}

\subsubsection{Copernicus}

Copernicus  observations of \epsori\ (star: 027)
were also  retrieved from the MAST archive. We selected scans 
that covered a large spectral range, and that included
\ionnl{O}{vi}{1032,1038}, \ionnl{S}{iv}{1062,1073} and
\ionnl{P}{v}{1118,1028}. We co-added those scans and generated
the spectrum using the IDL 
IUEDAC\footnote{http://archive.stsci.edu/iue/iuedac.html} 
library tools and normalized it using soft spline functions 
between selected nodes in the same sense as explained above 
for the optical data. After selecting, co-adding and normalizing, 
we obtained a spectrum with coverage of $\lambda \lambda$1000-1450.

\subsection{X-ray Data}

\subsubsection{Chandra}

\chandra\ observed \epsori\ on December, 12$^{th}$ 2003,
using the gratings of HETGS (High Energy Transition 
Grating Spectrometer) for 92 ks (PI: Wayne Waldron). 
The FWHM spectral resolutions are 23 m\AA\ and 
12 m\AA\ for Medium Energy Grating (MEG) and 
High Energy Grating (HEG) respectively. The effective area  
of these gratings  is significant for $\lambda \gtrsim$2 
\AA\ but falls strongly beyond $\lambda \gtrsim$ 16 \AA\ 
in the case of HEG and $\lambda \gtrsim$ 25 \AA\ for MEG. 
Because of the small number of counts collected in the HEG, 
we only used MEG data for this work. We reprocessed the data
using the CIAO version 4.6 tasks, following the standard 
threads as described in the CIAO documentation.
We combined the positive and negative first order MEG
spectra using \texttt{add\_grating\_orders} task.

\subsubsection{XMM-Newton}

\xmm\ observed \epsori\ on 6th March, 
2002 using the instruments: EPIC-MOS(1,2), EPIC-pn and 
the Reflection Grating Spectrometers (RGS1 and RGS2) 
(PI: Martin Turner). Our analysis 
was based only on the RGS data due to its high spectral
resolution (R$\sim$250 at 15 \AA). The exposure time 
for both spectrometers was 13 ks.

The data were reduced following standard procedures 
with SAS v13.5.0. We use the first order spectra 
of each spectrometer (RGS1 and RGS2) for the analysis. 
Together, RGS1 and RGS2 have a spectral coverage 
from 6 to 39 \AA.

\section{THE METHOD}\label{method}

The analysis was undertaken in two main stages. In the first 
stage we derived the photospheric and wind parameters using 
the standard method for massive stars and  the latest
version of \cmfgen. In the second stage, the X-ray 
emission is modeled. We then tested the consistency
of the adopted parameters with the X-ray line profiles
(mass-loss rates) and line ratios (abundances).

\subsection{Modeling Assumptions}

\subsubsection{Optical and UV analysis}

The calculations for our analysis were performed using
the code \cmfgen\ \citep{hillier98}. This code models a
spherical stellar atmosphere and stellar wind solving the
radiative transfer, radiative and statistical 
equilibrium equation system in non-LTE (NLTE) taking
into account line-blanketing effects. The transfer equation is   
solved in the co-moving frame (CMF). Table \ref{tatomic} describes 
the ionization states and number of full atomic levels and super-levels 
for each atomic species included in the models. 

The photospheric density structure is calculated 
by iteratively solving the hydrostatic equilibrium 
equation system below the sonic point as described by \cite{bouret12}.
Given an adopted mass-loss rate and the velocity profile $v(r)$,
the wind density structure is calculated using the continuity  equation. 
The adopted velocity profile is similar to that
commonly used for massive star winds \citep[e.g.][]{hillier03}:

\begin{equation}
 v(r)=\frac{2v_{tr}+(v_\infty-2v_{tr})(1-r_{tr}/r)^\beta}
      {1 + \exp{[(r_{tr}-r})/h_{eff}]},
\end{equation}

\noindent
where $v_{tr}$ and $r_{tr}$ are the transition velocity and
radius, between the photosphere and wind. The 
transition point is set as the radius where 
$v(r_{tr}) =0.75 v_s$ where $v_s{\sim}15$~\kms\ is the 
sound velocity. The term $h_{eff}=v_{tr}/(2(dv/dr)_{tr})$, 
is the scale height, $v_\infty$ is the terminal velocity 
and $\beta$ is the acceleration parameter.

Clumping is taken into account through the volume filling 
factor $f=\bar{\rho}/\rho(r)$, where $\bar{\rho}$ is the 
homogeneous (unclumped) wind density and $\rho$ is the density in
clumps. In this approach the clumps are assumed to be 
optically thin to radiation and the 
interclump medium is void \citep{hillier99}. The filling 
factor dependence with radius is defined by the relation: 
$f=f_\infty+(1-f_\infty)\exp(-v(r)/v_{cl})$, where $v_{cl}$
characterizes the velocity where clumping starts.

It is well known that we need to include the effect of
microturbulence on the photospheric and wind spectrum 
\citep[e.g.][]{howarth97}. For the formal solution, the depth dependence of 
the microturbulence velocity
was parameterized as in \cite{hillier03}: 
$\xi_{t} = \xi_{min} + (\xi_{max}-\xi_{min})v(r)/v_{\infty}$. 
Here, $\xi_{min}$ and $\xi_{max}$ are the photospheric and 
wind turbulence respectively. In this work we varied 
the $\xi_{min}$ from 10 to 20 \kms, a reasonable range 
for early B supergiant stars \citep{mcerlean98} and 
$\xi_{max}\sim$ 0.1-0.3 $v_\infty$. In the CMFGEN calculation
we used a microturbulent velocity that was independent of depth.

\subsubsection{X-ray Analysis}\label{axrays}

We assume that the X-ray emission comes from an ensemble of 
shock-heated regions within the wind (shock scenario). 
The shock scenario predicts regions distributed in the
wind where the plasma is strongly heated due to shocks 
caused by radiative instabilities \citep{owocki88,feldmeier97}. 

The current version of \cmfgen\ allows for X-ray emission 
from shocked regions in the wind using emissivity tables 
from \apec\ \citep{smith01} for different temperatures, 
but it doesn't take into account differences in abundances (generally), 
densities or the influence of the UV radiation on the populations of the
levels that give rise to the X-ray lines. 

Recently Zsarg\'{o} et al. (2015, in preparation)
developed a new version called \xcmfgen\ that makes possible
a consistent interaction between the code \apec\ and \cmfgen.
Here we will briefly  describe the main features of the method 
to compute the synthetic X-ray spectrum taking into account 
the wind radiation from \cmfgen.

As a convenient approximation, a set of $N_p$ plasmas, 
characterized by their temperature  (T$_X$), the radius
where the plasma emission starts (\Ro), and the X-ray filling factor ($f_X$),
are distributed within the wind. The filling
factor is parametrized as follows:

\begin{equation}
 f_{X-ray}(r)=
 \begin{cases} 
    f_X & \text{if } r \geq R_0 \\ 
    f_Xe^{-(v(R_0)/v(r))^2(R_0/r)^4} & \text{if } r < R_0,  
 \end{cases}
\end{equation}

\noindent
where $v(r)$ is the wind velocity profile and 
$f_X$ is the filling factor at infinity.
An iterative process is then performed to derive
the best estimates of the plasma parameters. 
A \xcmfgen\ run is done to calculate the stellar 
wind structure. Then, \apec\ is called in order to calculate 
a grid of plasma emissivities covering the wind densities 
and plasma temperatures. 
The relation between the plasma and wind density is set 
as: $n_{plasma}$=4$n_{wind}$ (adiabatic shock approximation). 
Here $n_{wind}$ corresponds to the density of the 
unclumped model. The relation between the plasma
and wind density only has a very minor influence on the analysis --
it's primary effect will be in the interpretation of the derived
filling factors (e.g.~what fraction of the wind is shocked).

The Astrophysical Plasma Emission Code (\apec) is a   
code for calculating level populations and emissivities 
for a hot plasma in collisional equilibrium. 
This code uses an extensive library of cross-sections 
(ATOMDB\footnote{www.atomdb.org}). In this work we use 
this library as it is described in \cite{foster12}.
Some changes were implemented in \apec\ in order to read the
\xcmfgen\ outputs as well as to use the corresponding abundances.
The main change was the inclusion of the radiative excitation
(i.e. UV photoexcitation) and deexcitation rates in the 
statistical equilibrium equations. The UV emission from
the wind affects the forbidden to intercombination ratio
R=f/i of the He-like triplets \citep{gabriel69,porquet01}.

Once \apec\ has calculated the grid of plasma emissivities, a new 
run of \xcmfgen\ is performed including the emissivities 
from \apec. This process yields a new radiation field that
is used in a second \apec\ computation. Usually, 
it is necessary for  only two loops (i.e.,
\xcmfgen\ - \apec\ - \xcmfgen\ - \apec\ - \xcmfgen) to get a consistent model.

A necessary condition to apply this method is that the X-rays
do not strongly affect the bulk of the cool wind. As noted previously by 
\cite{martins05} and \cite{macfarlane94}, this condition holds 
for early O supergiants. However, for late O and B stars X-rays
could change the wind ionization in a non-negligible way. An analysis
of that influence in the case of \epsori\ is presented 
in Section \ref{dxrays}. We conclude that for clumped wind
models this effect is low enough to apply this method. 
Thus, we applied the same fit procedure described 
by Zsarg\'{o} et al. (2015, in preparation) in their study of $\zeta$ Pup and 
did the X-ray analysis separately from the optical and UV analysis. 

We created a two-dimensional grid in
(T$_X$,R$_0$), varying each of them  between the following ranges: 
T$_X$=\scie{1.0-16.0}{6} K and R$_0$=1.2-6.1R$_*$. We then used 
\cmfflux\ to compute the X-ray spectrum for each (T$_X$,\Ro) 
combination. The starting filling factor is 
fixed at $f_X$=\scie{5}{-5}; the corresponding value for each 
plasma component is adjusted subsequently as described below.

The total X-ray spectrum results from adding the contributions
from each plasma component after it has propagated through 
the wind. We adjusted the  
plasma distributions (i.e. number of components, T$_X$,
R$_0$ and $f_X$) using the fitting package \xspec\ 
\citep{arnaud96} of HEASARC\footnote{http://heasarc.gsfc.nasa.gov/}.
A multiplicative model is included to take into account the effect of
interstellar absorption on the X-ray spectrum. 
The column density of neutral H used in this work is 
$\log [N(\textrm{\ionn{H}{i}})]$ cm$^{-2}$=20.48 \citep{diplas94}. 
As the molecular absorption in the direction of \epsori\ 
is low it doesn't affect the X-ray spectrum and hence is ignored.

Because of the small number of counts in the \chandra\ and \xmm\ 
observations, Poisson noise dominates, and we use 
the C-statistic \citep{cash79} to determine the quality 
of the fits.  While an arbitrary number of components 
can be included, the final models were typically composed 
of only four different temperatures.

The X-ray lines included in the analyses are listed in Table \ref{txlines}.

\begin{table}
\caption{\boldmath$\epsilon$~Ori Chandra and XMM-Newton lines. 
Data from ATOMDB.\label{txlines}}
\begin{tabular}[h!]{lccc}
\hline\hline
 Ion & Wavelength (\AA)  & Type & T$_{peak}$ [10$^6$ K] \\
\hline
 \ionn{Si}{xiv}  & 6.1822  & H-like   &  15.8\\
        \ionn{Si}{xiii} & 6.6479, 6.6866, 6.6866  & He-like  &  10.0\\
        \ionn{Al}{xii}  & 7.7573, 7.8070, 7.8721 & He-like  &  7.94 \\
        \ionn{Mg}{xi}   & 7.8503  & He-like  &  6.31\\
        \ionn{Mg}{xii}  & 8.4210  & H-like   &  10.0\\
        \ionn{Mg}{xi}   & 9.1687, 9.2297, 9.3143  & He-like  &  6.31 \\
        \ionn{Ne}{x}    & 10.2388 & H-like   & 6.31 \\        
        \ionn{Ne}{ix}   & 11.5440 & He-like  & 3.98 \\        
        \ionn{Ne}{x}    & 12.1339 & H-like   & 6.31 \\
        \ionn{Fe}{xvii} & 12.2660 & L-shell  & 6.31 \\
        \ionn{Ne}{ix}   & 13.4470, 13.5520, 13.6980 & He-like  & 3.98 \\
        \ionn{Fe}{xvii} & 14.2080 & L-shell  &  7.94 \\
        \ionn{Fe}{xvii} & 15.0140 & L-shell  &  6.31 \\
        \ionn{Fe}{xvii} & 15.2610 & L-shell  &  6.31 \\
        \ionn{Fe}{xvii} & 16.0040 & L-shell  &  7.94 \\
        \ionn{O}{viii}  & 16.0060 & H-like   &  3.16 \\
        \ionn{Fe}{xvii} & 16.7800 & L-shell  &  6.31 \\
        \ionn{Fe}{xvii} & 17.0510 & L-shell  &  6.31 \\
        \ionn{Fe}{xvii} & 17.0960 & L-shell  &  6.31 \\
        \ionn{O}{vii}   & 18.6270 & He-like  &  2.00 \\
        \ionn{O}{viii}  & 18.9670 & H-like   &  3.16 \\
        \ionn{O}{vii}   & 21.6020, 21.8040, 22.0980 & He-like  &  2.00 \\
        \ionn{N}{vii}   & 24.7790 & H-like   &  2.00 \\
        \ionn{N}{vi}    & 24.8980 & He-like  &  1.59 \\
        \ionn{C}{vi}    & 26.9900 & H-like   &  1.59 \\
        \ionn{C}{vi}    & 28.4650 & H-like   &  1.59 \\
        \ionn{N}{vi}    & 28.7870, 29.0810, 29.5340& He-like  &  1.58\\
        \ionn{C}{vi}    & 33.7370 & H-like   &  1.26\\
        \hline
\end{tabular}
\end{table}

\subsection{Stellar Parameters}

We used the optical spectrum to estimate the photospheric 
parameters such as the effective temperature, gravity and 
surface abundances. The method follows the analysis
route described by \cite{bouret12} \citep[see also][]{martins05,bouret13}. 
We adopted a radial velocity $v_r$=25.9$\pm$0.9 \kms\ \citep{evans67}.

A brief description of steps taken to derive the fundamental 
stellar parameters ($T_{eff}$, $\log g$, and CNO, Fe 
and Si abundances) follows.

\subsubsection{Luminosity}

We adopt two distances: the distance $d=$411.5$^{+245}_{-112}$ pc 
\citep{perryman97} from the old \hipparcos\  
catalog (HIP 26311)\footnote{http://vizier.u-strasbg.fr/viz-bin/VizieR-2},
and $d$=606$^{+227}_{-130}$ pc \citep{vanleeuwen07} 
who reanalyzed \hipparcos\  data.
The errors on the distance are approximately  30-50\%, which 
will yield an error on the luminosity of 60-100\%, and a similar, 
but somewhat smaller, uncertainty on the mass-loss rate.

A first estimation of luminosity was made based on 
$M_V$ and the bolometric correction (BC). The last one 
was taken from the relation between T$_{eff}$ and BC 
calculated by \cite{crowther06}(Fig.~4).
This relation comes from their sample of Galactic
B supergiants and from the SMC B supergiants reported 
by \cite{trundle04} and \cite{trundle05}. For this first
approach we used the temperature estimated by 
\citeauthor{crowther06} for $\epsilon$~Ori, 
$T_{eff}$=\scie{2.7}{4} K. The $M_V$ value was estimated
from the distance values from \hipparcos, the 
apparent visual magnitude $V$=1.69 \citep{lee68} and
the interstellar visual extinction $A_V$=3.1 \ebv,
with \ebv\ taken from the \citeauthor{crowther06}
study, \ebv=0.06. Thus, $M_V$=$-6.57$ for $d$=411.5 
pc and $M_V$=$-7.41$ for $d$=606.06 pc. The luminosity
is given by:

\begin{equation}
 \log(L_*/L_\odot)=(M_{B\odot}-M_V-BC)/2.5,
\end{equation}

\noindent
where $M_{B\odot}$ is the solar bolometric magnitude
($M_{B\odot}$=4.75).

Once the temperature and gravity were estimated, 
we used the UV flux, and the optical magnitudes ($UVB$) in 
order to re-estimate simultaneously the luminosity
and \ebv\ values.

\subsubsection{Gravity and Effective Temperature}

For estimating the effective temperature and gravity ($\log g$) 
a small grid of models encompassing \teff=\scie{2.5-2.9}{4} K 
and $\log g=2.8$ to 3.4, with steps of $0.1\times 10^4$ K and 0.1 dex 
respectively, were computed. The gravity was estimated by fitting the
wings of H$_\beta$, H$_\gamma$, H$_\delta$ and H$_\epsilon$
-- the lines were equally weighted for the
gravity analysis. Because of the mass loss rate of \epsori,
these Balmer lines are only very weakly influenced by wind emission.
In order to reduce  the degeneracy between $\log g$ and \teff, 
the  equivalent widths for \ionn{He}{i} and \ionn{He}{ii} 
lines were also utilized. The lines used were: 
\ionnll{He}{i}{4010, 4389, 5049} and \ionnll{He}{ii}{4201, 4542}. 
These lines show a weak dependence on the microturbulence.

For computing the effective temperature we used the 
ionization balance of both \ionn{He}{i} to \ionn{He}{ii} 
and \ionn{Si}{iii} to \ionn{Si}{iv}. We used the same grid described above to fit
the \ionnl{Si}{iv}{4090}, \ionnl{Si}{iv}{4115} and \ionnl{Si}{iii}{4554-76} lines
beside the \ionnll{He}{i}{4471, 4389, 4922} and the  \ionn{He}{ii} lines described above. 
UV lines, especially those belonging to \ionn{Fe}{iv}
(1550 - 1700 \AA), \ionn{Fe}{v} (1360 - 1385 \AA) 
as well as \ionn{Fe}{iii} (1800-1950 \AA), were used to check the result.

\subsubsection{Microturbulence}

To estimate the photospheric microturbulent velocity we calculated spectra for 
three values of  $\xi_{min}$ (10, 15 and 20 \kms) and fitted the equivalent
width of  the 4472, 5017 and 6680\,\AA\ \ionn{He}{i} lines,  
and the \ionn{Si}{iii} triplet at 4554-4576\,\AA. 
Our exploration of synthetic spectra shows that these 
lines depend strongly on the $\xi_{min}$ value and are 
weakly influenced by blending or abundance effects.

\subsubsection{Surface Abundances}

Once the temperature and gravity were constrained, we determined 
the surface abundances of the main species, namely, He, C, N, O, Si 
and Fe as follows:

\noindent
{\it Helium. }
As pointed out by \cite{mcerlean99,mcerlean98},
it is difficult to determine the He abundance in B stars primarily
because triplet and singlet lines of \ionn{He}{i} yield 
different He abundances. \cite{najarro06} showed that some \ionn{He}{i} 
singlets are influenced by a strong interaction of the He\,{\sc i} 
resonance transition at $\lambda\lambda$4923,5017 with UV \ionn{Fe}{iv} lines. 
This interaction directly affects \ionn{He}{i} 2p~$1$P$^{\rm o}$, which
influences the strength of optical singlet lines that makes them  
unreliable diagnostics. \citeauthor{najarro06} suggest using 
triplet lines for analysis. Based on these issues we decided 
to set the He abundance, expressed as $y=N[He]/N[He+H]$, to
the solar value $y$=0.091 for every one of our models.

\noindent
{\it Carbon. }
In the optical, we used lines of \mbox{\ionnll{C}{ii}{4267,6578-82}}, 
\ionnll{C}{iii}{4070,5696,4648-50} and \ionnll{C}{iv}{5802-12}.
In the case of \ionnl{C}{iii}{4648} and \ionnl{C}{iii}{5696},
\cite{martins12} showed that these lines have strong interactions
with the UV lines \ionnl{Fe}{iv}{538} and \ionnl{C}{iii}{538},
hence, they need be treated carefully. 
UV lines were used to check the abundance deduced from optical lines.
 We mainly used \ionnl{C}{iii}{1176}. The commonly
used line \ionnl{C}{iv}{1169} is not detected in UV data, since
it is blended with \ionnl{C}{iii}{1176}. The 
line \ionnl{C}{iii}{1247} is blended with \ionnl{N}{v}{1238-42}
emission and only the red emission can be used as a diagnostic.

\noindent
{\it Nitrogen. }
To estimate the nitrogen abundance we used the 
blend-free lines \ionnll{N}{ii}{3995, 4042-45, 4238-42, 
5002-06, 5677-81}, and \ionnll{N}{iii}{4380, 4635}. 
As a consistency check we also examined \ionnl{N}{ii}{5046} 
and \ionnll{N}{iii}{4098, 4512-18, 4643, 4868}. The former line
is blended with \ionnl{He}{i}{5049} while \ionnl{N}{iii}{4098} is 
blended with H$_\delta$. In the UV we used 
\ionnll{N}{iii}{1183-85} and \ionnll{N}{iii}{1748-52}. 
The \ionnl{N}{iv}{1718} line was excluded because 
it is affected by the wind and \ionnl{N}{v}{1240} 
is X-ray sensitive, and hence not suitable 
as an abundance diagnostic.

\noindent
{\it Oxygen. }
$\epsilon$~Ori shows a variety of \ionn{O}{ii} and \ionn{O}{iii} lines  ---
our abundance analysis was based on: \ionnll{O}{iii}{4368, 5594} and 
\ionnll{O}{ii}{4077, 4134, 4663}. Other lines that show abundance
dependence are those in the blend \ionnl{O}{ii-iii}{4415-18}. 
The UV lines commonly used for abundance diagnostics are 
\ionnll{O}{iv}{1338-43} and \ionnll{O}{iii}{1150-54} but they
show only a weak abundance dependence in the models. \ionnl{O}{v}{1371}
was not detected in the UV spectra of $\epsilon$~Ori.

\noindent
{\it Silicon. }
The silicon abundance was estimated using only \ionn{Si}{iii} and
\ionn{Si}{iv} lines since \ionn{Si}{ii} lines were not detected. 
The triplet \ionnll{Si}{iii}{4552-67-74} and \ionnl{Si}{iii}{5738}
were useful as well as \ionnll{Si}{iv}{4089, 4116}.
No UV lines were used for the analysis due to 
their strong dependence on the wind parameters.

\noindent
{\it Iron. }
We only used UV Fe lines for estimating its abundance. The
iron features \ionnll{Fe}{v}{1360 - 1385}, 
\ionnll{Fe}{iv}{1550 - 1700}, and \ionnl{Fe}{iii}{1800-1950}
were used. Again, it is important to have a reliable value 
for temperature to get an accurate abundance estimation.

\subsubsection{Rotation and Macroturbulence}

We account for the influence of rotation on the stellar spectrum in two ways.
First, we convolve a rotational broadening
profile with the synthetic spectrum.
This method assumes solid rotation of both the star and wind,
and that the line profile does not vary from the center to the limb \citep{gray08}.
Wind-free line profiles were used to estimate the projected
rotation velocity ($v_{rot} \sin i$).

In the second technique we calculate the synthetic spectrum using
the 2D code developed by \cite{busche05}. This code computes
the radiative transfer through the photosphere and wind using
an axysimmetric geometry and allows for rotation of the star. 
Recently, \cite{hillier12} studied the application of the code 
to optical line profiles of O stars.
They showed that while the convolution method that is commonly used to allow for
rotation is adequate for photospheric absorption lines it fails
for lines influenced by photospheric emission, or for lines with 
wind emission contribution. For the diagnostic lines H$\alpha$ 
and \ionn{He}{ii} $\lambda 4686$ it is important that the influence of 
rotation be correctly modeled.

Because the rotation rate of $\epsilon$~Ori is low we can use the same
simplifications as \cite{hillier12} -- we assume the rotation does 
not affect the density structure, temperature and surface gravity. 
We also assume that the star rotates as a solid 
body below $v(r)$=20 \kms\ and the angular momentum about the 
center is conserved above that value.

Finally, in order to compare the synthetic spectra  
with the optical observational data, it is necessary 
to take into account  instrumental broadening and macroturbulence.
Instrumental broadening was taken into account by convolving 
a Gaussian profile of 4.0 \kms\ from NARVAL spectral 
resolution. Macroturbulence broadening was included through a 
convolution assuming an isotropic Gaussian distributed 
velocity field. The value of the FWMH of such a distribution 
in \kms\ was estimated using the wings of isolated wind-free 
metallic lines.

\subsection{Wind Parameters}\label{wind}

The parameters that describe the stellar wind are those that
have to do with the wind velocity law ($\beta$ and $v_\infty$),
the mass-loss rate (\Mdot), the filling factor ($f_\infty$, $v_{cl}$)
and the wind turbulent velocity ($\xi_{max}$). All of these
influence  spectral features in the UV, optical, and X-ray band. Some line profiles are strongly
influenced by two or more of the parameters. For example,
the strength of H$\alpha$ is highly sensitive to both the mass-loss rate and 
filling factor, and its profile shape is sensitive to $\beta$.
Thus, these wind parameters need to be determined simultaneously
using several spectral features from different ions and species.

The value of \Mdot\ is first constrained using the H$\alpha$ 
strength for $f_\infty$= 1.0, 0.1, 0.05 and 0.01
($f_\infty$=1.0 means smooth wind). A consistency
check was then done using the UV line profiles of \ionnll{C}{iv}{1548,1551}, 
\ionnl{N}{v}{1240}, \ionnll{Si}{iv}{1394,1403}
and \ionn{C}{iii} $\lambda 1776$, which showed only a
weak dependence on mass-loss rate (in our models) when compared with H$\alpha$.

The clumping factor ($f_\infty$) is determined using
\ionnll{S}{iv}{1062,1073}, \ionnll{P}{v}{1118-28} and \ionnl{N}{iv}{1718}. 
The parameter $v_{cl}$ 
has a slight influence on the H$\alpha$ shape, so its value
is adjusted only to improve the line profile 
once the \.{M} is estimated from line strength. \cite{hillier03} 
suggested that its value should be less than 100 \kms. Likewise,
\cite{bouret03} \citep[see also][]{bouret05} suggested 
that clumping should start close to the photosphere 
($v_{cl}\sim$30 \kms). For the models in this work, 
we used $v_{cl}$ values from 20 to 100 \kms.

The velocity profile parameters ($v_{\infty}$, $\beta$) were 
constrained using UV P~Cygni line profiles and the  H$\alpha$ 
line profile. The H$\alpha$ profile shape is strongly 
sensitive to $\beta$ -- we used values from 1.0 to 2.4 
to find the best profile. Once we alter $\beta$ it is necessary to re-tune
the mass-loss rate to match the H$\alpha$ intensity.

In the same sense, $v_\infty$ was estimated using the blue 
wing of the P~Cyg profile of \ionnll{C}{iv}{1548,1551}
and the \ionnll{Si}{iv}{1394,1403} profile. 
As the shape of that wing is also affected by the wind 
turbulence $\xi_{max}$  we tune  $v_\infty$ and 
$\xi_{max}$ simultaneously.

\subsection{X-ray -- Wind Parameters}

As X-rays propagate through the wind they can be absorbed
at the same time photoionize the gas. The effect on an X-ray
line is an attenuation of the profile on the red side,
because the red-shifted photons from the rear hemisphere
encounter a longer path length through the wind than
the blue-shifted photons from the front hemisphere
\citep{owocki01,macfarlane91}. 
Since the X-ray opacity varies with wavelength
the influence on line profiles varies with wavelength.
Furthermore, the strength of the effect  depends primarily on 
the wind column density and hence the mass-loss rate. Thus, with a 
known opacity distribution, it is possible to estimate
the mass-loss rate from the observed X-ray lines \citep{cohen14}.

In this work we utilize X-ray cross-sections 
from \cite{verner95}. The spatial variation in opacity 
is determined by the velocity law and mass-loss rate.
We use the same values
estimated from the optical and UV data.
Once the best fit is attained, the line profiles are
checked for consistency. Previous \.{M} estimations 
from X-ray lines used individual line profiles 
\citep{oskinova06,cohen14}; in this work we use models 
for the whole X-ray spectrum. A similar approximation 
was developed by \cite{herve13}, but they
used a fiducial radial dependence for opacity.

The effects of $v_\infty$ and  $\beta$ on X-ray profiles have 
been studied by \cite{cohen10}. Here, we use the same values 
obtained from the optical and UV analysis. We show below that 
$\beta$ can influence R$_0$, and this gives us the opportunity 
to check our results against the instability wind models 
\citep{feldmeier97,runacres02,dessart03,dessart05} relating the 
wind acceleration with the shock formation region.

\subsection{X-ray -- Abundances}

Once global best fit to the X-ray spectra is obtained, the abundances
are adjusted so as  to improve the fit of line ratios. 

To compute the observed line strength ratios, the 
line fluxes were measured using Gaussian profiles 
and the statistical tools from \xspec. The free parameters 
for each fit were the normalization factor and 
the line width ($\sigma$ of each Gaussian). 
The line center is generally fixed at its laboratory value,
but when necessary it was shifted to match the peak 
of the observed line -- it was not included as a free 
parameter in the fit. We will specifically focus on the 
CNO line ratios. Recently Zsarg\'{o} et al. (in preparation) 
showed that good choices are:

\begin{equation}\label{eqrationo}
 R(N/O)=\frac{Flux(\mbox{\ionn{N}{vii} } Ly\alpha)}
 {[Flux(\mbox{\ionn{O}{viii} } Ly\alpha) + Flux(\mbox{\ionn{O}{vii} } He-Like)]},
\end{equation}

\noindent
and

\begin{equation}\label{eqratiocn}
 R(C/N)=\frac{Flux(\mbox{\ionn{C}{vi} } Ly\alpha)}
 {[Flux(\mbox{\ionn{N}{vii} } Ly\alpha) + Flux(\mbox{\ionn{N}{vi} } He-Like)]},
\end{equation}

\noindent
that have a weak dependence on temperature and are good choices 
to test the consistency of our abundance values from the optical and 
UV analysis. New abundances can be estimated when the observed
ratios are compared with the ones from our models.

\begin{table*}
\begin{threeparttable}
\caption{Stellar parameters of \epsori\ from optical and UV data.\label{tpara}}
\begin{tabular}{cccccccccc}
\hline\hline
Log (L$_*$/L$_\odot$)\tnote{d} & T$_{eff}$ & $\log g$ & R$_*$ & M$_*$
& $\xi_{min}$ &$v\sin i$ & $v_{macro}$ & \ebv & {\it d}\\
& 10$^3$ [K] &  & \rsun & M$_\odot$ & \kms & \kms & \kms & mag  & pc\\
\hline
5.59$^{+0.51}_{-0.20}$ & 27.0$\pm$0.5 & 3.0$\pm$0.05 & 28.6 & 30.0 & 15-20 & 40-70\tnote{c} 
&70-100 & 0.091 $\pm$ 0.01 & 412\tnote{a} \\
5.92$^{+0.32}_{-0.18}$ & 27.0$\pm$0.5 & 3.0$\pm$0.05 & 42.0  & 64.5 & 15-20 & 40-70\tnote{c} 
&70-100 & 0.091 $\pm$ 0.01 & 606\tnote{b} \\
\hline
\end{tabular}
\begin{tablenotes}
 \item[a] HIPPARCOS catalog \citep{perryman97}
 \item[b] HIPPARCOS catalog \citep{vanleeuwen07}
 \item[c] Value computed using the convolution method.
 \item[d] Errors from distance uncertainties.
\end{tablenotes}
\end{threeparttable}
\end{table*}

\section{RESULTS}\label{results}

\subsection{Stellar Parameters}

The main stellar parameters that we estimated for \epsori\ are 
shown in Table~\ref{tpara}. Each row in the table corresponds 
to the parameters from the two different distances from the
\hipparcos\ catalogs. The luminosity from the lower
distance lies between those calculated by \cite{crowther06}
($\log(L_*/L_\odot)$=5.44) and \cite{searle08} 
($\log(L_*/L_\odot)$=5.73) and is close to other estimates 
reported in the literature \citep[e.g.][]{lamers74,kudritzki99,mcerlean99}.
 The luminosity computed with the
larger distance is the highest value reported and makes \epsori\ 
one of the most luminous B-supergiants in the Galaxy. This
value is almost twice the calibrated value found by \cite{searle08}
for galactic B0Ia stars (They used jointly their sample and 
that of \cite{crowther06}.). A similar statement applies to the mass-loss
rate and stellar radius (mass). Thus the distance of \cite{vanleeuwen07} makes
\epsori\ an unusual B0Ia supergiant. 
For the analysis, we prefer to use the lower value in order to compare
with the previous work on \epsori\ that were undertaken
assuming a distance closer to that of \citeauthor{perryman97}.

\begin{figure}
 \includegraphics[trim=0 0 -50 0,clip,width=\linewidth]{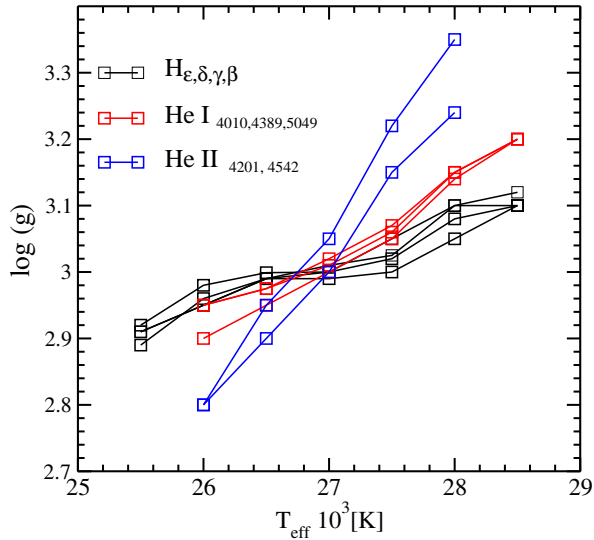}
 \caption{Constraints on \teff\ and $\log g$. Shown are fits to the Balmer line wings (black),
 the equivalent width of \ionn{He}{i} (red) and \ionn{He}{ii} (blue) lines.}
 \label{fgrav}
\end{figure}

Figure \ref{fgrav} shows the analysis of Balmer line wings and \ionn{He}{i-ii} lines. 
We concluded that $\log g$=3.0. We also found that for values greater than 
3.05 and lower than 2.95 it is not possible to fit the Balmer
and \ionn{He}{i-ii} lines consistently. Thus, we estimate an error 
of 0.05 for $\log g$.  

\begin{figure*}
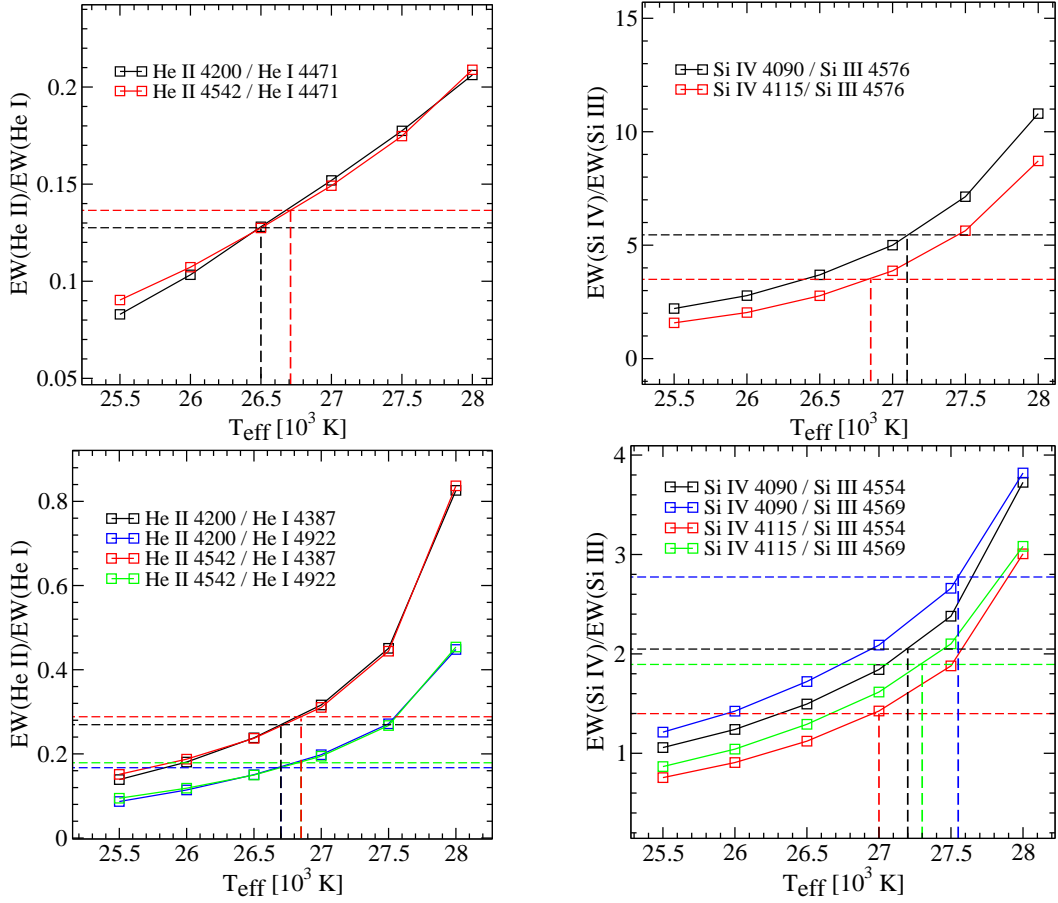

 \includegraphics[clip,width=0.36\linewidth]{ftempa.eps} \hspace{1cm} 
 \includegraphics[clip,width=0.35\linewidth]{ftempb.eps} \\
 \includegraphics[clip,width=0.36\linewidth]{ftempc.eps} \hspace{1cm}  
 \includegraphics[clip,width=0.35\linewidth]{ftempd.eps} 
 \caption{Ionization balance for \ionn{He}{i-ii} and \ionn{Si}{iii-iv}. The ratio of
          equivalent widths (EWs) for selected pairs of He and Si lines are 
          shown for models with $\log g$=3.0 and T$_{eff}$~=\scie{2.55-2.8}{4}~K. 
          Each square represents the model that matches the observed ratio 
          of the equivalent widths (EWs) of the lines. The horizontal dashed 
          lines show the equivalent width ratios 
          measured from optical data, while the vertical dashed lines show
          the projection of the matching model EW to the temperature axis. 
          We estimated the measured EW ratio errors. The maximum is around
          1 \%, which would influence our lower temperature limit 
          by $\lesssim$100 K.
          }\label{ftemp}
\end{figure*}

Figure \ref{ftemp} shows the analysis for the ionization balance 
of \ionn{He}{i-ii} and \ionn{Si}{iii-iv}.      
The ionization balance of \ionn{He}{i-ii} yields 
an effective temperature approximately $2.65 \times 10^4$ to $2.68 \times 10^4$\,K  while the
\ionn{Si}{iii-iv} ionization balance yields a value $\sim$ $2.68 \times 10^4$\, to $2.75  \times 10^4$\,K. 
From these results we concluded that the effective temperature
of \epsori\ is T$_{eff}$=27000$\pm$500\,K.        
In the UV, \ionnl{Si}{iii}{1312} shows a strong dependence on
temperature. This line yields a slightly lower temperature of
T$_{eff}$=26200 K. Nevertheless, \ionn{Fe}{iv-v}
lines and the \ionnll{O}{iv}{1342,1343} lines support a value of 27000\,K.
The \ionnll{Si}{iii-iv}{1417,1416}  lines provide a temperature estimate  of 27500\,K.

\begin{table*}
\begin{threeparttable}
\caption{Wind parameters of \epsori\ from optical and UV data. \label{twpara}
The mass-loss rate for both \hipparcos\ distances is given.}
\begin{tabular}{ccccccccc}
\hline\hline
Model & $f_\infty$ & \Mdot/$\sqrt{f_\infty}$\tnote{a} & \Mdot/$\sqrt{f_\infty}$\tnote{b}
& $v_{\infty}$ & $\beta$ & $\xi_{max}$ & $v_{cl}$ \\
           &  & \sunyr & \sunyr & \kms &  & \kms & \kms \\
\hline
A & 0.1  & \scie{1.55}{-6}  & \scie{2.70}{-6} & 1800 & 2.0 & 200 & 40  \\
B & 0.05 & \scie{1.66}{-6}  & \scie{2.80}{-6} & 1800 & 2.0 & 200 & 30  \\
C & 0.01 & \scie{1.75}{-6}  & \scie{2.90}{-6} & 1800 & 2.0 & 200 & 20  \\
D & 1.0  & \scie{1.40}{-6}  & \scie{2.55}{-6} & 1800 & 2.0 & 200 & --  \\
\hline
\end{tabular}
\begin{tablenotes}
 \item[a] HIPPARCOS catalog \citep{perryman97} ($d$=412 pc)
 \item[b] HIPPARCOS catalog \citep{vanleeuwen07} ($d$=606 pc)
\end{tablenotes}
\end{threeparttable}
\end{table*}

\begin{figure*}
 \centering
 \includegraphics[width=0.45\linewidth]{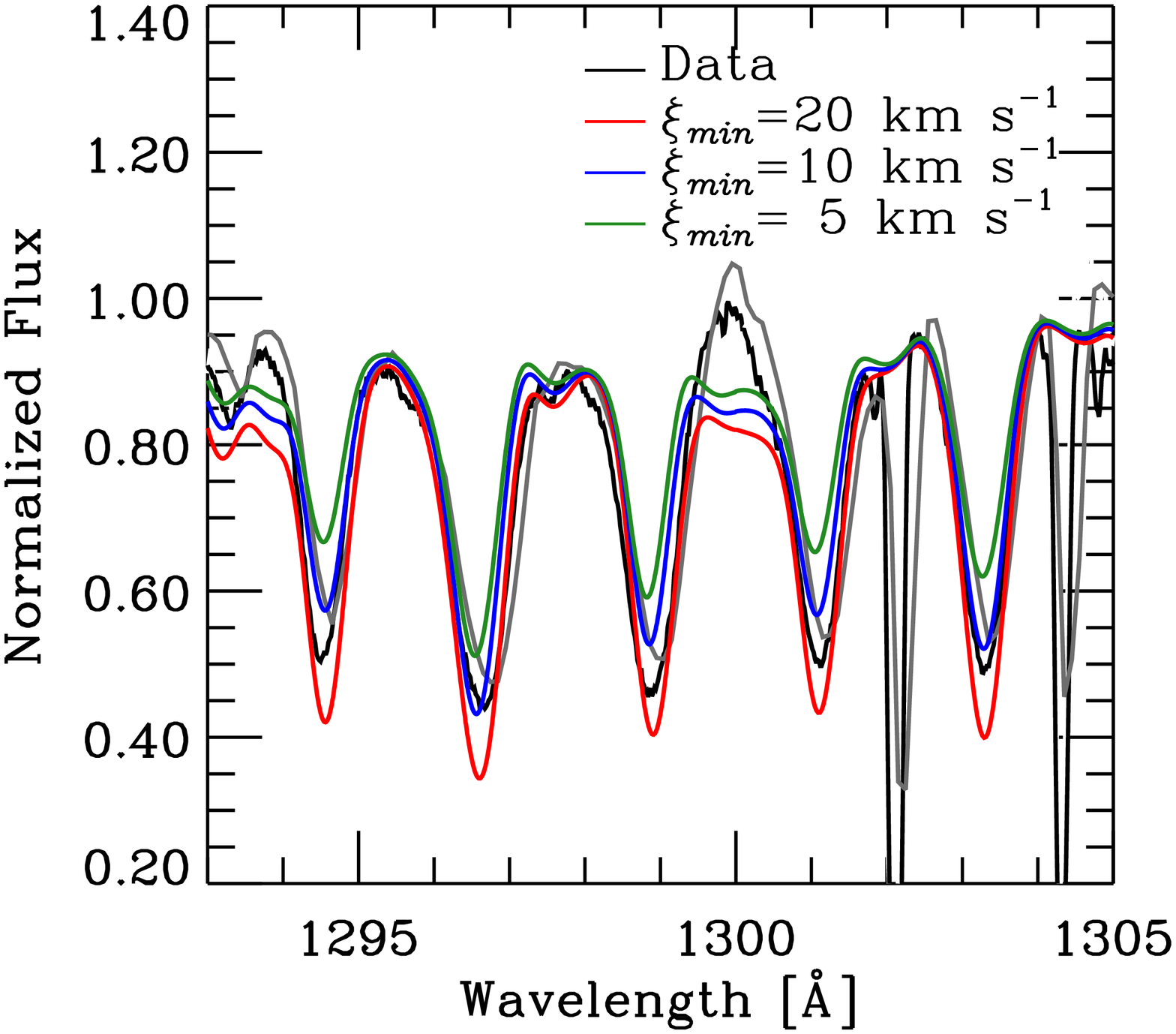}
 \includegraphics[width=0.45\linewidth]{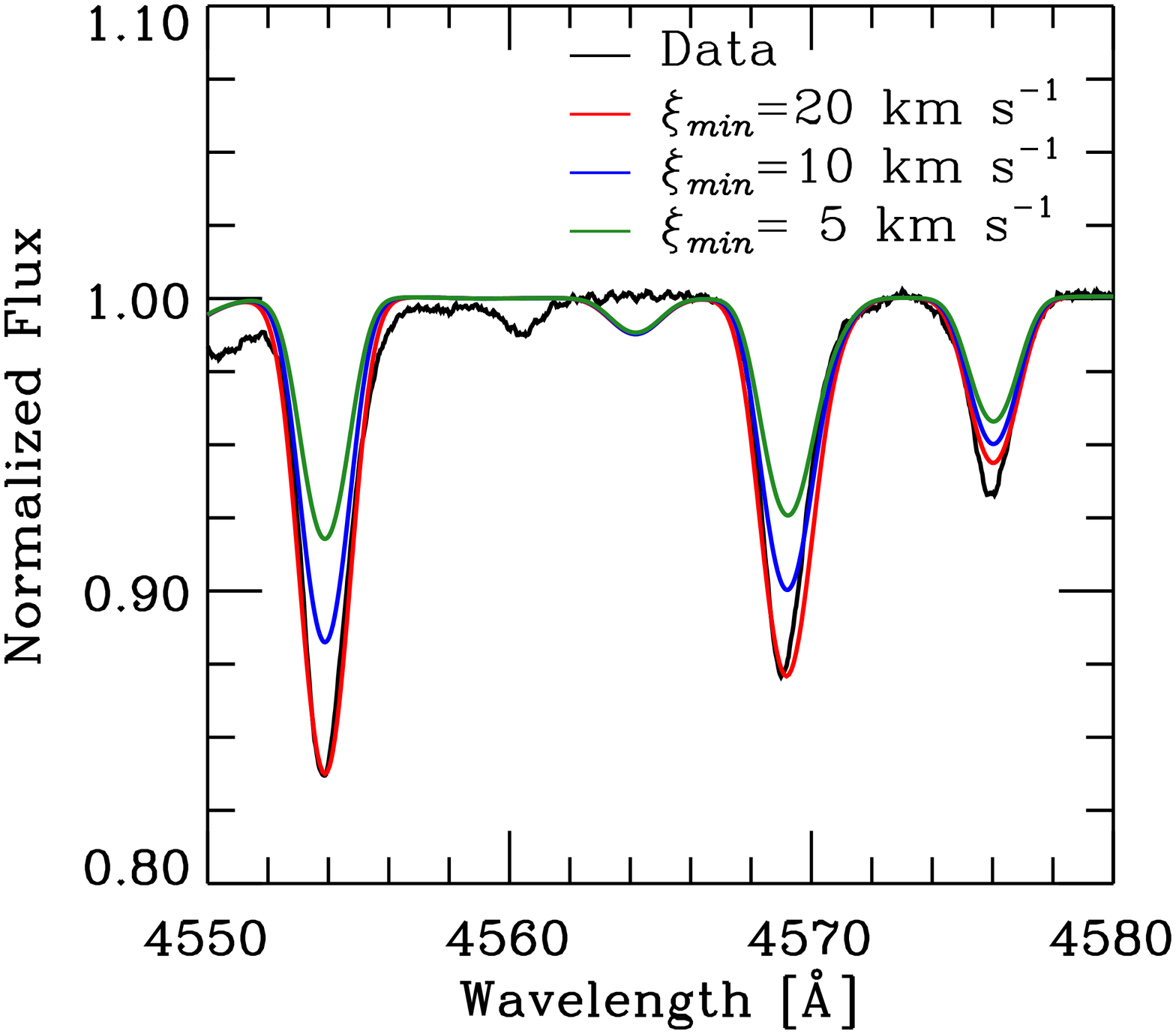}
 \caption{Dependence of UV and optical \ionn{Si}{iii} lines on photospheric
          microturbulence ($\xi_{min}$). In the left panel, the black line represents
          the GHRS data and the gray line \copernicus\ data.
          Profiles for three photospheric
          microturbulent velocities are shown: 20 \kms\ (red), 
          10 \kms\ (blue) and 5 \kms\ (green).}\label{fturb}
\end{figure*}

Photospheric microturbulence velocity ($\xi_{min}$) was estimated to be between 15 
and 20 \kms\ using \ionn{He}{i} lines. The silicon triplet 
(4554-4576 \AA) suggests a $\xi_{min}$ close to the lower limit, 15 \kms\ 
while the UV \ionn{Si}{iii} multiplet indicates a microturbulence closer to
10\,\kms (Fig.~\ref{fturb}). The variations do not change the reported limits
for temperature and gravity.

Figure \ref{fturb} shows the line profiles for 
optical and UV \ionn{Si}{iii} lines. The models 
correspond to three values of turbulence velocity:
20 \kms\ (red), 10 \kms\ (blue) and 5 \kms\ (green).
We found divergence between the UV and optical
silicon lines -- the UV (\ionnll{Si}{iii}{1290-1310}) complex
suggests $\xi_{min}\sim$10 \kms\,
while the optical triplet \ionnll{Si}{iii}{4550-4580}
requires a value of $\xi_{min}\sim$20 \kms.
Furthermore, the \ionnll{Fe}{iv}{1450-1500} and 
\ionnll{Fe}{iv}{1600 -1700} lines rule out $\xi_{min}$
lower than $\sim$10 \kms.

The discrepancy between UV and optical \ionn{Si}{iii} lines
can be overcome by increasing the silicon abundance by 
a factor of $\sim$1.2 with $\xi_{min}$=15 \kms. However, 
such a value is incompatible with other
optical lines (e.g. \ionnl{N}{iii}{4098}  and \ionnll{C}{iii}{4648-4654}).

The rotation and macroturbulence velocities were estimated
simultaneously from fitting the line shapes. We estimated values of 
40-70 \kms\ for rotation and 70-100 \kms\ for $v_{macro}$.
That rotation value is lower than previously 
reported ($v\sin i =91$\,\kms\ by \cite{howarth97}).
However upon close examination we found this rotation rate
accurately fits some optical line profiles, such as those belonging to 
\ionnl{Si}{iii}{4554-4576}. Rotation is analyzed in  \S~\ref{rotation}.

\subsection{Wind Parameters}

We calculated models for four 
different values of filling factor ($f_\infty$). 
The best fitting parameters for each of these 
models are shown in the Table \ref{twpara}. 
We use capital letters as reference for the subsequent
analysis. The best synthetic spectra for each filling
factor from UV to optical together with data are 
shown in Appendix A.

\subsubsection{Mass-loss rate}

Because of the strong sensitivity of the H${\alpha}$ line profile 
to the mass-loss rate, it is possible to attain high precision 
for this parameter through fitting its line 
strength. We found \Mdot/$\sqrt{f_\infty}{\approx}$\scie{1.6}{-6} \sunyr\ for \epsori\ ($d$=412 pc).  
The mass-loss rate for each filling factor is shown in Table \ref{twpara}.
As expected, \.{M}$/\sqrt{f_\infty}$ is almost independent of $f_\infty$, and
is lower than the value \Mdot$\sim$\scie{2}{-6}\ \sunyr  estimated by
\cite{searle08} and \cite{crowther06}. The main reason for the discrepancy
is their adopted lower values for $\beta$ (1.1 and 1.5, respectively).
A lower $\beta$ value yields lower densities 
in the H$\alpha$ formation region, and hence a higher mass-loss rate is required
to fit the line emission. As these values were computed using mainly 
H$\alpha$, its variability will increase the uncertainty of 
the mass-loss rate. \cite{morel04} and \cite{thompson13} reported
H$\alpha$ variability in its shape profile and strength. The results
shown here are based on data collected on a specific date,
and don't show H$\alpha$ variability during observations. 
We changed the mass-loss rate in order to obtain the 
strongest and the weakest line from \citeauthor{thompson13}'s 
profile. The corresponding values are \scie{4.8}{-7} and 
\scie{2.8}{-7} \sunyr\ respectively (assuming $\beta=2$, $f_\infty=0.05$ 
and $d=412$\,pc). The variation is approximately 30\% about 
our derived mass-loss rate of \scie{3.6}{-7} \sunyr.
This shows that small changes on \.{M} yield
larges changes in the H$\alpha$ profile.
A full variability analysis is necessary to improve our 
conclusions about \Mdot\ for \epsori, but this kind of 
analysis is beyond the scope of this paper.

In our analysis we did not include the 
infrared (IR) and radio spectral regions, but we can compare our
results with those from previous studies. \cite{blomme02} used
radio fluxes from 3.6  and 6.0 cm and computed a
\Mdot$\sim$\massrate{1.66}{-6} assuming a smooth model. This is
consistent with our estimate, but only if clumping persists into the radio
region (our model fluxes at 6.0 cm is $\sim$0.7 mJy, while the mean of reported
\citeauthor{blomme02}'s ones is $\sim$0.74$\pm$0.13. mJy.). Models by \cite{runacres02} 
indicate that structure can persist into the radio region and may influence radio
diagnostics. On the other hand \cite{puls06} found a difference between stars
with strong and weak winds. For strong winds their analysis suggested that 
wind clumping declined into the radio region while for weak winds they found similar
clumping in the inner and outer wind. In future work radio and millimeter fluxes
should be incorporated into the analysis.

Using IR observations \cite{repolust05} obtained
\Mdot$/\sqrt{f_\infty}$=\massrate{3.7}{-6} while \cite{najarro11} derived
\Mdot$/\sqrt{f_\infty}$=\massrate{2.0}{-6} (for the comparisons we scaled the
reported values to our distance 
and $v_\infty$). The \Mdot\ derived by \citeauthor{najarro11}
is consistent with our estimate while that of  \citeauthor{repolust05}
is a factor of  $\sim$ 2 higher.

UV lines are less sensitive than H$\alpha$ 
to changes in \Mdot/$\sqrt{f_\infty}$. Therefore, 
we use UV lines only to confirm our H$\alpha$ values. 
Moreover, UV line strengths also depend 
on ionization structure and some of them also on the X-ray 
emission (e.g. \ionn{N}{v} and \ionn{C}{iv}).
The X-ray independent and non-saturated UV line 
\ionnl{C}{iii}{1176} confirmed the \.{M}(H$\alpha$)
values. \ionnll{Si}{iv}{1394-1402} was not 
well reproduced, likely because of vorosity-porosity effects, 
so it was disregarded in the analysis (\S~\ref{discussion}).

\subsubsection{Velocity profile}

\begin{figure}
 \centering
 \includegraphics[angle=-90,trim=0 0 0 150,width=\linewidth,clip]{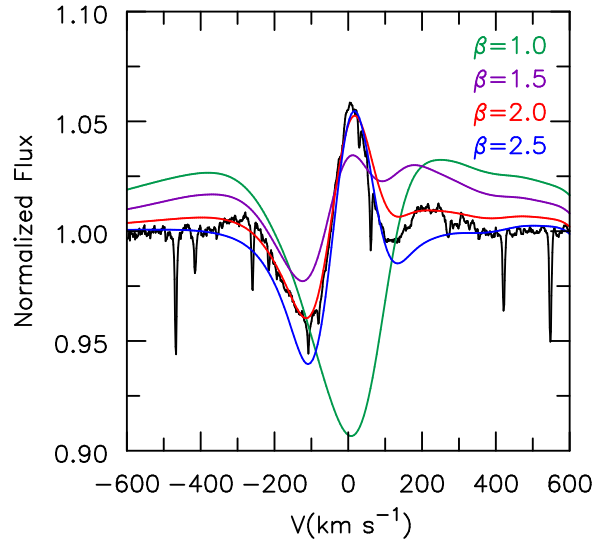}
 \caption{H$\alpha$ profiles for different $\beta$ values: 
          1.0 (green), 1.5 (purple), 2.0 (red) and 2.5 (blue). Black
          line represents the optical data. Each line profile was
          computed with the \Mdot\ that matches the observed line 
          strength.}\label{fbeta} 
\end{figure}

The wind acceleration parameter ($\beta$) strongly affects
the H$\alpha$ profile \citep{hillier03}. The H$\alpha$ profiles 
calculated for $\beta$=1.0, 1.5 2.0 and 2.2 are shown in Figure~\ref{fbeta}.
From this figure, we concluded that the better $\beta$ value is around 2.0 
(values lower that 1.8 or higher that 2.2 were unable to reproduce the profile). 
This value is higher than others previously calculated 
for early B-supergiants \citep[e.g.][]{kudritzki99,crowther06,searle08}. 
For each $\beta$ we explored a wide range for \Mdot/$\sqrt{f_\infty}$,
however we were unable to match the observed line profile for low $\beta$ values.

\citeauthor{crowther06} ($\beta$=1.5) used two sets of optical 
data for \epsori, but the line shape was not reproduced 
in either case. They obtained high velocity wing emission 
that is not detected, and they were unable to reproduce the blue 
absorption observed in H$\alpha$ profile.

\citeauthor{searle08} ($\beta$=1.1) do not show their H$\alpha$ profile; 
we tried to fit this line with their $\beta$ value by tuning  
other wind parameters but we were unable to reproduce the 
line shape with such a low $\beta$ value.

In the above works other Galactic early-type B-supergiant stars 
were also analyzed. The derived $\beta$ values are between 1.0 and 1.5.
For Small Magellanic Cloud (SMC) B0-B1 supergiants, \cite{trundle04} also found
$\beta$ values lower than ours.  \cite{evans04} analyzed
two B0Ia stars within a sample of supergiants
from the Magellanic Clouds: AV235 and HDE 269050. 
They found $\beta$ parameters of 2.50 and 2.75, respectively.

Spectral variability of H$\alpha$ could be one reason for the discrepancy 
between our values and those derived from previous works.
\cite{thompson13} reported  profile variability on a time scale
of weeks. This variability is still not understood 
\citep[see also][]{martins15}. An alternative to be investigated
in the future is a different radial dependence for the filling factor.
\cite{runacres02} predicted that the clumping factor ($f_{cl}=1/f$) grows 
until 10-50 R$_*$ and then falls in the outer wind regions.
However, \cite{puls06} shows that for denser winds high
clumping factors are present close to the stellar surface.
\epsori\ doesn't show a high mass-loss rate, but a different
clumping distribution should be investigated in combination
with other $\beta$ values.

The $v_\infty$ and $\xi_{max}$ parameters were estimated together using the blue 
absorption wing of UV lines, especially \ionnl{C}{iv}{1550} and 
\ionnl{C}{iii}{1175}. We found that the value of $v_{\infty}$ lies between
1700 and 1800 \kms, while the wind turbulence velocity is $\xi_{max}{\sim}$200 \kms.

The values calculated here are between the previous ones calculated by
\cite{crowther06} and \cite{searle08} (1600 and 1910 \kms\ respectively).
The \ionnl{C}{iii}{1175} profile rules out a value of 1600 \kms\ for $v_\infty$, 
since a value of $\xi_{max}\gtrsim$300 \kms\ is necessary in order 
to fit the blue absorption wing. Such a high value 
for $\xi_{max}$ distorts the \ionnl{Si}{iv}{1400} 
profile and the \ionnl{C}{iv}{1550} blue absorption wing.

\subsubsection{Clumping}

\begin{figure*}
 \centering
 \includegraphics[width=0.35\linewidth]{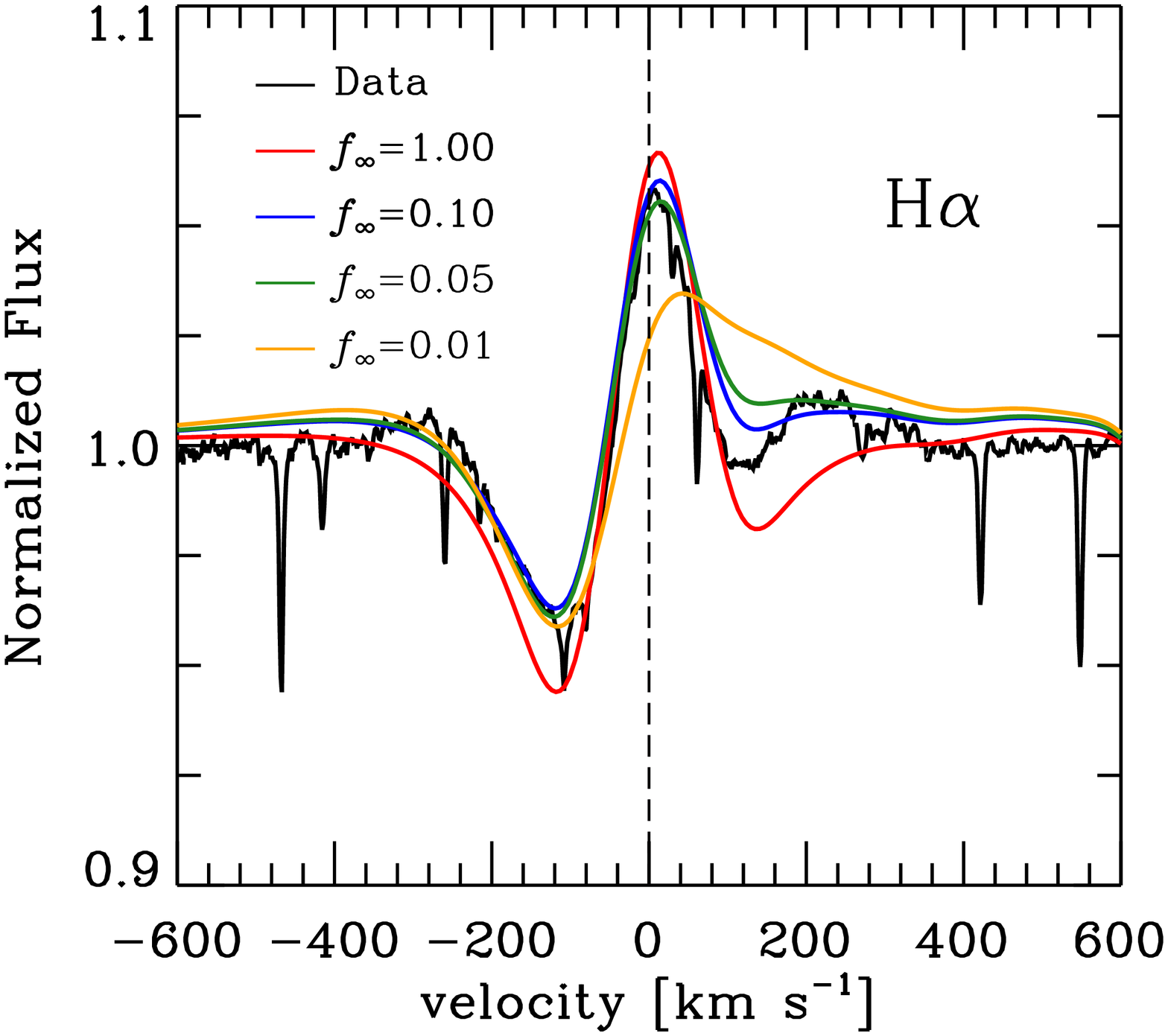}
 \includegraphics[width=0.35\linewidth]{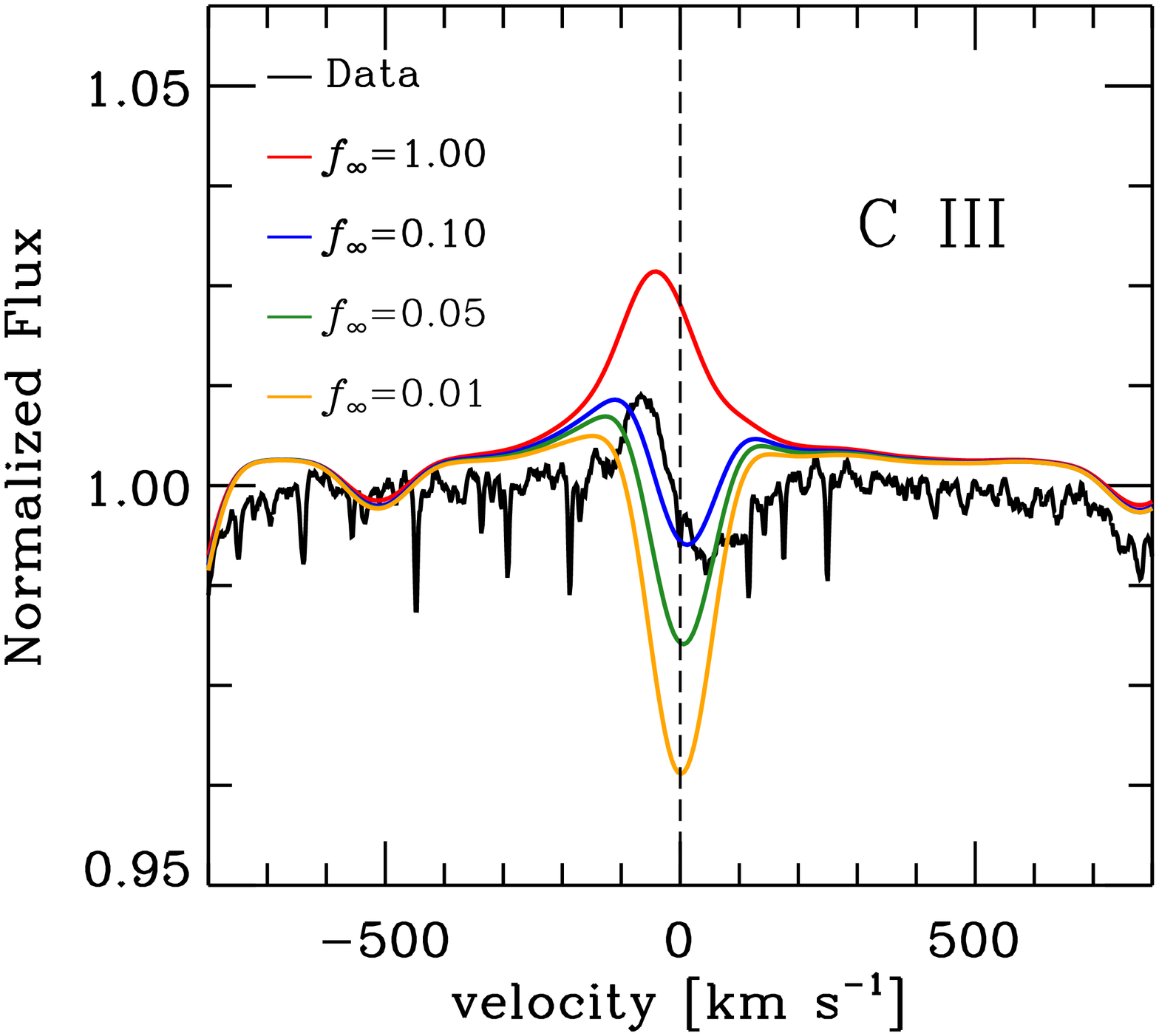} \\
 \includegraphics[width=0.35\linewidth]{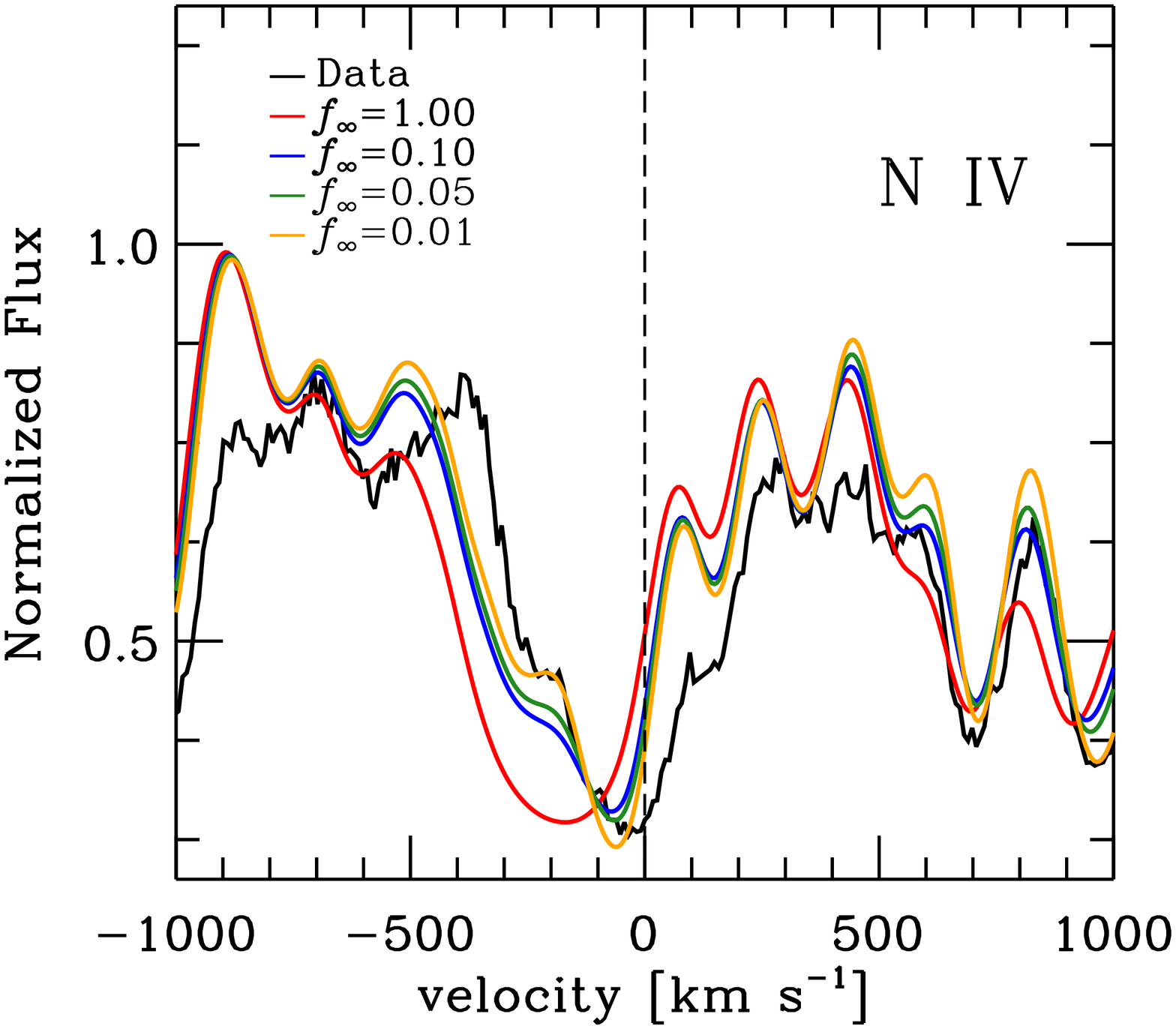}   
 \includegraphics[width=0.35\linewidth]{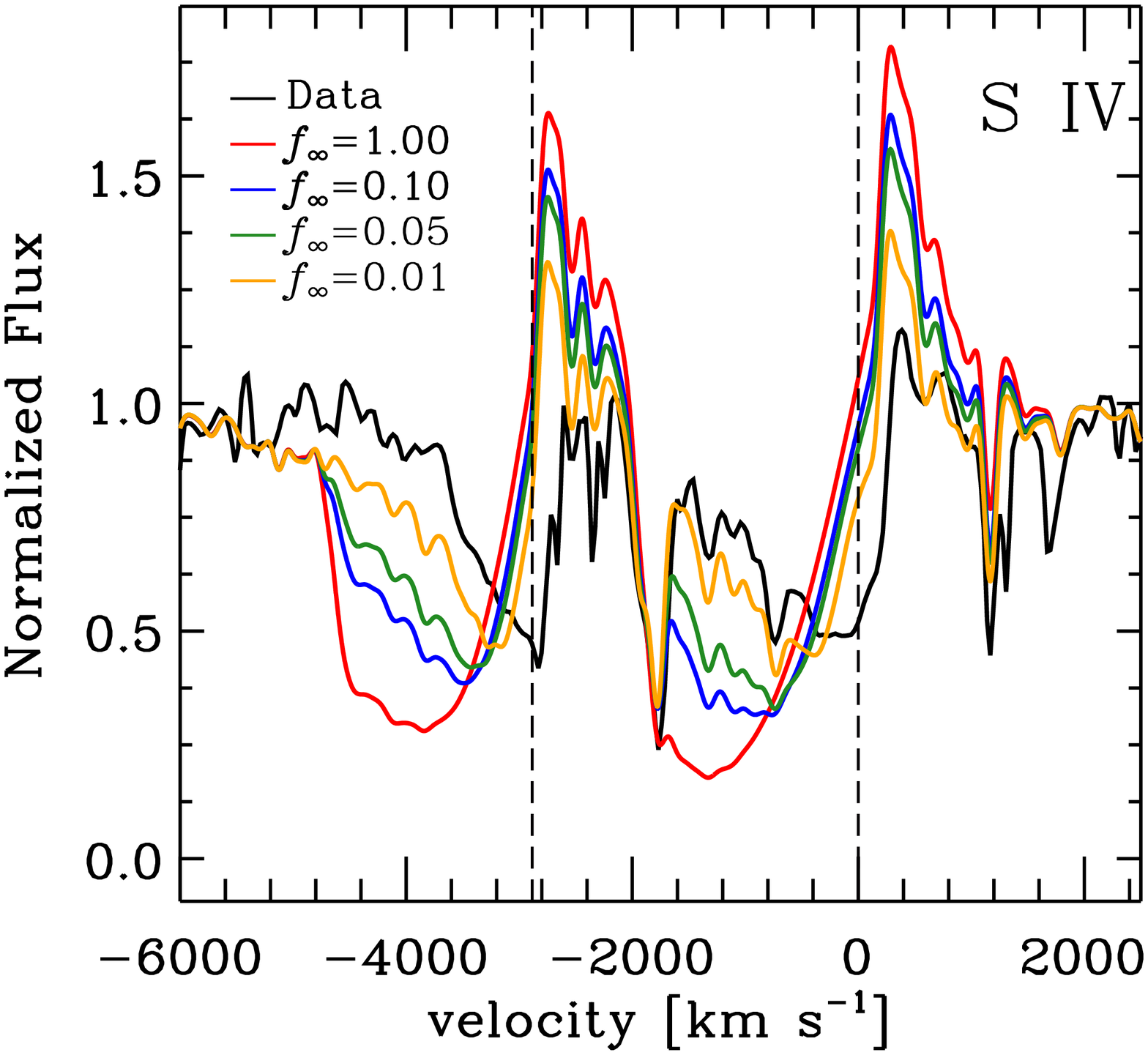} 
 \caption{Shown is the effect of clumping on selected optical and UV lines. Upper panels
  correspond to the optical lines H$\alpha$ and \ionnl{C}{iii}{5697},
  while the lower ones correspond to UV lines \ionnll{S}{iv}{1062,1073}
  and \ionnl{N}{iv}{1718}.
  } \label{fclump}
\end{figure*}

Besides the mass-loss rate, we also changed $v_{cl}$ 
(the point where the clumping starts) aiming to 
optimize the H$\alpha$ profile. The best values 
for $v_{cl}$ for each $f_\infty$ are shown in
Table \ref{twpara} and are between 20 and 40 \kms. 
Figure \ref{fclump} shows the effect 
of $f_\infty$ on different  lines in the optical and UV. We 
found that \ionnll{S}{iv}{1062,1073} and \ionnl{N}{iv}{1718}
yield a low value for $f_\infty$ ($\leq$0.01). The same 
was found for \ionnll{P}{v}{1118-28} (not shown 
in the figure). On the other hand, the optical lines yield a 
moderate value between 0.05 and 0.1. These discrepancies 
cannot be corrected by altering the mass-loss rates without 
spoiling the H$\alpha$ profile. It is possible to alter 
the abundances until the model fits the lines, but the derived values
for a reasonable value for $f_\infty$ are too low to be plausible, especially for sulfur.

\subsection{Abundances from UV and Optical}

We chose model ``A'' to estimate the abundances 
of CNO, Si and Fe. The lines selected for estimating the abundances 
are not affected by wind emission, hence it is irrelevant which 
model, from Table \ref{twpara}, we choose. 
Solar abundances for Si and Fe fit the optical and UV spectra 
to within 10\%. Because of this, we fixed their abundances 
to solar values and concentrated our analysis on CNO abundances.
Nevertheless, as it was noted above, a silicon
abundance of 1.2 Si$_\odot$ helps to reconcile the fit to the UV and optical
\ionn{Si}{iii} multiplets.

\begin{figure*}
 \centering
 \includegraphics[angle=-90,width=0.45\linewidth]{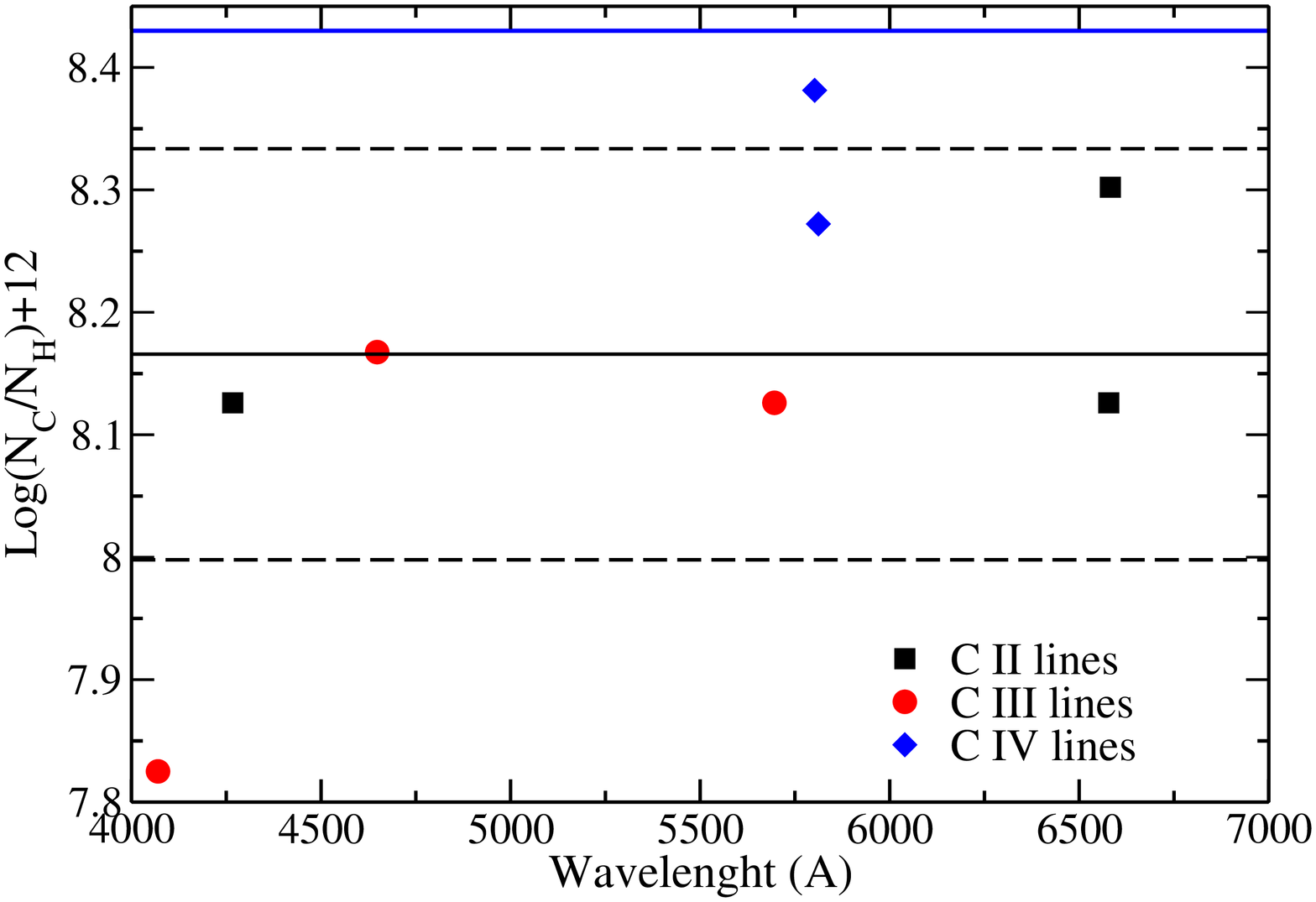} \\ 
 \includegraphics[angle=-90,width=0.45\linewidth]{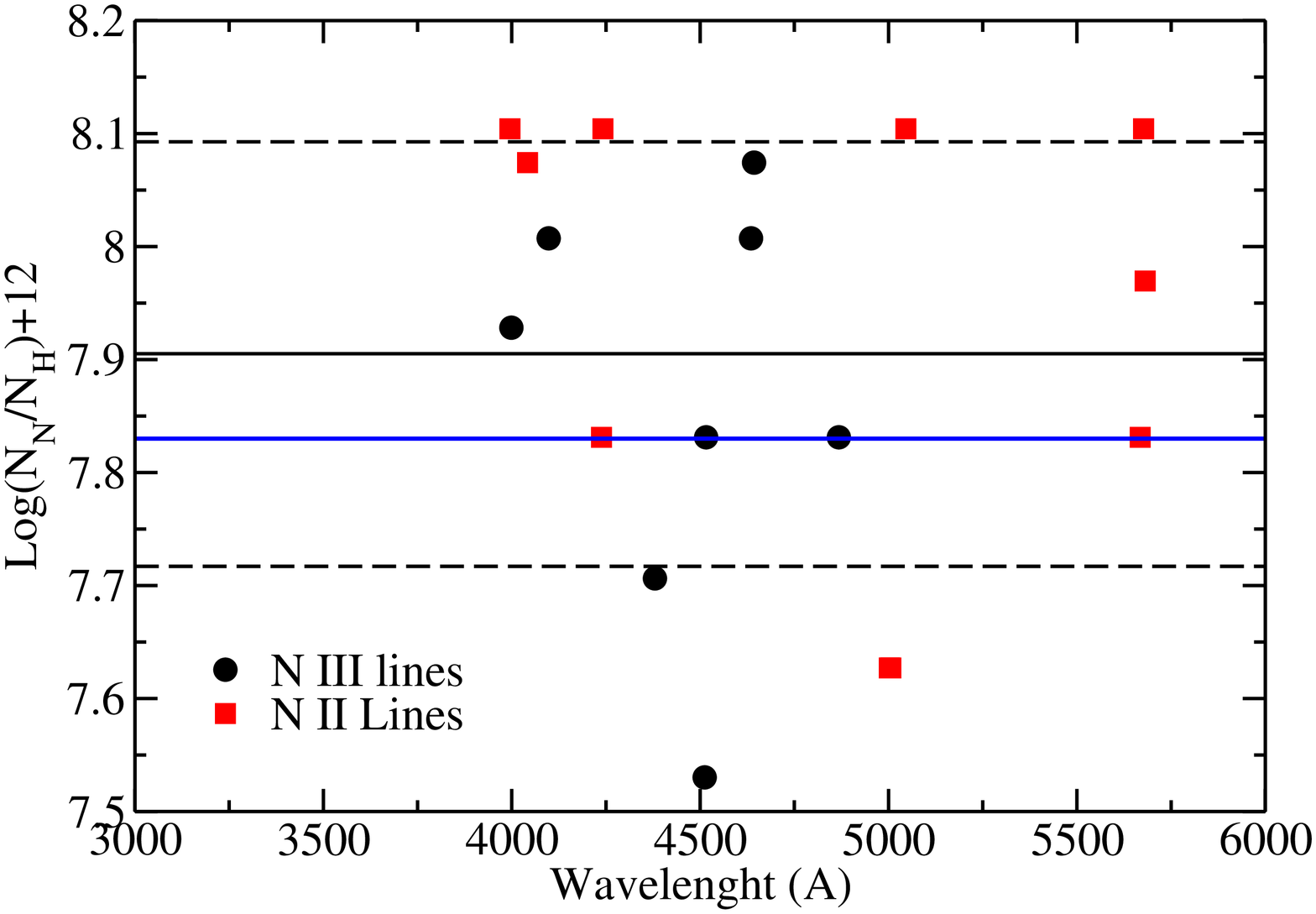} \\
 \includegraphics[angle=-90,width=0.45\linewidth]{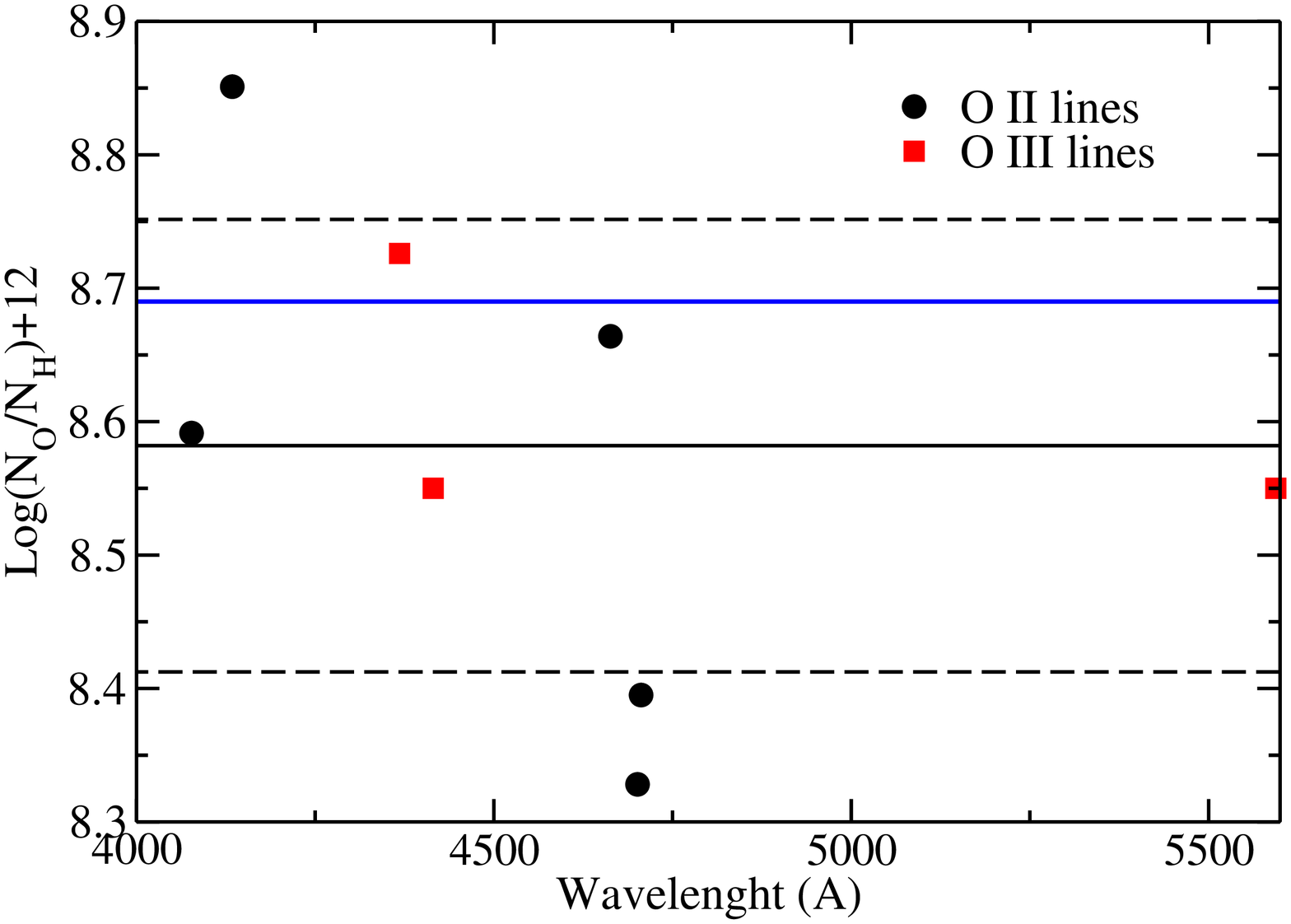}   
 \caption{Abundances for CNO. The dots show the abundance that fits each
          line for each ion and species. The black line is the
          mean of those abundances, the dashed lines are the mean
          $\pm$ the standard deviation and the filled blue line 
          represents the solar value.}\label{fabun}
\end{figure*}

Figure \ref{fabun} shows three panels. In each of these panels
the y-axis shows the abundance estimate obtained from each transition
and the x-axis shows the line wavelength. 
The numerical abundances are relative to hydrogen, expressed
as: $\log(N_X/N_H)+12$. The filled black line represents the
mean value (simple average) while the dashed lines represent the 
standard deviation of those measurements. 

Table \ref{tabund} shows the mean CNO abundances and their standard
deviations,  the N/C and N/O abundance ratios, and the corresponding solar values.
These ratios are calculated as: $[$N/(C,O)$]$= $\log(N_N/N_{(C,O)}
)_*-\log(N_N/N_{(C,O)})_\odot$. The other three columns show
previous abundance estimates from \citeauthor{crowther06} (C06)
and \citeauthor{searle08} (S08).

\begin{figure*}
 \centering
 \includegraphics[width=0.4\linewidth]{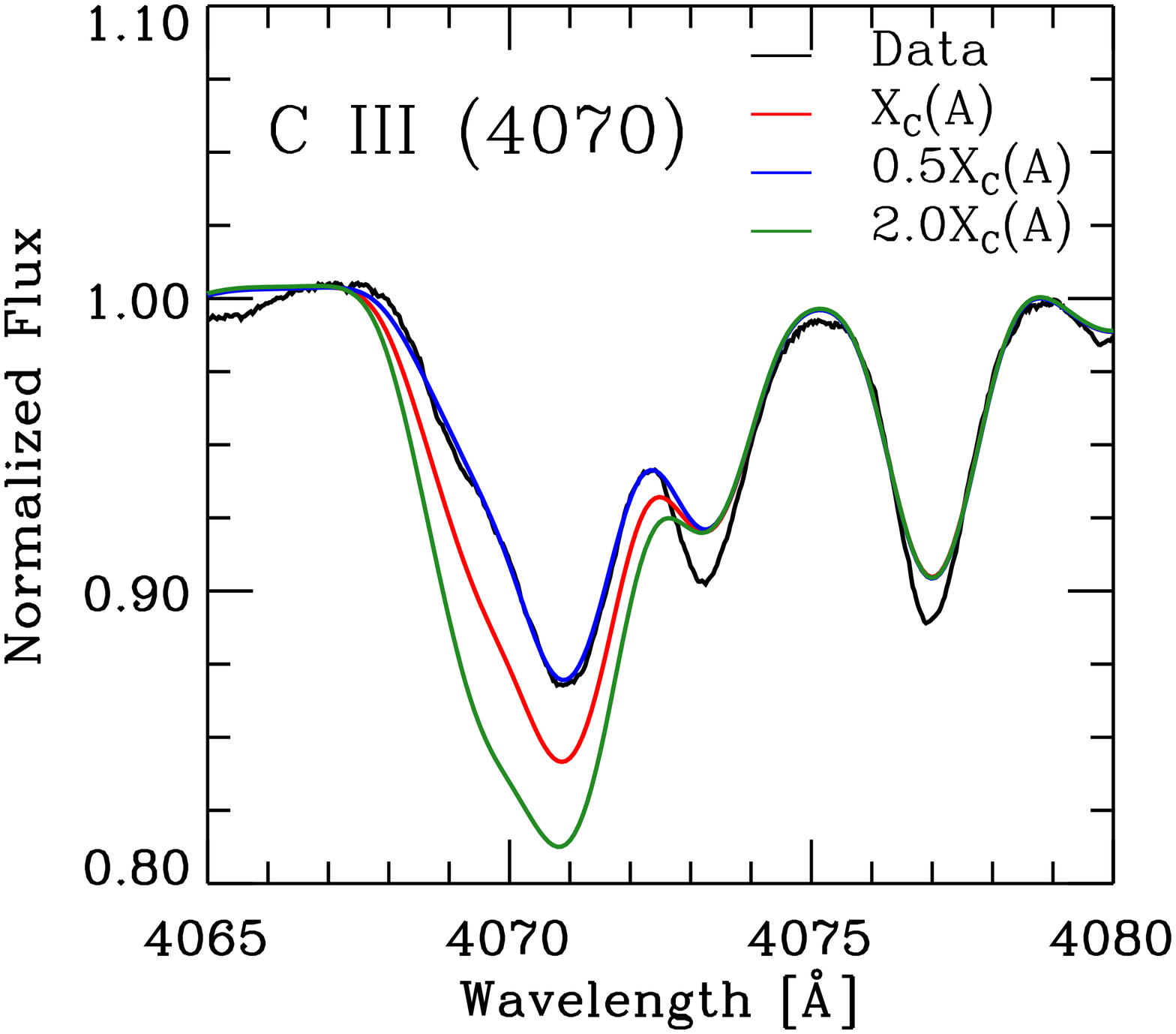}
 \includegraphics[width=0.4\linewidth]{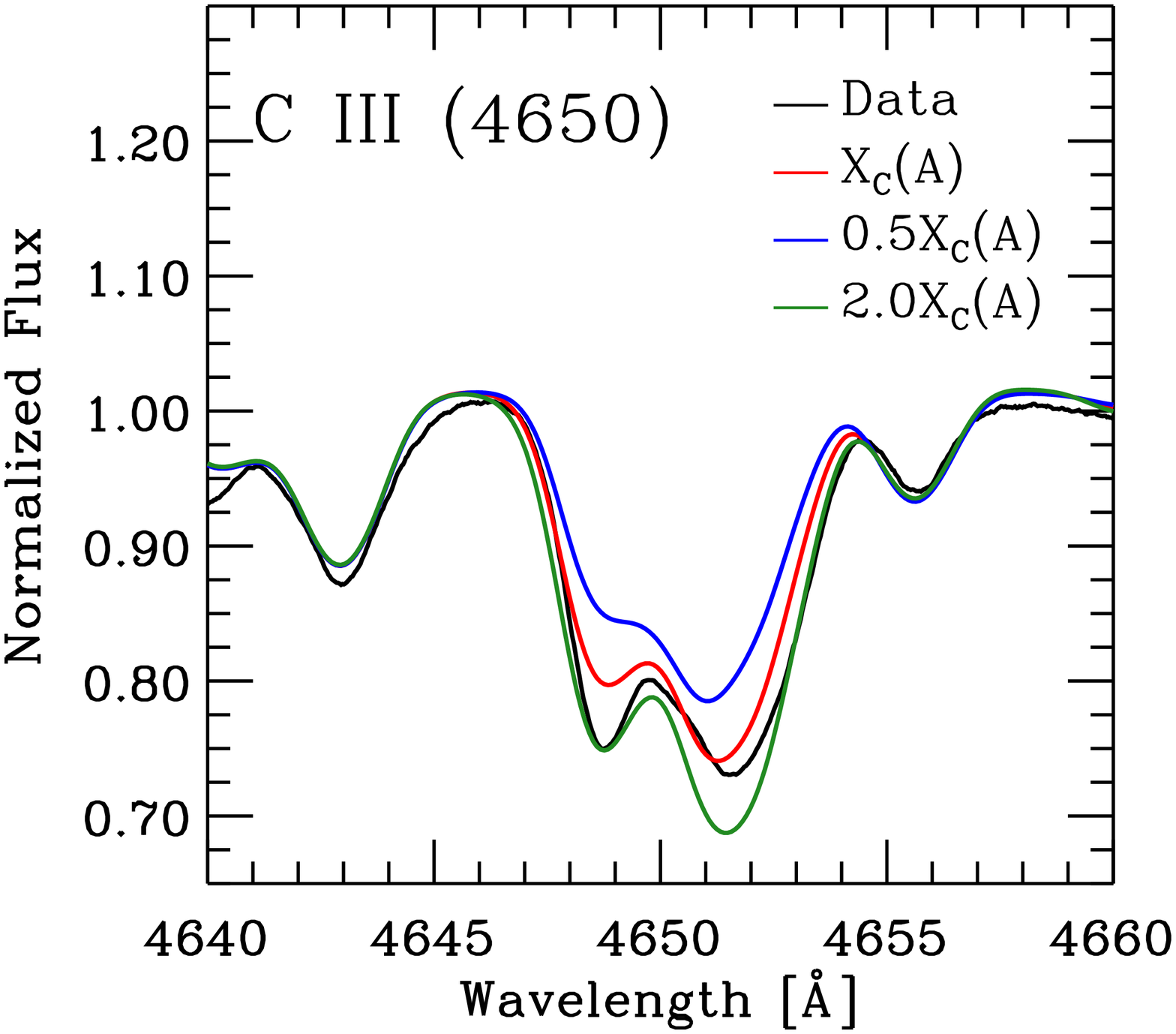} \\
 \includegraphics[width=0.4\linewidth]{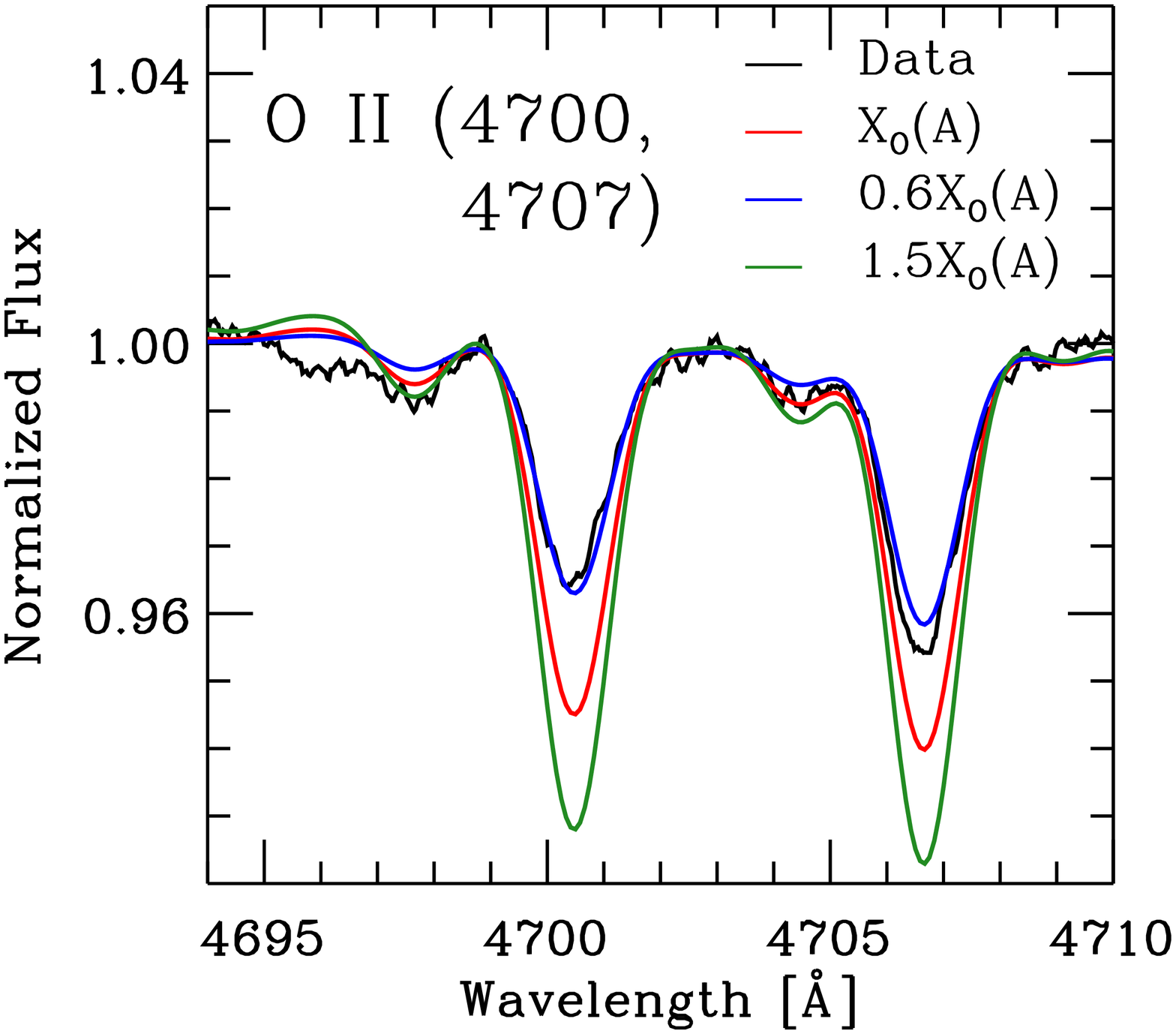}   
 \includegraphics[width=0.4\linewidth]{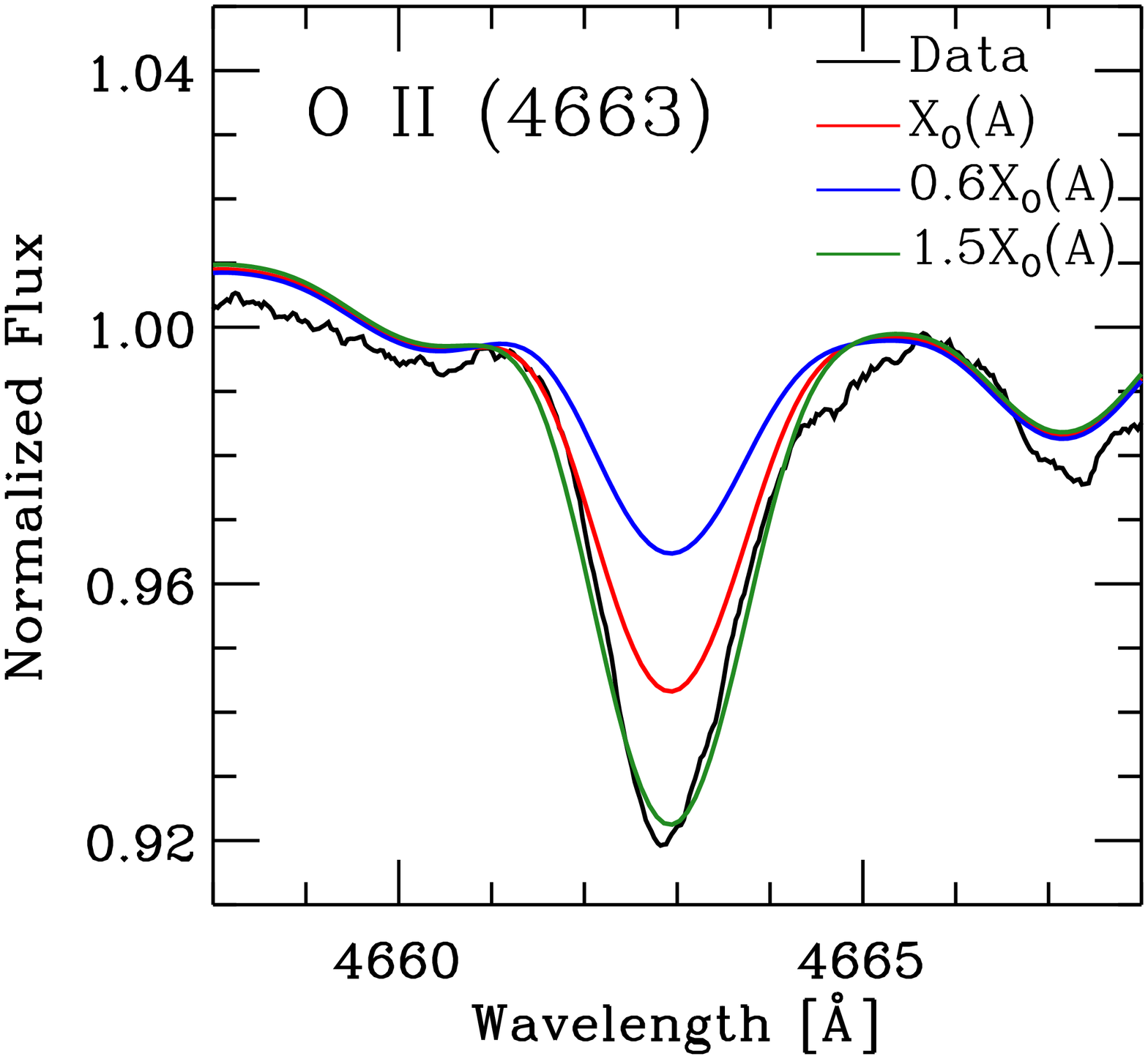} 
 \caption{Abundance discrepancy from lines of the same ion:
 \ionn{C}{iii} (upper panels) and \ionn{O}{ii} (lower panels).
        \ionnl{C}{iii}{4070} requires half the abundance needed 
        by \ionnl{C}{iii}{4650} while the \ionnl{O}{ii}{4663} abundance is 2.5 
        times larger than that found from \ionnll{O}{ii}{4700,4707}.}
 \label{fabunl}
\end{figure*}

Our values for $[$N/(C,O)$]$ show a small nitrogen enhancement and 
carbon and oxygen depletion. This provides evidence for 
CNO processed material at the star surface. 
Nevertheless, when compared with C06 and S08, our values are closer to
the solar ones.  

\begin{table}
\caption{CNO abundances from optical and UV data. 
              $[$N/C,O$]$=$\log(N_N/N_{C,O})_*-\log(N_N/N_{C,O})_\odot$ \label{tabund}}
\begin{tabular}{cccccc}
\hline\hline
Species & This Work & $\sigma$(dex) & S08 & C06 & Solar \\ 
\hline
C         &   8.16   & 0.16 &  7.66  & 7.95 & 8.43  \\
N         &   7.90   & 0.18 &  7.31  & 8.15 & 7.83  \\
O         &   8.58   & 0.17 &  8.68  & 8.55 & 8.69  \\
Si        &   7.51   & 0.04 &  7.51  & 7.51 & 7.51  \\
Fe        &   7.50   & 0.04 &  7.50  & 7.50 & 7.50  \\
$[$N/C$]$ &   +0.33  & --   &  +0.26 & +0.8 & 0.00  \\
$[$N/O$]$ &   +0.15  & --   &  -0.49 & +0.5 & 0.00  \\
\hline
\end{tabular}
\end{table}

The abundance determinations typically scatter within 
a  factor of 2 either side of the mean, as
shown in  Figure \ref{fabun}. Figure \ref{fabunl} shows 
the sensitivity of several lines to the abundance. For instance, 
\ionnl{C}{iii}{4070} requires half of the abundance predicted 
by \ionnl{C}{iii}{4650}. Similarly the \ionnll{O}{ii}{4663}
abundance is 2.5 times larger than that obtained with \ionnll{O}{ii}{4700,4707}.  

The possible causes of the abundance discrepancies are many ---
complex NLTE effects, deficiencies in the atomic models, blending,
issues related to microturbulence and macroturbulence. 
For example, it was pointed out by \cite{nieva06} that NLTE effects can cause
discrepancy between the abundances estimated using \ionn{C}{ii} lines
4267 \AA\ and 6587-82 \AA. We didn't find strong discrepancies
among these lines --- the difference is not larger than 0.2 dex.
Likewise, complex NLTE effects could affect the abundance estimates
of other lines such as \ionn{C}{iii} \citep{martins12} and 
\ionn{N}{ii-iv} \citep{rivero11,rivero12a,rivero12b}.

\subsection{Wind Parameters from X-rays}

We performed the fit procedure as described in Section~\ref{axrays},
using the \chandra\ and \xmm\ data simultaneously. The results for
each of the models described above (A, B, C and D) are shown in Table
\ref{txmodel}. For each model we found that low temperature 
($\sim$1-3$\times$10$^6$ K) plasmas  are necessary to account for 
the  \ionn{N}{vii} Ly$\alpha$, \ionn{N}{vi} He$\alpha$ and 
\ionn{C}{vi} Ly$\alpha$ lines, as well as the
He-like line of \ionn{Ne}{ix} He$\alpha$. Similarily, a hot component
is needed to account for \ionn{Si}{xiii},
\ionn{Mg}{xii} and \ionn{Mg}{xi} lines. The \ionn{Si}{xiv}~Ly$\alpha$ 
at 6.18 \AA\ is not detected by \chandra, and hence 
the hottest plasma temperature is less than \scie{1}{7}~K.
These temperatures are consistent with the distribution of heating-rates computed by 
\cite{cohen14b} for \epsori. They show that \epsori\ has low heating rates for T$\gtrsim$10$^7$ K.

\begin{figure*}
 \centering
 \includegraphics[width=\linewidth]{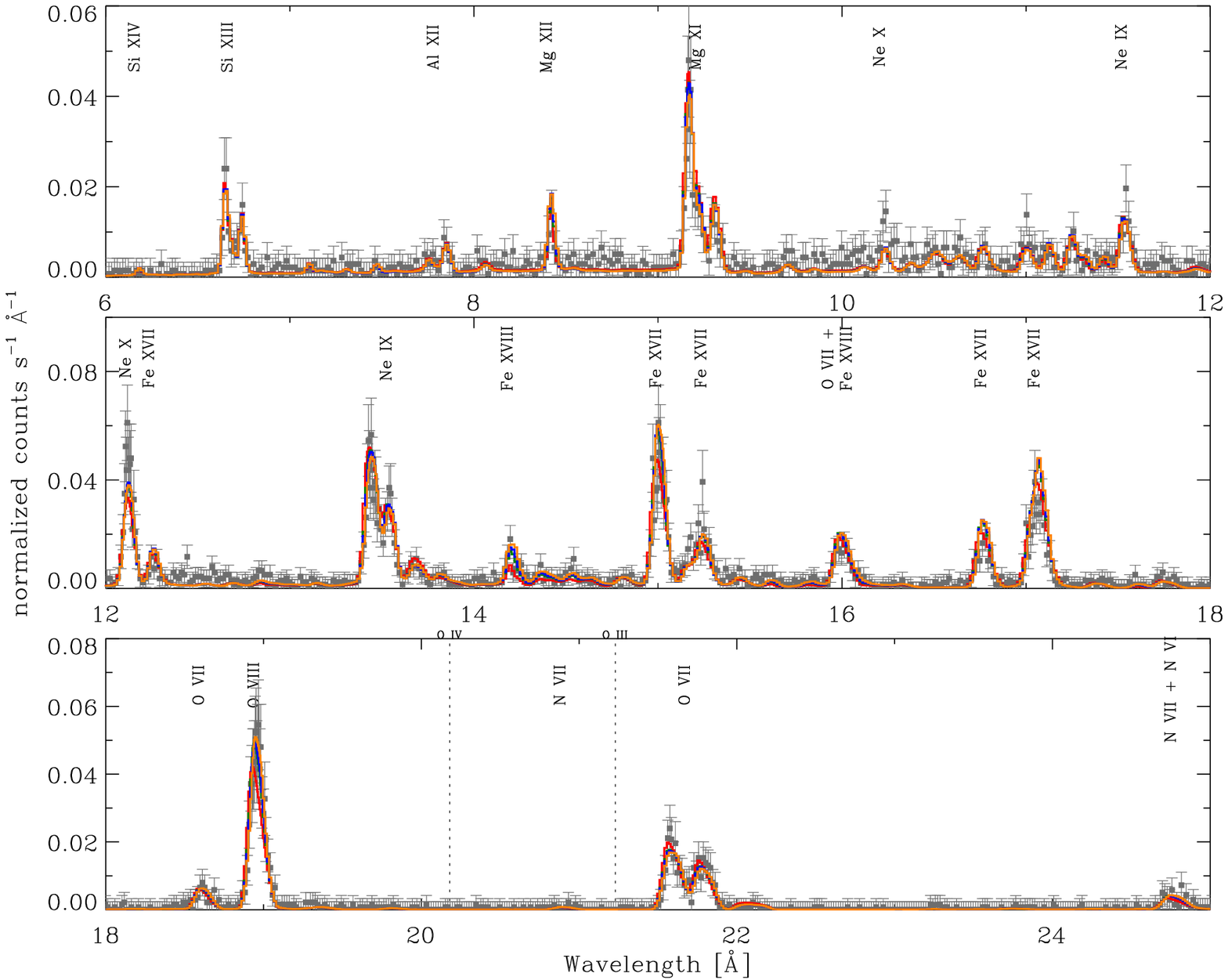}
 \caption{\chandra\ data (gray squares with error bars) and best 
          X-ray models for A(green), B(blue), C(orange), 
          and D(red). Dots lines show the K-shell ionization
          threshold for \ionn{O}{iii} and 
          \ionn{O}{iv}. The models are indistinguishable
          except for a few lines.}\label{fchandra}
\end{figure*}

\begin{figure*}
 \centering
 \includegraphics[width=\linewidth]{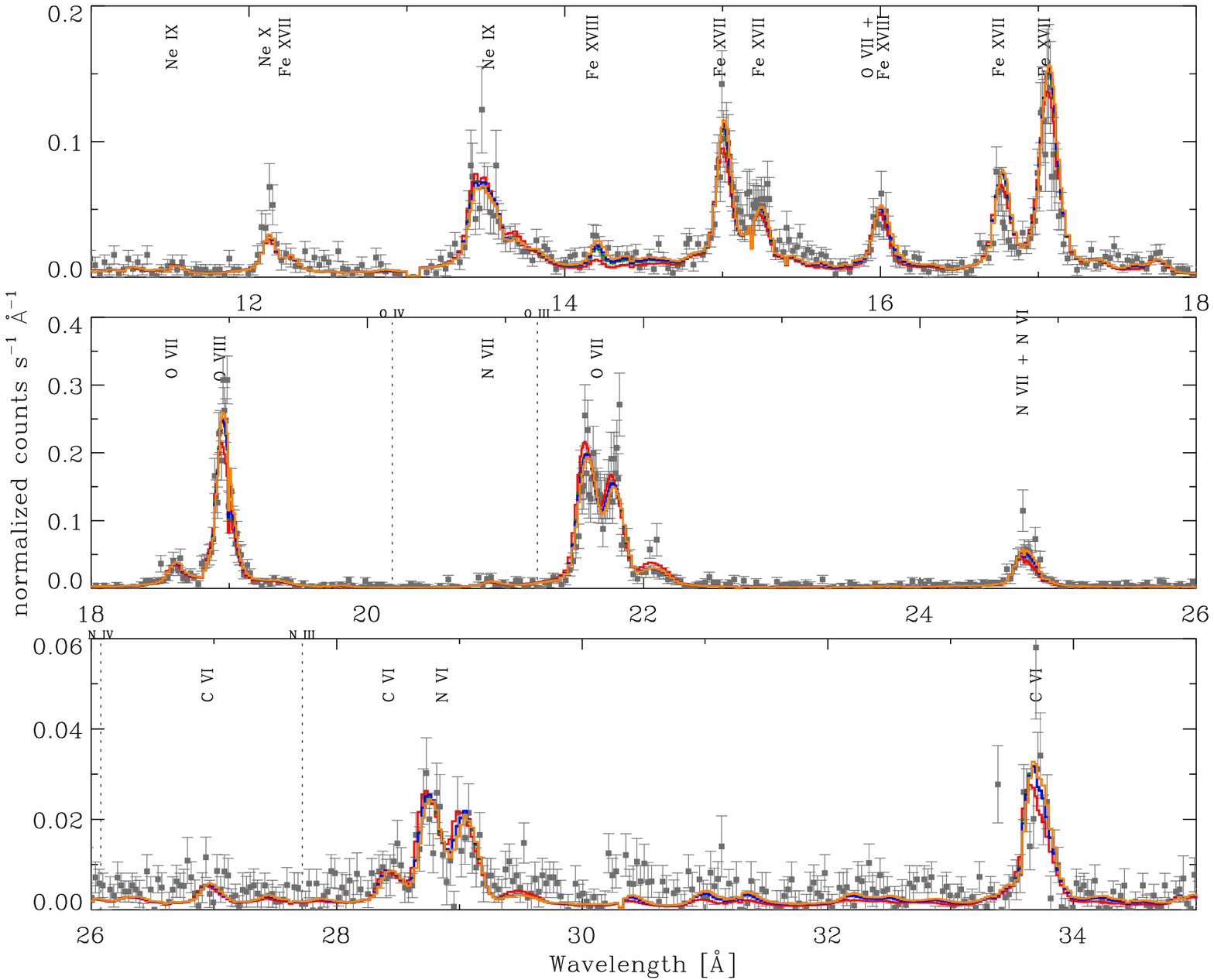}
 \caption{\xmm\ data (gray squares with errors bars) and best X-ray 
          models for A(green), B(blue), C(orange), 
          and D(red). Dots lines show the K-shell ionization
          threshold for \ionn{O}{iii}, \ionn{O}{iv}, 
          \ionn{N}{iii} and \ionn{N}{iv}. }\label{fxmm}
\end{figure*}

Figures \ref{fchandra} and \ref{fxmm} show the \chandra\ and
\xmm\ data with the models A (green), B (blue), C (orange) and D(red), 
respectively. The fits are reasonable for most lines, except for
\ionnl{Ne}{x}{12.13} which is too weak in all our models.
The fit of this line is strongly coupled to the fitting of the 
\ionn{Mg}{xi-xii} lines  --- better \ionn{Ne}{x} line profiles were 
obtained when those lines were excluded from the fits 
but the model Mg lines are too strong. The whole fit improves
if we lower the Mg abundance approximately 10\%. 
On the other hand,
\cite{drake05}  and \cite{cunha06} have suggested that
the currently accepted solar abundance of neon might be 
underestimated. We computed models increasing the Ne abundance
by 40\%, although the \ionnl{Ne}{x}{12.13} profiles improve,
the emission in the Ne He-like triplet is overestimated.

Table \ref{txmodel} shows the spatial distribution of the 
emitting plasmas. Typically the coolest plasma is found to
exist at larger radii than the hottest plasma.
Specifically, for 10$^6$\,K plasma \Ro\ is around 4-4.9 R$_*$, 
for the \scie{2-3}{6}\,K plasma R$_0$ is between
$\sim$3-4.7 R$_*$, and for the \scie{7}{6} K plasma 
R$_0\sim$2.1-2.9\,R$_*$. We note that the onset radii
are above the height where clumping starts 
($v_{cl}$=20-40\kms=$v(r{\sim}$1.12-1.23 R$_*$)).
The same result was found by \cite{cohen11} in their
analysis of HD 93129A -- the estimated onset radii were 
larger than R$_{cl}$.

\begin{table}
\caption{Plasma parameters for the best models that
              fit \chandra\ and \xmm\ data simultaneously.\label{txmodel}}
\begin{tabular}{ccccccc}
\hline\hline
T$_X$    & R$_0$ & $f_X$ & \hspace{.5cm} &
           \Tx    & \Ro & $f_X$ \\
           10$^6$ K & R$_*$ & 10$^{-3}$ & \hspace{.5cm} &
           10$^6$ K & R$_*$ & 10$^{-3}$\\
\hline
\multicolumn{3}{c}{model A}  & &  \multicolumn{3}{c}{model B} \\
\hline
1.0    &   4.84   & 26.93  & \hspace{.5cm} &  1.0    &   4.45   & 50.67 \\
2.0    &   4.56   & 7.22   & \hspace{.5cm} &  2.0    &   4.73   & 14.70 \\
3.0    &   3.05   & 4.76   & \hspace{.5cm} &  3.0    &   2.95   & 8.31 \\
7.0    &   2.60   & 0.34   & \hspace{.5cm} &  7.0    &   2.63   & 0.72 \\
\hline
\multicolumn{3}{c}{model C}  & & \multicolumn{3}{c}{model D}\\
\hline
1.0    &   3.89   & 123.0 & \hspace{.5cm} &  1.0    &   6.10   & 12.79  \\
2.0    &   4.72   & 62.03 & \hspace{.5cm} &  2.0    &   3.40   & 1.111  \\
3.0    &   2.77   & 24.58 & \hspace{.5cm} &  3.0    &   3.85   & 1.66  \\
7.0    &   2.89   & 4.26  & \hspace{.5cm} &  7.0    &   2.14   & 0.029  \\
\hline
\end{tabular} 
\end{table}

The present analysis allows for contributions
from different plasmas (\Tx\ and \Ro) to each X-ray 
line. Here, the principal lines in the 
\chandra\ and \xmms\ wavelength range are fitted
consistently with the same model. Thus, we found that 
short wavelength lines have flux 
contributions from hot plasmas, unlike long wavelength lines  
that mainly have contributions from colder plasmas.
The same trend was found by \cite{herve13} (their Fig. 4)
in their analysis of the RGS spectrum of $\zeta$ Pup.

The findings indicate that the shortest wavelength lines
start to be emitted from regions close to the stellar surface in the wind, 
while the longer wavelength lines originate from the outer regions.
This result is confirmed by line width measurements --
lines with a higher wavelength have larger line widths 
(Tables \ref{tcstat} and \ref{tcstat2}).

\begin{figure}
 \centering
 \includegraphics[width=0.95\linewidth]{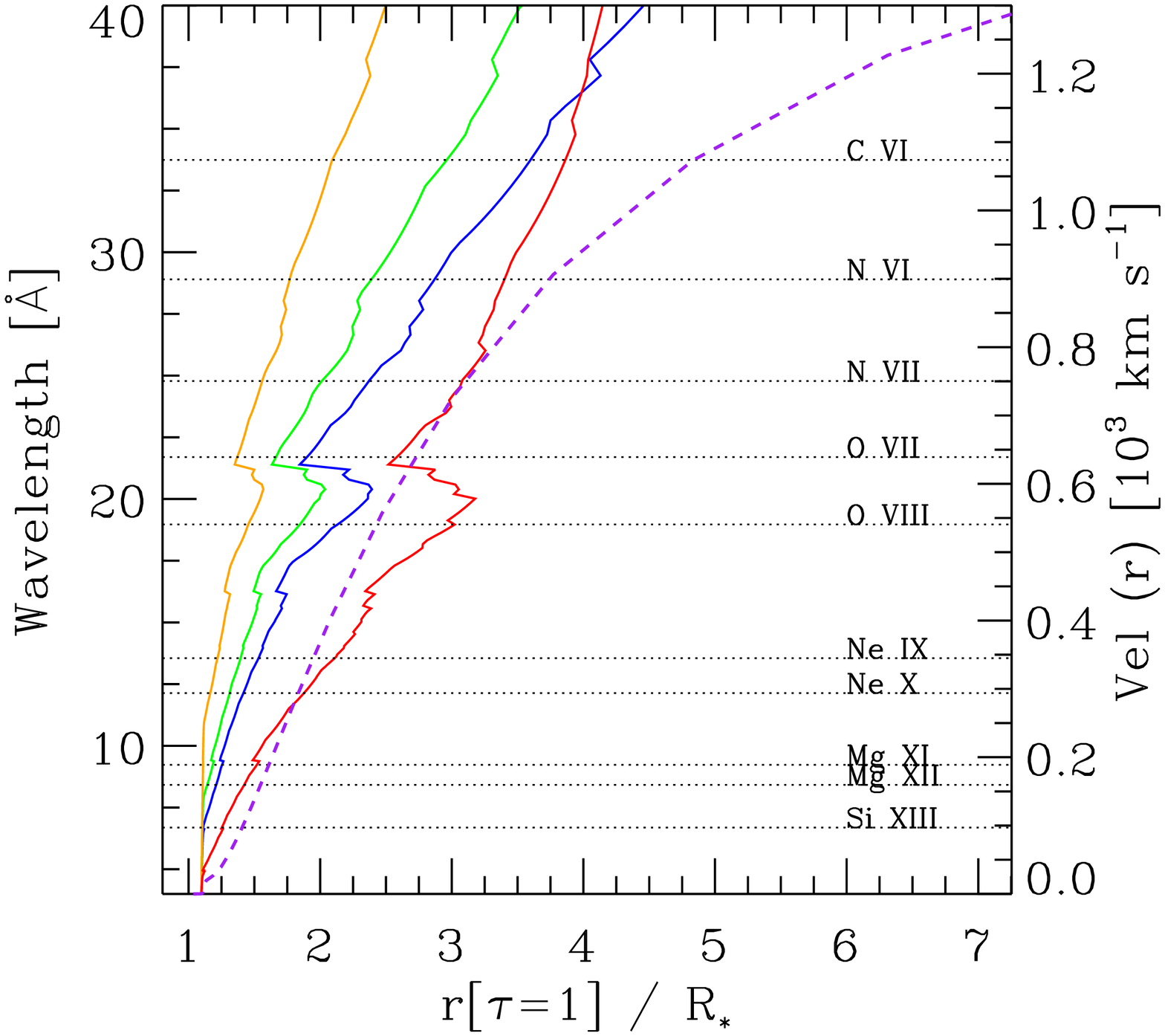}
 \caption{The radius where the continuum optical depth
 is equal to 1 ($R_{\tau_X=1}$) for wavelengths
 from 4 to 40 \AA. The curves are as follows: model A (blue), B (green), C (orange) and D(red).
 The dashed  purple line shows the velocity profile through the wind
 while the black dashed lines show the  wavelengths 
 for the main H-like and He-like lines. As we lower 
  \Mdot\ the radius at  which $\tau_X=1$ is also reduced. The less steep increase of $R_{\tau_X=1}$
 for the smooth model at longer wavelengths ($\lambda > 30\,\AA)$ is caused by
 ionization of He$^+$ to He$^{++}$. See Section \ref{discussion}.}
 \label{ftaux}
\end{figure}

\cite{waldron07} argued that this trend, found in some supergiants,
doesn't indicate that the cooler shocked plasma is formed only  
in the outer wind regions. Rather, the inner cool gas is not seen due to optical
depth effects. In other words, the wavelength dependence of 
continuum X-ray optical depth means that long-wavelength  photons produced
close to the surface are absorbed by the optically thick wind.
\cite{cohen14} emphasized optical depth effects as an explanation. 
They didn't find a significant trend in their sample of O stars (\epsori\ included). However data for 
some stars is suggestive of longer wavelength lines from
lower-temperature plasma being formed farther out in the wind.
Figure \ref{ftaux} illustrates this point.
It shows curves of the radius where the continuum 
X-ray optical depth equals unity, $r(\tau_x=1)$, and the
wavelengths of  the main H-like and He-like lines.
Utilizing Figure \ref{ftaux} in conjunction with Table \ref{txmodel}  it
is apparent that every ``observed'' plasma exists
above the $r(\tau_x{=}1)$ where the transmission factor is high
for each wavelength, as was pointed out by \cite{leutenegger10}.
Despite the high \Ro, we estimate that $\sim 70$\% of the emitted X-rays
are absorbed by the wind -- the X-ray absorption is very
signifcant for $\lambda\gtrsim$12 \AA.

\begin{figure*}
 \centering
 \includegraphics[angle=90,width=0.40\linewidth]{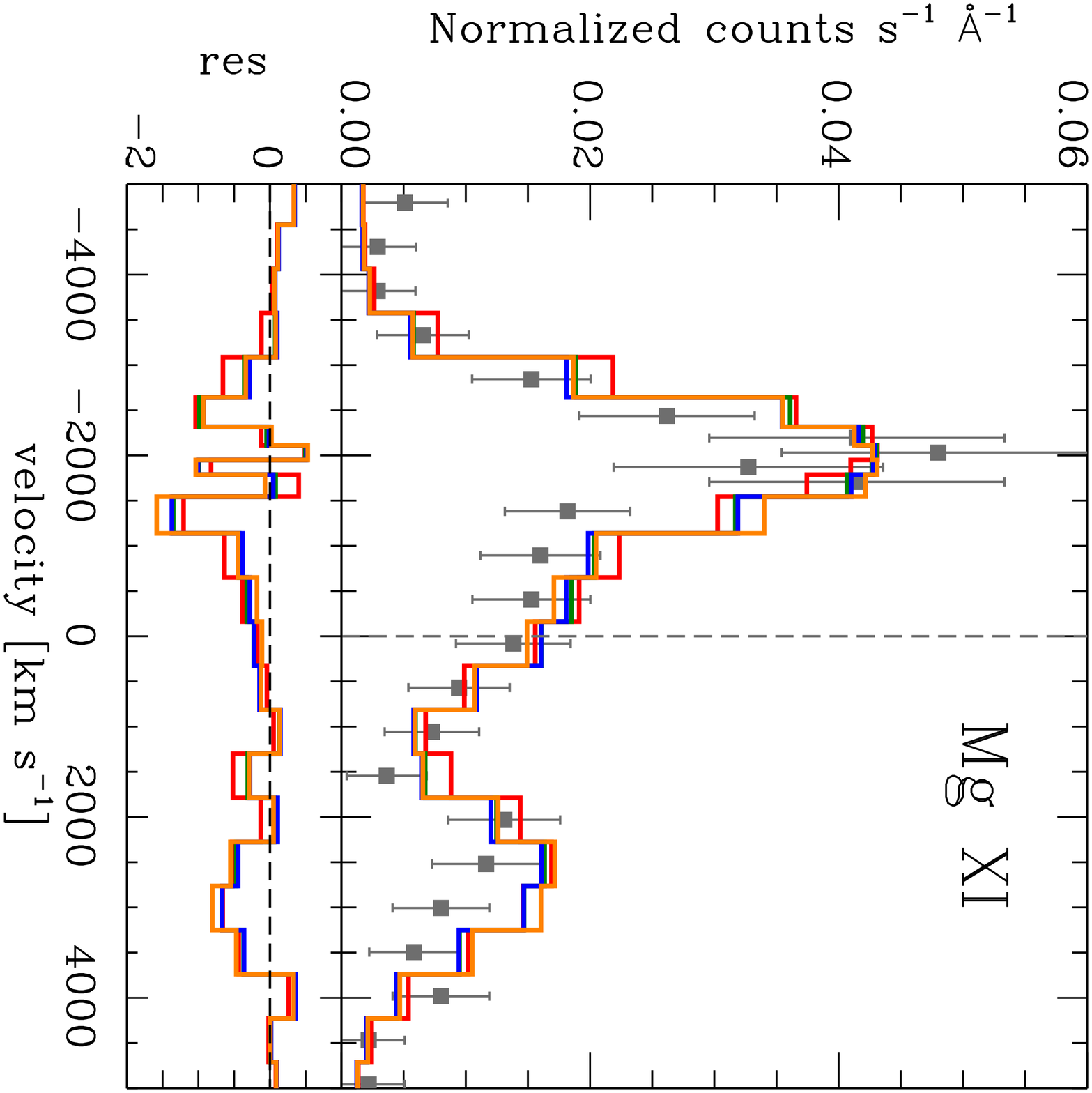}
 \includegraphics[angle=90,width=0.40\linewidth]{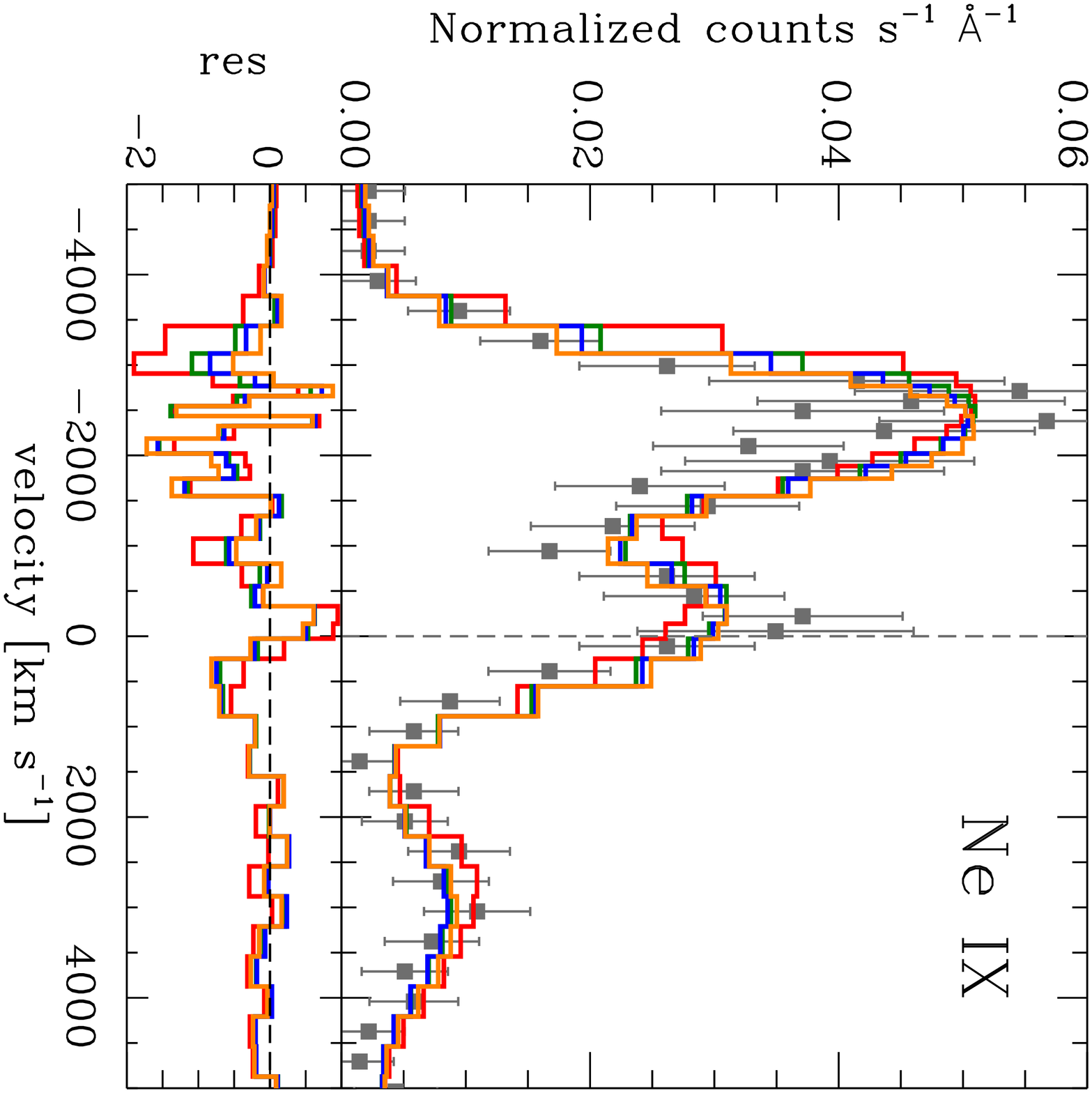} \\
 \includegraphics[angle=90,width=0.40\linewidth]{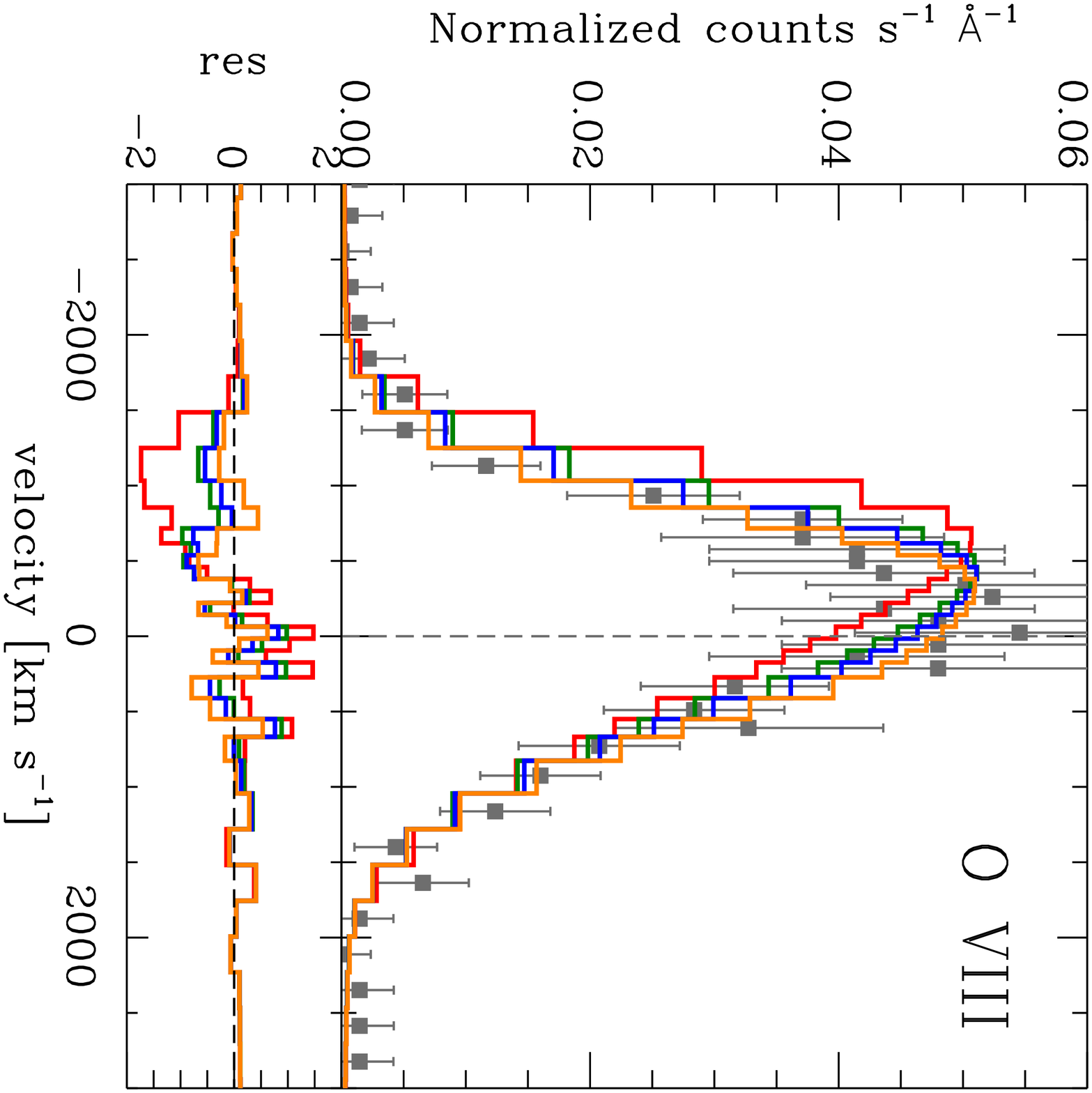}   
 \includegraphics[angle=90,width=0.40\linewidth]{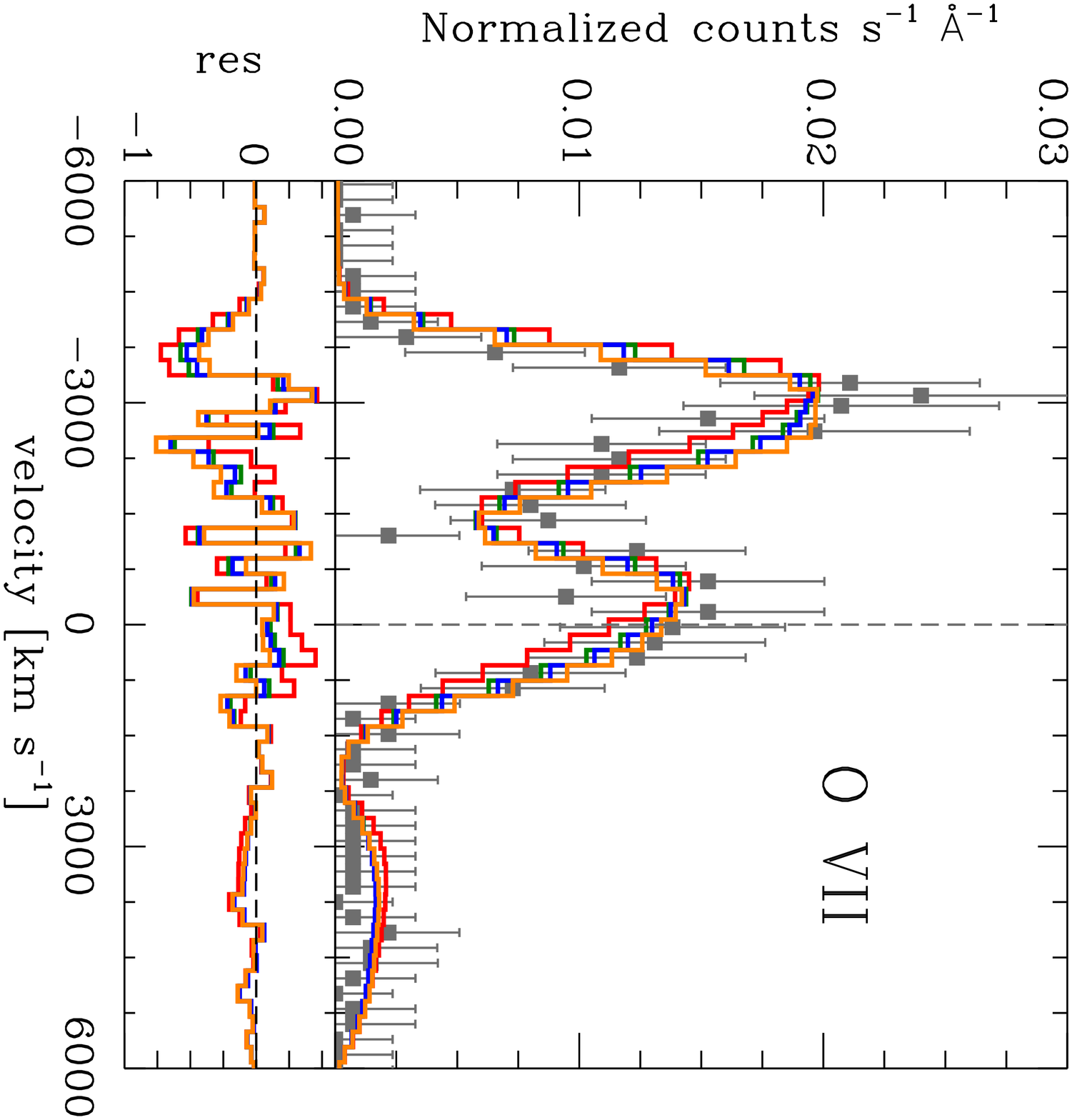} 
 \caption{Observed \chandra\ (squares) and synthetic line profiles
          corresponding to model A (blue), B (green), C (orange)
          and D (red). Note the primary difference among the models is
          the notable line asymmetry for the highest \Mdot\ model, especially
          for the longer wavelength lines where wind opacity is higher.
          Lower panels show the residuals.}\label{flchandra}
\end{figure*}

\begin{figure*}
 \centering
 \includegraphics[angle=90,width=0.40\linewidth]{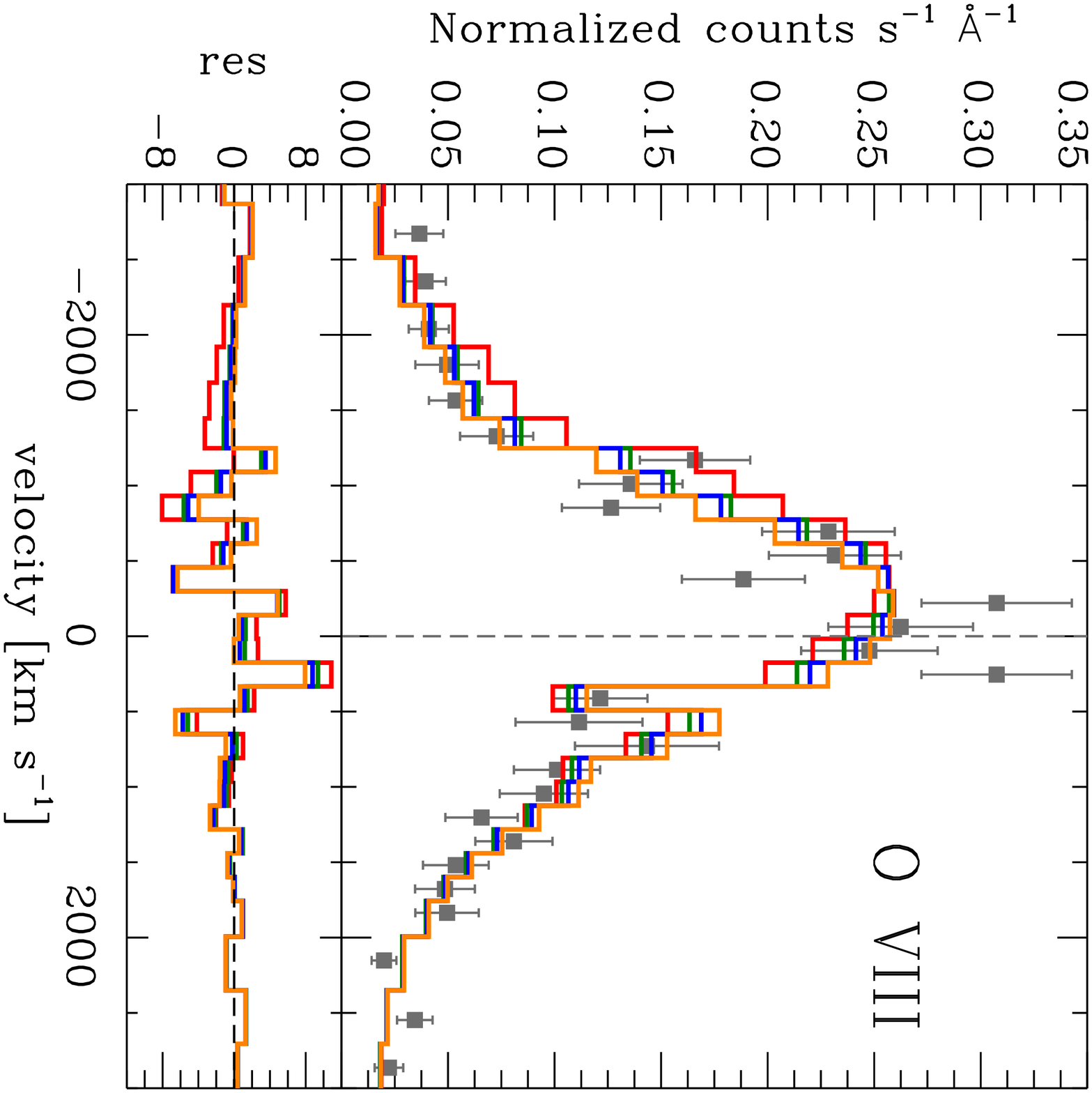}
 \includegraphics[angle=90,width=0.40\linewidth]{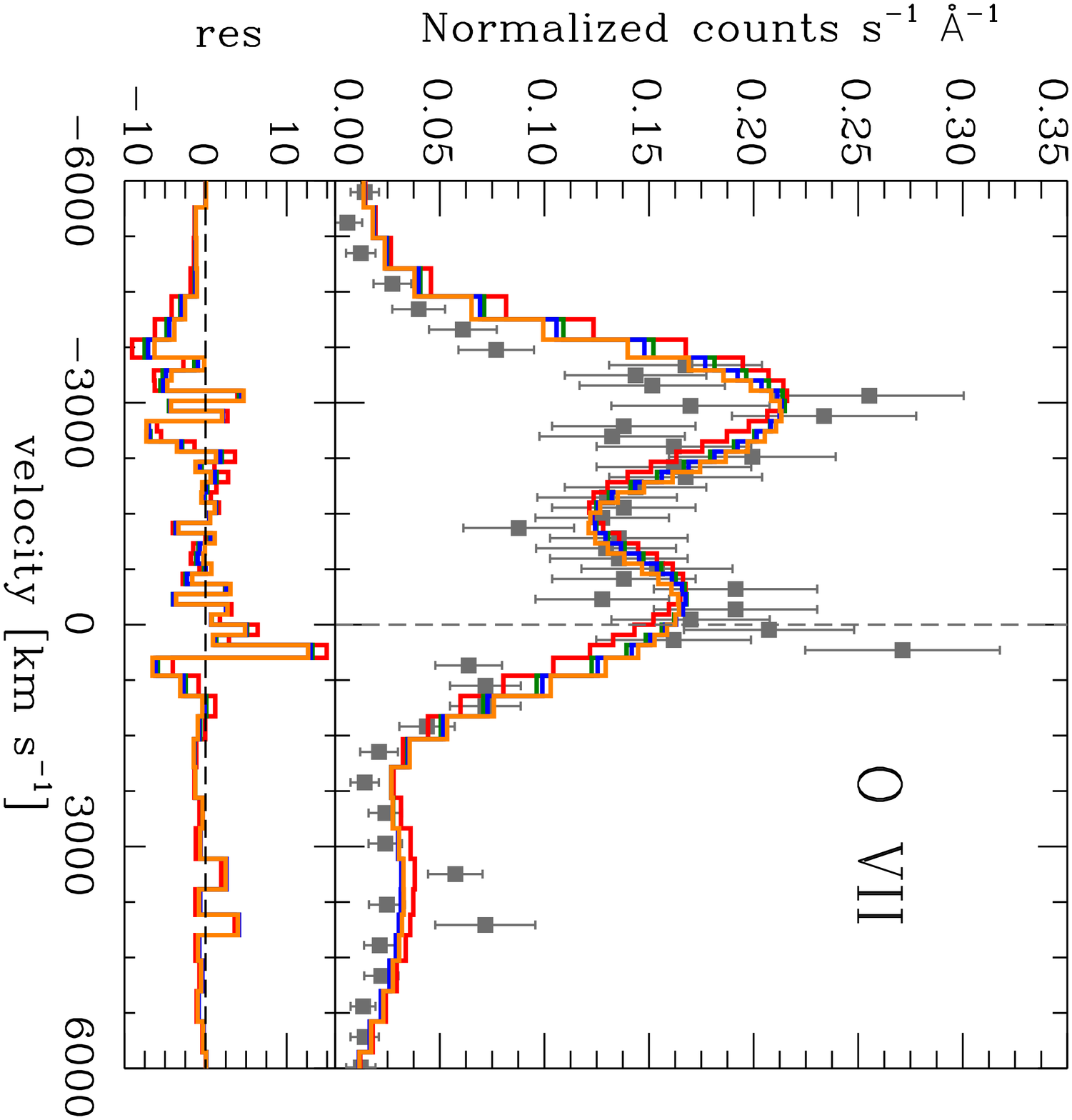} \\
 \includegraphics[angle=90,width=0.40\linewidth]{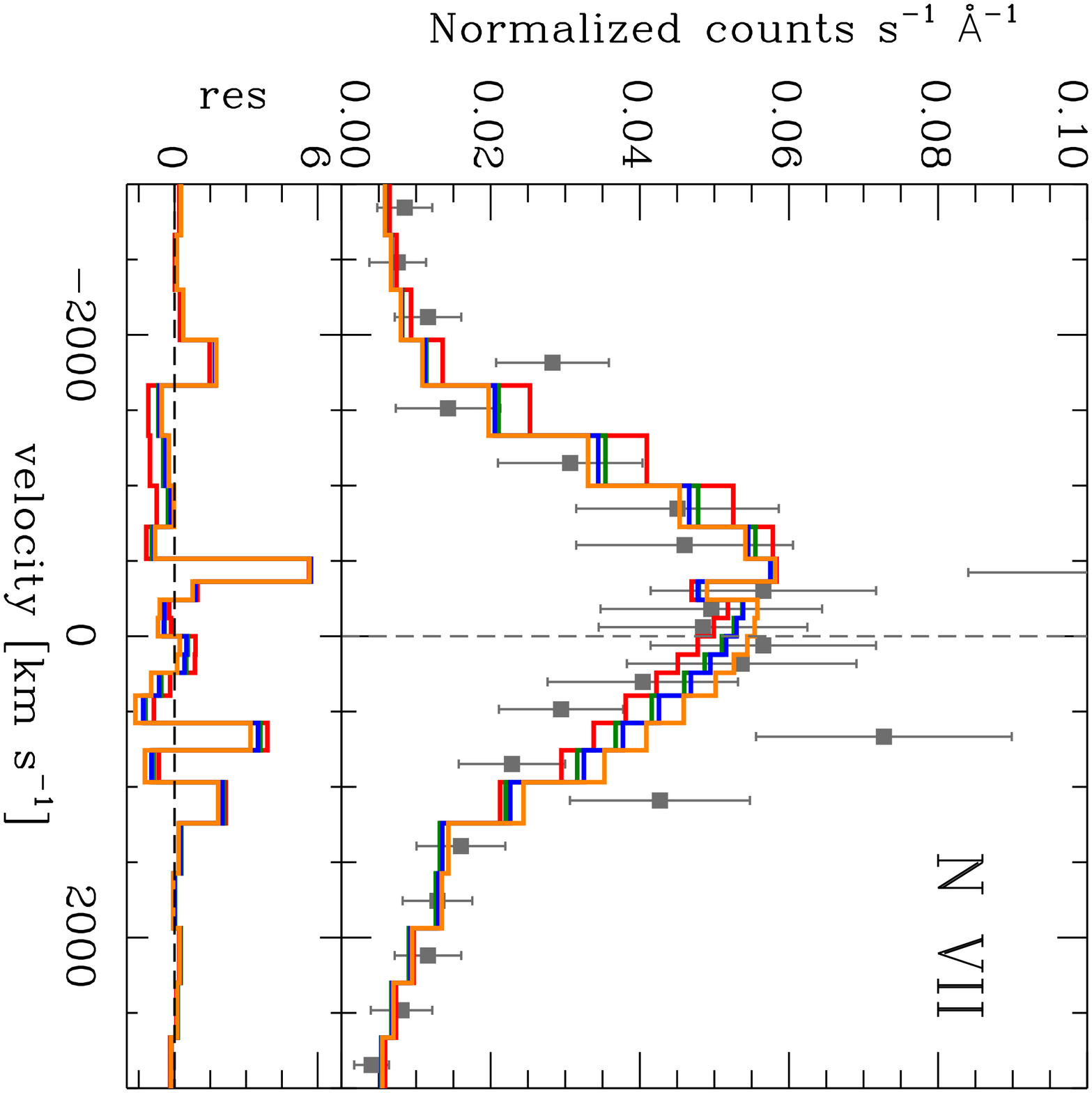}   
 \includegraphics[angle=90,width=0.40\linewidth]{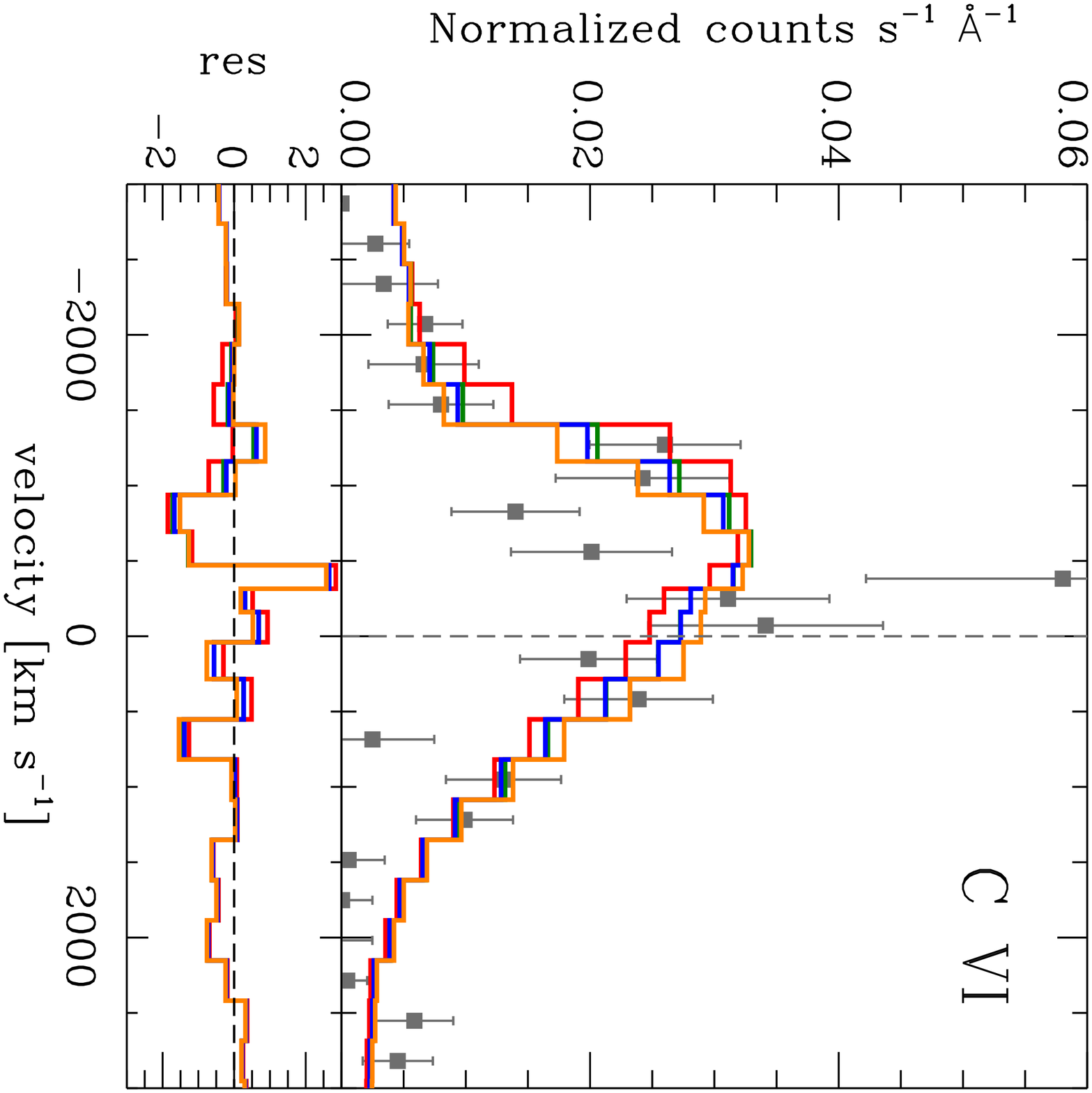} 
 \caption{Observed \xmms-Newton  (squares) and synthetic line
           profiles  corresponding to model A (blue), B (green), C (orange)
          and D (red). Lower panels show the residuals.}\label{flxmm}
\end{figure*}

Figures \ref{flchandra} and \ref{flxmm} show the calculated line 
profiles for some H-like and He-like X-ray lines together 
with data from \chandra\ and \xmms. 
H-like profiles are centered at the red component of the doublet, 
whilst He-like profiles are centered at the red component of
the intercombination doublet. A visual inspection shows that 
the observed lines are 
slightly blue-shifted, and that every model reasonably reproduces 
the profiles, with the exception of model ``D'' (red line).
This is especially clear when we look at the \ionn{O}{viii}  profile.

\begin{table*}
\begin{threeparttable}
\caption{Statistics ``{\it C}'' from fits of H-like (Ly$\alpha$) and He-like 
              \chandra\ lines. The values in the parentheses are the
              number of bins used for each line fit.\label{tcstat}
              }
\begin{tabular}{lcccccccc}
\hline\hline
Ion & \ionn{Si}{xiii}  & \ionn{Mg}{xii} & \ionn{Mg}{xi}
            & \ionn{Ne}{x} & \ionn{Ne}{ix} & \ionn{O}{viii} & 
            \ionn{O}{vii} & \ionn{N}{vii} \\
            & (35) & (22) & (50) & (30) & (80) & (50) & (145) & (66)\\
\hline
$\sigma$\tnote{a}  & -- & 350 & -- & 610 & -- & 706 & -- & 765 \\
C(A)  &  31.39 & 17.25 & 55.93 & 23.91 & 67.19 & 54.36  & 132.66 & 79.42\\
C(B)  &  31.66 & 17.81 & 56.62 & 24.66 & 68.22 & 50.69  & 131.47 & 79.22\\
C(C)  &  34.75 & 18.63 & 58.89 & 27.15 & 70.07 & 48.82  & 133.50 & 80.20\\
C(D)  &  28.27 & 15.91 & 54.76 & 31.93 & 78.95 & 88.40  & 153.49 & 78.92\\
\hline
\end{tabular}
\begin{tablenotes}
 \item[a]{The HWHM for each line in \kms\ of the Gaussian fit to the line profile.}
\end{tablenotes}
\end{threeparttable}
\end{table*}

\begin{table*}
\begin{threeparttable}
\caption{Statistics ``C'' from fits of H-like (Ly$\alpha$) and He-like 
         XMM lines \label{tcstat2}}
\begin{tabular}{lccccccc}
\hline\hline
Ion & \ionn{Ne}{x} & \ionn{Ne}{ix} & \ionn{O}{viii} & 
\ionn{O}{vii} & \ionn{N}{vii} & \ionn{N}{vi} &  \ionn{C}{vi}  \\
& (20) & (68) & (38) & (88) & (46) & (106) & (45)\\
\hline
$\sigma$ \tnote{a} & 610 & -- & 706 & --& 765 & -- & 880 \\
C(A)  & 18.46  & 64.61 & 58.56 & 109.60 & 56.51 & 164.83 & 69.97  \\
C(B)  & 18.32  & 64.11 & 57.28 & 107.53 & 56.52 & 166.93 & 69.81  \\
C(C)  & 18.37  & 64.10 & 57.79 & 104.16 & 57.47 & 165.10 & 71.16  \\
C(D)  & 19.64  & 68.66 & 73.24 & 126.13 & 56.61 & 159.85 & 71.82  \\
\hline
\end{tabular}
\begin{tablenotes}
 \item[a]{The HWHM for each line in \kms}
\end{tablenotes}
\end{threeparttable}
\end{table*}

In order to clarify which models match the observations we calculated the
{\it ``C''} statistics for each H-line and He-like line model profile. To do this
we allow the flux for each model profile to be a free-parameter, and
choose the scaling that minimizes the {\it C} statistic.
The results are shown in Tables \ref{tcstat} and \ref{tcstat2} 
for \chandra\ and \xmms\ data respectively. From the values, 
we confirm that the smooth wind  model (model ``D'') can 
be ruled out based on the fit to \ionn{O}{viii} and \ionn{O}{vii}
($\Delta C{\gg}$6.63 when compared to the minimum among the other
values of the same line), although the
\ionn{Ne}{ix}, \ionn{N}{vii} and \ionn{C}{vi} lines also have 
inferior-C statistics for model D. The other models, and other lines, 
have similar values for the {\it C} statistics, 
and can be considered statistically indistinguishable\footnote{In our
first approach we scaled each model profile to have the same peak value.
While the {\it C} statistics were higher than those tabulated in  Tables \ref{tcstat} and \ref{tcstat2},
the same trends are seen.}

From the results described above we conclude that the mass-loss
rate required to fit the X-ray spectrum of \epsori\ is
$\lesssim$ \massrate{4.5}{-7}, which confirms the UV and 
optical results that the \epsori\ wind is clumped. Furthermore,
these values of mass-loss rate agree with those reported
by \cite{cohen14} for the same star using exclusively 
\chandra\ data. \cite{cohen14} computed two values for 
the mass-loss rate of \epsori: \massrate{6.5}{-7} and 
\massrate{2.1}{-7}. 
The former comes from excluding from their analysis 
the lines (3 lines from 10) which possibly are affected 
by resonance scattering \citep{leutenegger07}, while the 
lower value results from including all of the lines of
their spectrum. As pointed out above R$_0$ values move the 
X-ray line formation region out of the absorbed part of the inner 
wind, making it unnecessary to include the resonance scattering to
make more symmetric lines. Including resonance scattering effects in
the analysis could have an influence on the computed $\beta$ and R$_0$ values,
but not in strong way due to the low density of \epsori's wind 
(which is less than that of $\zeta$ Pup). 

Models using a different wind acceleration parameter ($\beta=1.0, 1.2$) 
were also calculated. Wind and photospheric parameters from 
model ``A'' were adopted. 
For short wavelengths ($\lambda{\lesssim}$14 \AA) the model 
yields profiles that cannot be distinguished from the 
$\beta$=2 model. But for longward wavelengths 
a better fit is provided  by the $\beta$=2 model. Specifically,
the models fail to reproduce the He-like \ionn{O}{vii} line, 
and the Ly$\alpha$ lines of \ionn{N}{vii} and \ionn{O}{viii}.
For this set  of lines the model profiles are too asymmetric 
when compared with the data. Quantitatively, the ``$C$'' 
values increase $\sim$20-100\% for models with $\beta$=1-1.2 when 
compared with $\beta$=2 profiles (e.g. $\Delta C{\sim}45$ 
in the case of Ly$\alpha$ \ionn{O}{viii}). In the case of 
He-like \ionn{N}{vi} and Ly$\alpha$ \ionn{C}{vi} 
lines the increase in ``$C$'' is approximately 5-10\% 
($\Delta C{\sim}3.0{-}6.0$).

The plasma temperatures were fitted again as described 
in Section \ref{axrays}. The lower temperatures are the same 
as those calculated above, but a better fit was achieved by switching 
the hottest plasma to \scie{8}{6} K. The onset radii are closer than 
those from the model with $\beta$=2 because models with lower $\beta$
yield higher velocities in the region from R=1.1-6.1 R$_*$.   
The coolest plasma (10$^6$ K) has $R_0= 2.16 R_*$, the plasma 
of \scie{2}{6} K has $R_0=6.1 R_*$, while the plasmas of \scie{3}{6} 
K and \scie {8}{7} K have $R_0$ values of 1.9 R$_*$ and
1.4 R$_*$ respectively.

\subsection{Abundances from X-rays}

The metal abundances, relative to hydrogen, could be
estimated using the continuum emission relative
to line emission. But in the case of \xmms\ data,
most of the spectral range is contaminated by
line emission. However, a narrow region around
the nitrogen lines shows weak or no lines. The modeled
continuum in this region is almost 50\% below that
observed. To fit the continuum
it would be necessary to both scale the model emission measures and
to lower the metal abundances by approximately 
$\sim$50\% in oder to preserve the fit to the lines. By limiting the fit
to nitrogen lines region a good match for both the continuum
and those lines can be found. However, most of continuum comes 
from the hottest plasma (T$_X{=}$\scie{7}{6} K) which 
spoils the fit to the shortward wavelength spectrum. Furthermore,
as it's shown in the figures \ref{fchandra} and \ref{fxmm}
the shortward continuum is fitted fine by our models. 
This discrepancy could be due to a lack of 
pseudo-continuum weak lines in \apec\ as suggested 
by Zsarg\'{o} et al. (in preparation).

From the reasons exposed above, it was not possible 
to derive metal abundances relative to hydrogen --
only the relative abundances among metals. We searched 
for some trends by measuring the fluxes for the main 
X-ray metal lines from the data, and compared them to the 
corresponding model fluxes.

The observed fluxes were calculated by fitting Gaussian profiles for
each H-like line and He-like triplet. 
Also, we calculated the fluxes for Fe lines around 
15 \AA\ and 17 \AA. Each of these blends
(two lines around 15 \AA\ and two more around 17 \AA) 
were treated separately, but their fluxes were added in order
to compare them with the model values whose fluxes were estimated taking
into account the whole blend.

\begin{figure}
 \centering
 \includegraphics[width=0.95\linewidth]{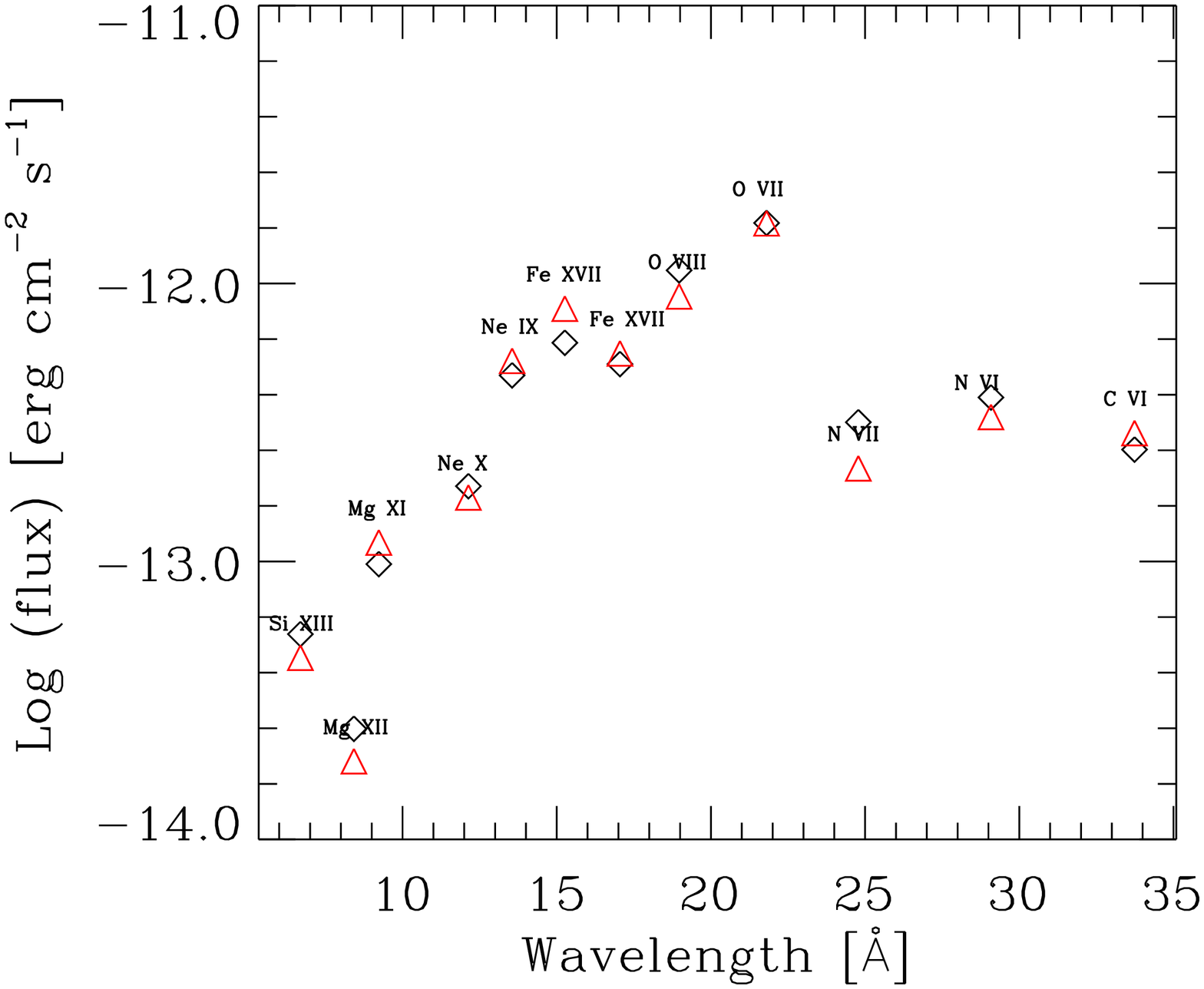}
 \caption{X-ray line fluxes. The observed 
fluxes are shown as black diamonds while the values for
model ``A" are red triangles. }\label{fxfluxes}
\end{figure}

Figure \ref{fxfluxes} shows the measured X-ray line fluxes 
and the ones from model ``A''. There are not strong trends that would 
suggest changes in abundances. The model line fluxes for \ionn{N}{vii}
Ly$\alpha$ and \ionn{N}{vi} are too small compared with 
observations -- increasing the nitrogen abundance by 30\% 
would bring them into better agreement.

\begin{figure*}
 \centering
 \includegraphics[width=0.45\linewidth]{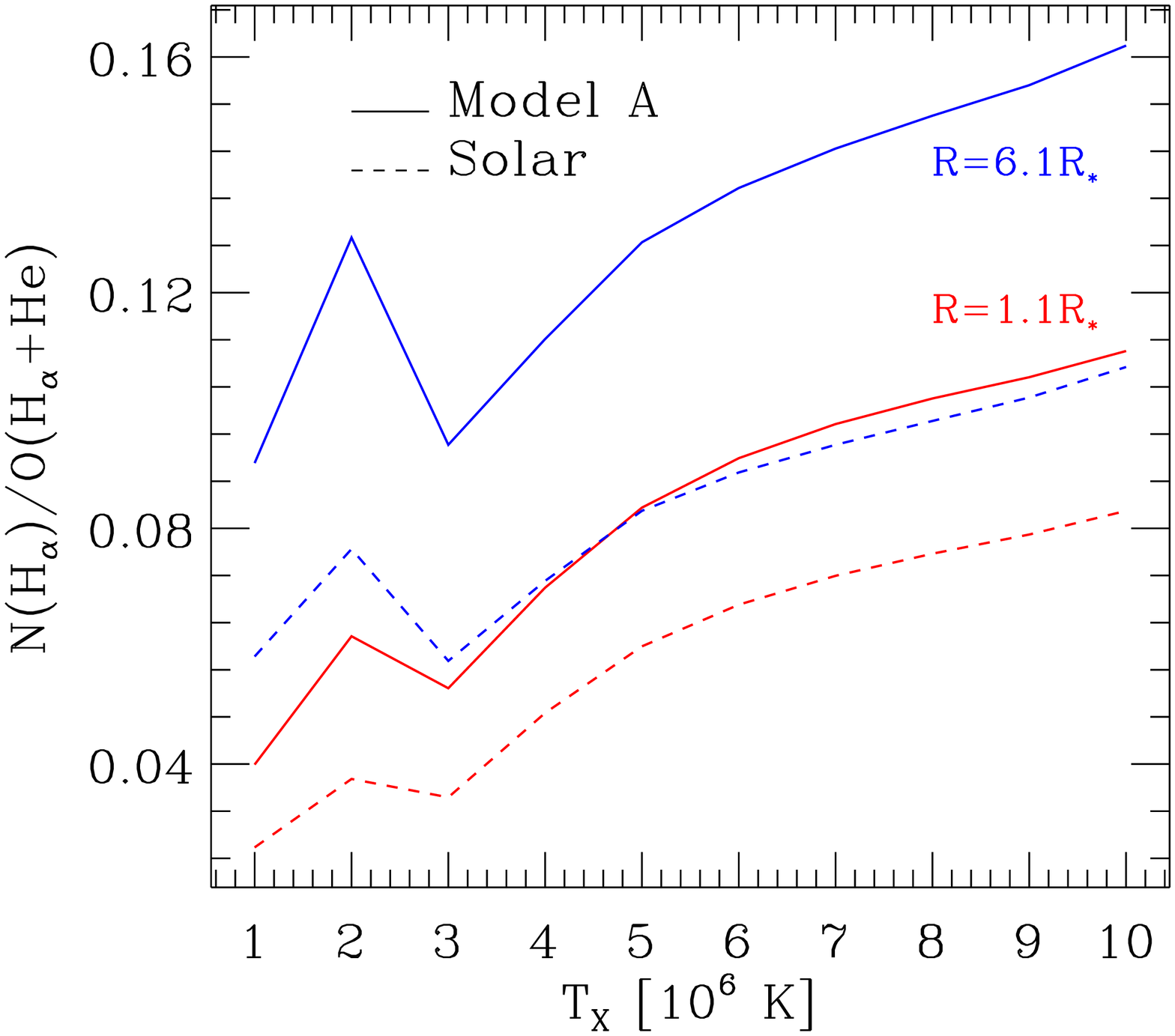}
 \includegraphics[width=0.45\linewidth]{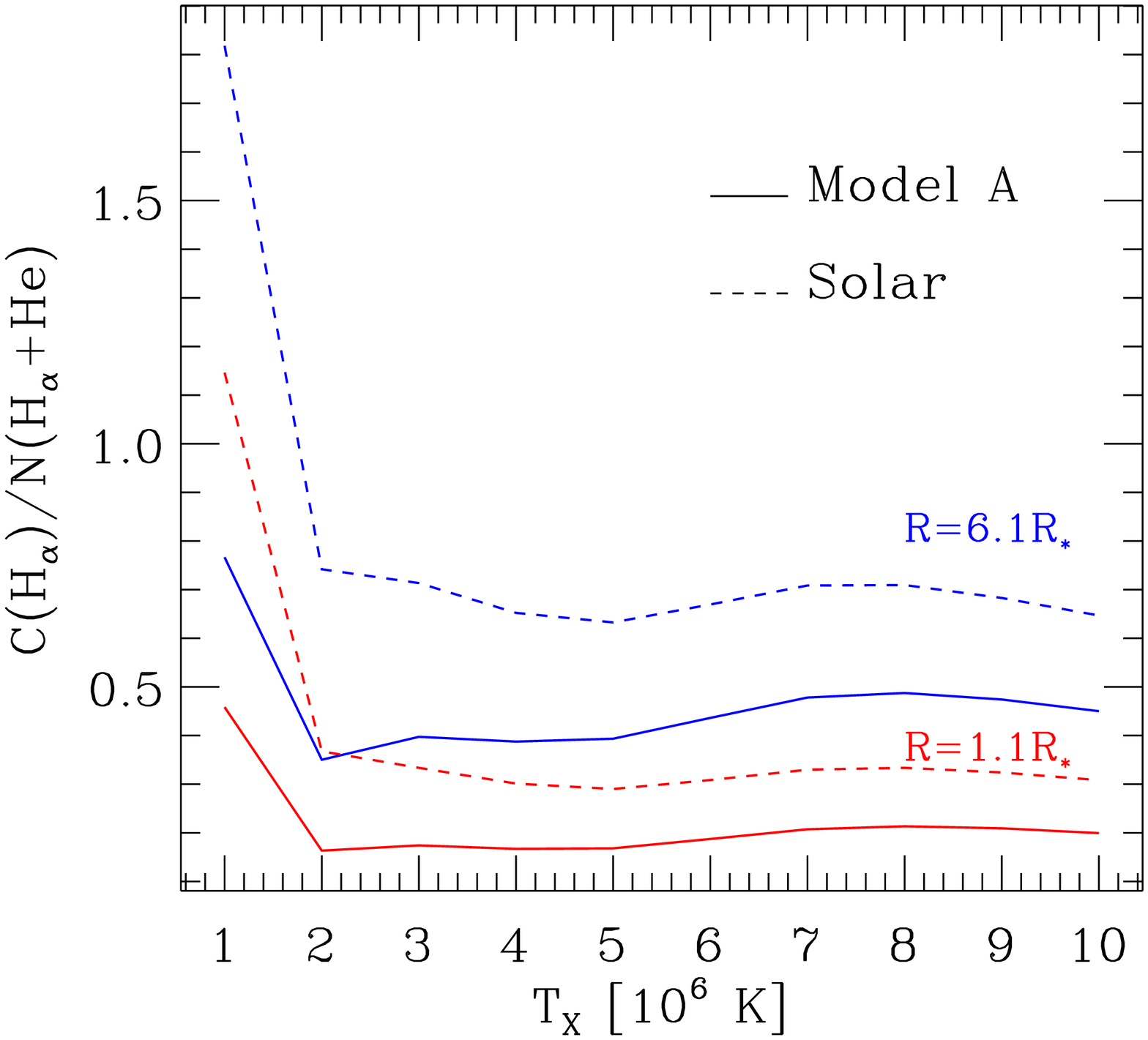}   
 \caption{Temperature behavior of  line ratios calculated with
          Equations \ref{eqrationo} and \ref{eqratiocn} with the
          temperature of emitting plasma (T$_X$).
          Blue lines correspond to models in which the 
X-ray emitting plasma starts at \Ro=6.1R$_*$ while the
red lines correspond to models with \Ro=1.6R$_*$.}\label{flratio}
\end{figure*}

In order to quantify the consistency of the abundances we calculated
lines flux ratios that show a low temperature dependence.
We used the ratios expressed in  
Equations \ref{eqrationo} and \ref{eqratiocn}. The dependence 
of these flux ratios  on temperature  is shown in Figures 
\ref{flratio}a and  \ref{flratio}b for a grid of models with
the same abundances as in model ``A'' (solid lines) and in a model
with solar abundances from \cite{asplund09}(dashed lines). 
For each set of abundances we also examined the effect of the
starting X-ray emission radius (i.e., \Ro). 
The line ratios depend on \Ro\ due to the strong dependence of the opacity on wavelength.
As longer wavelength lines are less attenuated at higher \Ro, 
the N(H$\alpha$)/O(H$\alpha$+He) ratio will increase with \Ro. Also, higher temperatures
favour the presence of \ionn{N}{vii}, while \ionn{O}{viii} + \ionn{O}{vii} is 
almost constant. This highlights the need for a reliable emission model 
if abundances are to be derived from X-ray data.  

Table \ref{txabund} shows the values of these ratios 
for model ``A'', the same model but with 
an nitrogen abundance enhanced by a factor of 1.3 
compared with observation (with errors in parentheses).
The enhanced nitrogen model improves the agreement 
between the observed and model values. Such a small
correction is within the error estimated for the N
abundance determined from optical and UV data. We
conclude that the relative abundance values derived 
from the X-ray analyses are statistically consistent with those 
derived from UV and optical analyses, and indicate a modest 
enhancement of N relative to both C and O.

\begin{table}
\caption{CNO abundances from X-ray data.
Errors on measured values are in parentheses.\label{txabund}
}
\begin{tabular}{ccccc}
\hline\hline
 & Measured & Model A & Model A 1.3 N & Solar  \\
\hline
R(N/O)    &   0.125 (0.008)   & 0.10 &  0.12  & 0.06  \\
R(C/N)    &   0.357 (0.020)   & 0.52 &  0.39  & 0.98  \\
\hline
\end{tabular} 
\end{table}

\subsection{Interclump Medium}\label{icm}

\cite{zsargo08} pointed out that it is necessary
to include the interclump medium radiation in order to reproduce
the line strength for \ionn{O}{vi} in $\zeta$ Pup.
They  showed that when the progenitor of the superion 
(e.g. \ionn{N}{iii} for \ionn{N}{v}) is the dominant ion in the wind, 
the line optical depth does not depend on wind density, 
and hence the interclump medium,
with its larger volume, can contribute significantly to the strength
of superion line profiles. We find that this is also the case for $\epsilon$~Ori.

\begin{figure*}
 \centering
 \includegraphics[width=0.45\linewidth]{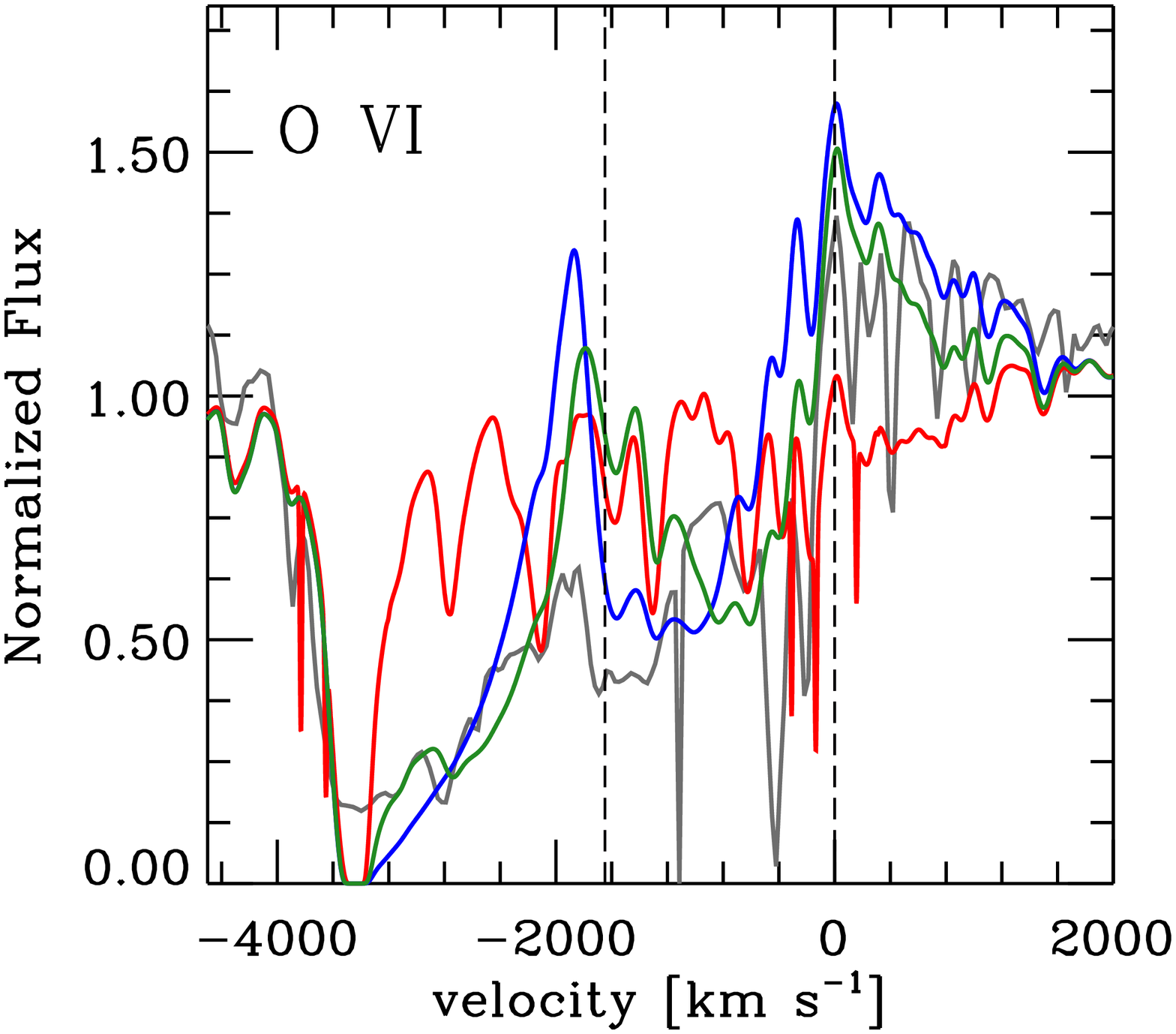}
 \includegraphics[width=0.45\linewidth]{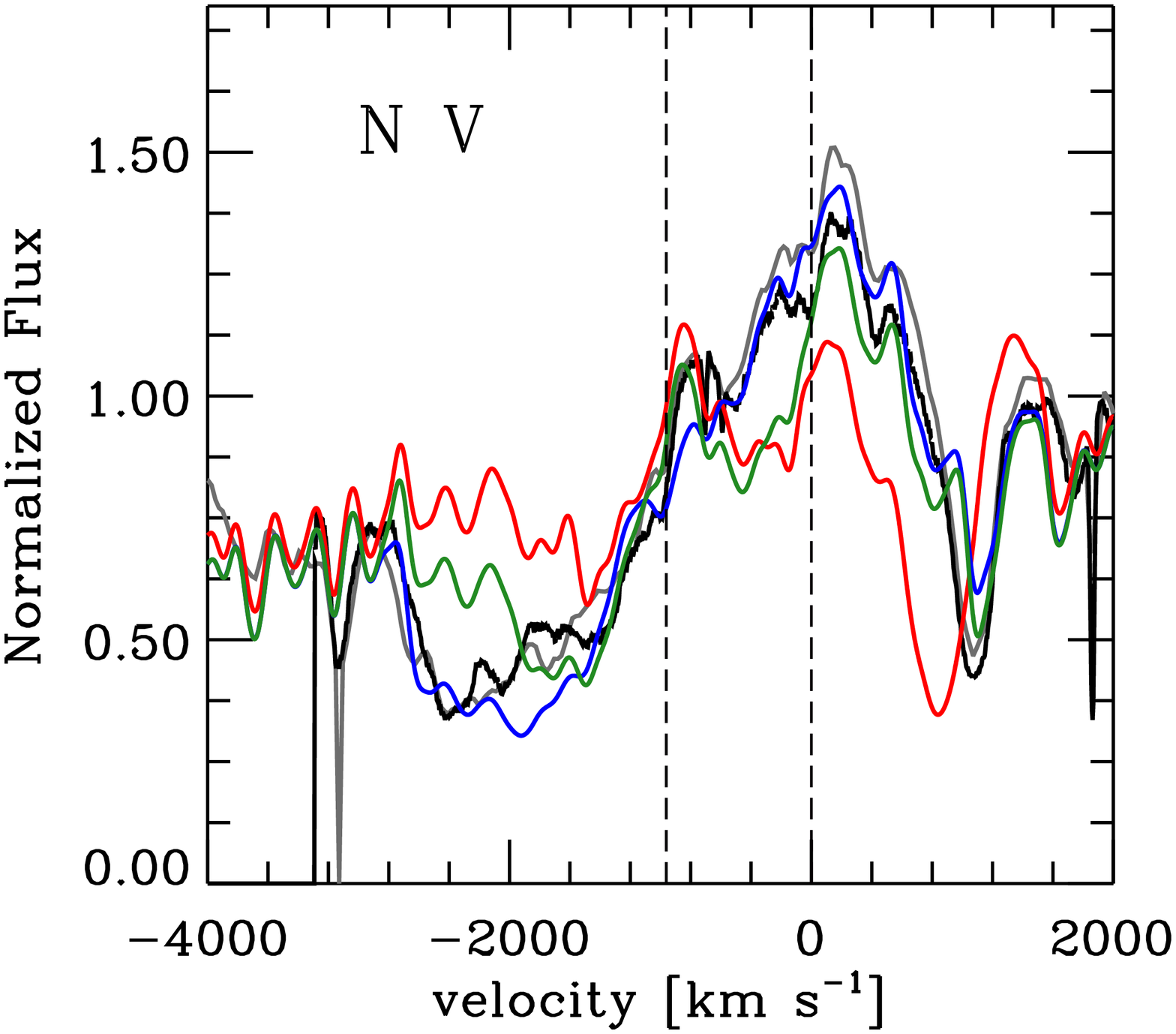}   
 \caption{The effect of including the emission of interclump medium.
          The left panel shows line profiles for \ionnll{O}{vi}{1031,1037}
          and the right panel line profiles for \ionnll{N}{v}{1238,1242}.
          The black line corresponds to \ghrs\ data, the gray line to \copernicus\
          data and the red line to the clumped model ``A''. The blue line corresponds
          to emission from the ICM with a density contrast  of 100, and the
          green line to a density contrast of 200. A much better fit to the profiles is obtained
          when we allow for the interclump medium.}\label{ficm}
\end{figure*}

Figures \ref{ficm}a and \ref{ficm}b show that our clumped model doesn't reproduce 
the observed line profiles for \ionnl{N}{v}{1238, 1242} and \ionnll{O}{vi}{1031, 1037}
(red line) in $\epsilon$~Ori. Using the method of \cite{zsargo08}, 
we calculated the emission from a tenuous medium between
the clumps. This medium occupies most of the wind volume,
with a filling factor $f_{ICM}\sim$0.9-0.95. For this
exercise we use the same radiation field from the clumped model --   
this field is independent of the ICM. We scaled the
wind density by two factors, such that
$\delta_{ICM}$=$\alpha\times\delta_{cl}$, where the $\delta_{cl}$ is the 
density in clumps, with $\alpha_1$=0.1$\times f_\infty$ 
and $\alpha_2$=0.05$\times f_\infty$, and assumed that the new 
component is smooth due to its high 
filling factor. We calculated
this model independently from the clumped model and 
the hot plasma, but using the same hydrostatic 
structure.

Figures \ref{ficm}a and \ref{ficm}b show the line profiles
calculated for two ICM models corresponding to model ``A'' 
with a density contrast between clumped and ICM of 
0.1$\times f_\infty$=0.01 (blue line) and 
0.05$\times f_\infty$=0.005 (green line). These values
mean that less than 10\% of the wind mass is in the interclump medium.
The figures show that the ICM models
predict line profiles quite similar to those observed
for the \ionn{N}{v} and \ionn{O}{vi} lines, while
the clumped model (red line) fails to reproduce the profiles.
We also tried models with even more tenuous interclump media,
but the superion line profiles started to weaken.

\begin{figure}
 \centering
 \includegraphics[angle=-90,width=0.95\linewidth]{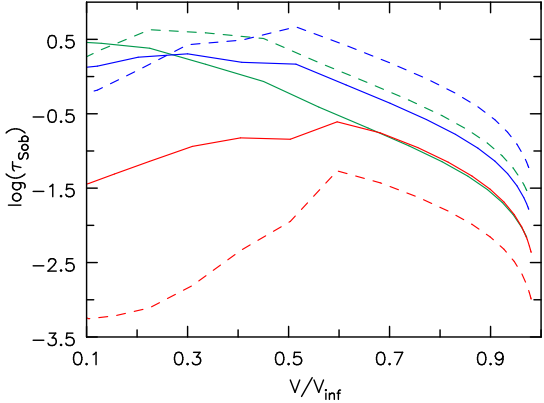}   
 \caption{Sobolev line optical depth for \ionn{N}{v} $\lambda$1238 \AA\ (solid) and  
           \ionn{O}{vi} $\lambda$1032 \AA\ (dashed).
          The red lines correspond to the clumped model ``A'', 
          while the blue and green lines are for the interclump medium models with 
          density contrasts of 100 and 200 respectively.}\label{ficmt}
\end{figure}

Figure \ref{ficmt} shows the Sobolev optical depth of the
\ionn{N}{v} $\lambda$1238 \AA\ and \ionn{O}{vi} $\lambda$1032 \AA\ 
transitions for the clumped model (red), and the
ICM models with a density contrast of 100 (blue) and
200 (green). It is evident that there is a large increase the in the
optical depth in both lines in the ICM models.
In their analysis of $\zeta$ Pup, \cite{zsargo08} found 
that the flux of the \ionn{N}{v} line has contributions 
from the clumps as well as from the ICM. 
In the case of $\epsilon$~Ori, we found that most of 
the line flux comes from the ICM for both the \ionn{O}{vi} 
and \ionn{N}{v} lines. The importance of the ICM
medium for these lines arises because of ionization effects.

The ICM has only a very weak influence
 on other lines, due to its much lower density. The ICM
medium is important for the \ionn{O}{vi}  and \ionn{N}{v} lines
because they arise from impurity species whose population
is almost independent of density in the wind if they arise from Auger ionization,
and if \ionn{O}{iv}  and \ionn{N}{iii} are the dominant ions.
Its influence on $\rho^2$ dependent lines, such
as H$\alpha$, is insignificant as was already pointed out by \cite{sundqvist11},
and as confirmed by our own calculations.
 \cite{sundqvist11} notes that macro clumping does lead to a slight weakening
 of H$\alpha$.

\subsection{Rotation}\label{rotation}

We chose  model ``A'' in order to study in more detail the influence 
of rotation on the spectrum. We re-calculated 
the spectrum using the two-dimensional method of \cite{busche05} 
as described in Section \ref{method}, with the same value for
the $v\sin i=70$\kms\ estimated from the optical data. 

\begin{figure}
 \centering
 \includegraphics[angle=-90,width=0.95\linewidth]{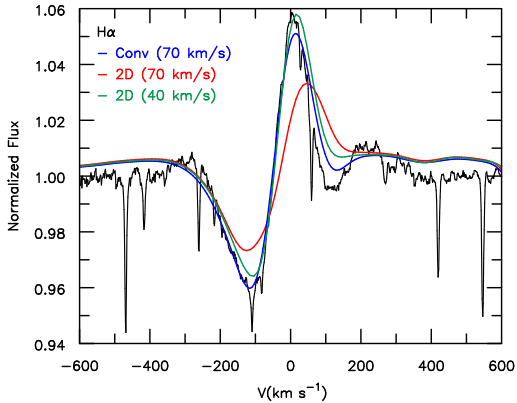}   
 \caption{H$\alpha$ profiles from model ``A''. Rotation is accounted
          for using different methods: convolution with a 70 \kms\ kernel
          (blue), 2D profile with $v\sin i$=70 \kms\ (red) and 
          2D profile with $v\sin i$=40 \kms\ (green) (see text
          for details).}\label{frotH}
\end{figure}

Figure \ref{frotH} shows the H$\alpha$ profile calculated using the
convolution method (solid blue line) and the two-dimensional radiative
transfer method (solid red line). There is a notable difference 
between the line profiles. The 2D profile shows weaker absorption and emission,
as well as a more extended red wing emission. This strong difference, while
at first surprising for such a low rotation rate (70 \kms), can be explained from the high value of $\beta$.
A high value for this parameter yields a more extended 
region where the H$\alpha$ line
is produced, and where differential rotation can affect the line profile. 
The main reason for that difference is  the dependence of profile
shape on the impact parameter as was pointed out by \cite{hillier12}.
This dependence redistributes the fluxes through the line, yielding 
wider wings emission and lower flux at the line core for both the absorption
and emission components. As a consequence, a weaker P~Cyg profile is
produced.

\begin{table}
\centering
\caption{Rotation parameter $v\sin i$, calculated
using the fast Fourier transform method for the listed optical
lines.\label{trotation} }
\begin{tabular}{lc}
\hline\hline
\multicolumn{1}{c}{Line} & $v\sin i$\\
Ion $\lambda$ [\AA] & \kms \\
\hline
\ionn{C}{ii} 4267   &   46 \\
\ionn{N}{ii} 3996   &   33 \\
\ionn{Si}{iii} 4569 &   39 \\
\ionn{Si}{iii} 4576 &   39 \\
\ionn{O}{ii} 4592   &   42 \\
\ionn{O}{ii} 4597   &   39 \\
\ionn{O}{ii} 4663   &   56 \\
\ionn{He}{i} 4389   &   56 \\
\ionn{He}{i} 4471   &   55 \\
\ionn{He}{i} 4715   &   38 \\
\ionn{He}{i} 5017   &   32 \\
\hline                
\end{tabular}
\end{table}

In order to reduce the effect described above, we reduced the
rotation parameter ($v\sin i$) to 40 \kms\ and re-calculated the
H$\alpha$ profile using the 2D code. The resulting profile 
is shown in Figure \ref{frotH} (green line). 
It is clear that this profile reproduces better 
the line shape than the former with $v\sin i$=70 \kms. 
The rest of the optical lines, such as \ionnll{Si}{iii}{4453-4576}, 
are well reproduced with 40 \kms, but with a macroturbulence of $\sim$90-100 \kms\ 
and $\xi_{min}$=15 \kms. However, such a low rotation 
had not been reported until recently -- \cite{martins15} 
found the same rotation value for \epsori\ in their variability study 
of OB stars. Their value is based on the fast fourier 
transform method \citep[see][]{simon07,gray08}.

We used the same method to estimate $v\sin i$ based
on isolated absorption lines in the optical.
The list of lines and the corresponding rotation value are shown in
Table \ref{trotation}. All of the
$v\sin i$ values are below 70 \kms, with
an average of 43 \kms, almost identical to the value
of 40 \kms\ reported by \citeauthor{martins15}

\section{Discussion}\label{discussion}

In this paper we have presented the results of a 
multi-wavelength spectroscopic analysis of \epsori. 
We conclude that the main photospheric and wind 
parameters computed from optical and UV data 
are completely compatible with those from an X-ray
analysis. We concluded that clumped models
with $f_\infty{<}0.1$ ($f_\infty{\sim}0.1$) are able to reproduce the whole spectrum
from optical to X-rays. Nevertheless, there are some points 
that arose from the analysis that need to be 
discussed more deeply.

\begin{figure*}
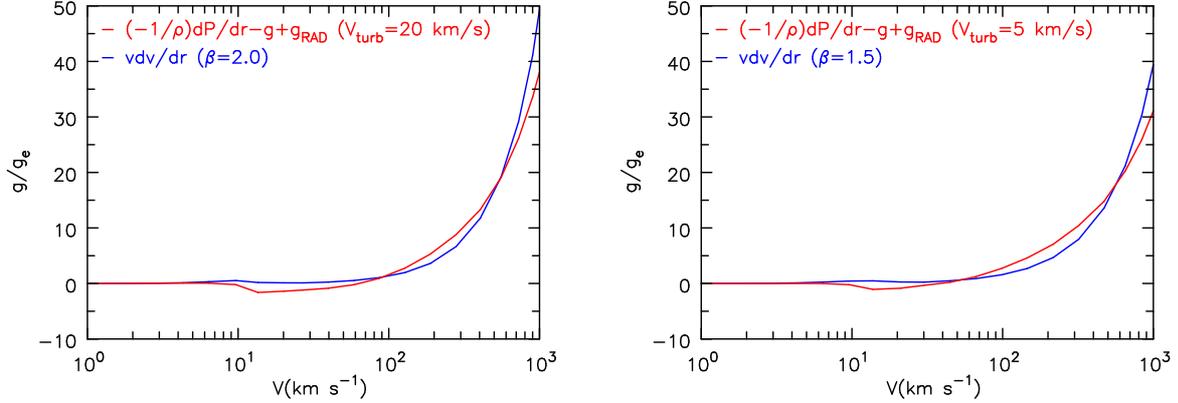

 \centering
 \includegraphics[angle=-90,width=0.45\linewidth]{fconsista.ps}
 \includegraphics[angle=-90,width=0.45\linewidth]{fconsistb.ps}
 \caption{Comparison between velocity profile and acceleration
 terms from momentum equation \eqref{eqmomentum} divided by the
 radiative acceleration of free electrons ($g_e$). Left panel
 corresponds to model ``A'' with acceleration parameter
 $\beta$=2.0 and $\xi_{min}$=20 \kms. Right panel
 corresponds to the same model, but with acceleration parameter
 $\beta$=1.5 and $\xi_{min}$=5 \kms.
 }\label{fconsist}
\end{figure*}

\subsection{Hydrodynamical consistency}

The high value for $\beta$ derived here
presents some problems for consistency with the momentum
wind equation. The radiative force is dominated by momentum 
transfer in optically thick lines. The strength of this force is 
proportional to a power of the velocity profile 
slope ($dv/dr$), which is steeper when $\beta$ is low. 
This means that a high $\beta$ value will yield 
lower radiative force that may not be able to drive the 
wind, especially around the sonic point.
Further, a larger value of $\beta$ implies that the gravity
and the radiative force are almost equal.

The momentum equation for a spherically symmetric wind is:

\begin{equation}\label{eqmomentum}
 v\frac{dv}{dr}=-\frac{1}{\rho}\frac{P_{g}}{dr}-g+g_{rad},
\end{equation}

\noindent
where $P_g$ is the gas pressure, $g=GM_*/r^2$ and
$g_{rad}$ is the total radiative force. 
Figure \ref{fconsist} shows that the right side (red) 
and the left side (blue) of Equation~\eqref{eqmomentum}  
for the model ``A'' (left panel) disagree around the 
sonic point ($v_s{\sim}$17 \kms). This also happens for models B and C. 
It is easy to see that the radiative force is insufficient 
 to match the velocity profile until approximately 
$\sim$100 \kms. The discrepancies between the 
curves are approximately 40\%, but around the sonic point 
they reach 100\%. In the outer wind there are also discrepancies
but these can probably be overcome by changes in the velocity law. However, given the assumptions of our modeling (simple radial variation
of the filling factor, and the neglect of porosity and vorosity), we do
not consider this to be a worthwhile exercise.

Some alternatives have been suggested by \cite{bouret12} to
overcome this problem. One is to increase the mass-loss
rate and the filling factor in order to increase the
radiative force and to conserve the spectral 
fitting. But in the case of \epsori, as we described
above, a filling factor $>$0.1 is incompatible with the
X-ray and UV spectra. Even the smooth model doesn't meet 
the momentum equations around the sonic point.

As discussed by \cite{bouret12}  and \cite{lucy10}, 
one solution to this problem is to reduce the microturbulent velocity
in the vicinity of the sonic point ($\xi_{min}$)
and to decrease the $\beta$ value. 
A lower $\xi_{min}$ will reduce the blocking of continuum radiation 
coming from the inner atmosphere by the blue line wings, hence
boosting the force at the sonic point. A lower $\beta$
will also increase the line radiative force at the sonic point.

The right panel of  Figure \ref{fconsist} shows the
results for a model with the same parameters of model ``A,''
but with radiative force
calculated for $\beta$=1.5 and $\xi_{min}$=5\,\kms.
The mass-loss rate for this model was adjusted 
to fit the emission strength of H$\alpha$.
The figure shows that even though the curves approach 
each other around the sonic point the hydrodynamical  
equation is still not satisifed\footnote{The hydrostatic equation 
is not exactly satisfied at 10\,\kms\ in these models as 
we neglected clumping when the hydrostatic equation was 
solved. The slight departure of $f$ from unity (e.g., 0.8) 
in this region is sufficient to cause the observed discrepancy.}. 
Furthermore, lowering the turbulence
velocity contradicts the values derived from the 
spectral analysis.

Another option is to increase the number of 
levels taken into account in atomic models 
as well as to include more species in the computation. 
To investigate this point we calculated a model increasing the number of 
levels for Fe ions and including Ne, Cl, Ar, and Ca 
with solar abundances. However, we found that these 
changes did not improve the momentum balance
for the default wind law with $\beta$=2 and a microturbulence
of $20\,$\kms. 

The discrepancies may be due to some simplifications that are commonly assumed for modeling 
stellar winds. Microturbulence and macroturbulence are not well
understood, as well as their influence on the wind dynamics, especially
around the sonic point. Further, variability of line profiles in
\epsori\ shows that the assumption of a steady-state wind is invalid.
Indeed, the difficulty of driving the wind through the
sonic point in \epsori\ may be explicitly linked with the observed variability
of H$\alpha$.

Recent studies of variability in \epsori\ have been 
undertaken by \cite{prinja04}, \cite{thompson13} and
\cite{martins15} based on H$\alpha$, \ionn{Si}{iii} and \ionn{He}{i} line profiles. 
The detected variability has time scales of a few days
and those works linked wind variability with 
photospheric activity that could be caused by stellar 
pulsations or small scale magnetic fields.

\subsection{\ionn{S}{iv} profiles}

As for phosphorus, sulfur has a relatively low abundance, and hence its resonance lines commonly appear unsaturated.  As a consequence these lines can provide useful  constraints on the mass-loss rate, especially
if S$^{3+}$ is the dominant ion in the wind (as it is for \epsori). The lower left panel of Figure~\ref{fclump} shows the  dependence on filling factor of the  theoretical profiles for the \ionnll{S}{iv}{1062,1073} doublet.\footnote{The $\lambda 1073$ line is a blend of two lines at 1072.96\,\AA\
($gf=0.168$) and 1073.51\,\AA\ ($gf=0.0156$) which have the same lower level. The $\lambda 1062$ line has a $gf=0.1$ (Kramida, A., Ralchenko, Yu., Reader, J. and NIST ASD Team 2014. NIST Atomic Spectra Database (version 5.2). Available: http://physics.nist.gov/asd. National Institute of Standards and Technology, Gaithersburg, MD).}
 As is readily apparent from the figure, we are unable to reproduce the observed profile shape even with the lowest $f_\infty$ value (0.01). The sulfur lines show weak  absorption and almost no emission in their P~Cygni
profiles, in stark contrast to the models that show strong emission and blue absorption. Particularly striking is the absence of absorption at high velocities.

\begin{figure}
 \centering
 \includegraphics[width=0.8\linewidth]{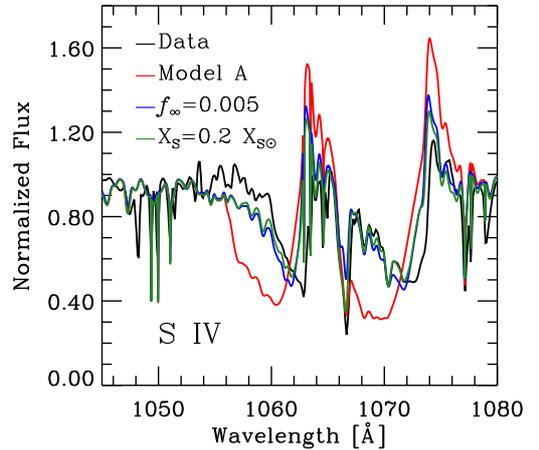}
 \caption{Comparison of the observed line profiles of 
          \ionnll{S}{iv}{1062,1073} with synthetic profiles computed using
          the model ``A'' parameters (red), high clumped model (blue)
          and low sulfur abundance model (green).
          The black line corresponds to \copernicus\ data.}\label{fsulfur}
\end{figure}

As  shown in Figure~\ref{fsulfur}, we can only ``fit" the sulfur doublet 
if we use an absurdly low abundance ($\sim$0.2 S$_\odot$ with  
$f_\infty$=0.1) or filling factor ($f_\infty{\sim}0.005$).  Even with this 
abundance, the calculated line profiles show more 
emission than the data. Moreover, an abundance of 0.2 S$_\odot$ 
is incompatible with other UV and optical sulfur lines 
(e.g., \ionnl{S}{iv}{1098-1100,4486} and \ionnl{S}{iii}{4254}). 
This problem is quite similar to that  reported previously with 
\ionnll{P}{v}{1117,1128} in O stars \citep[e.g.][]{hillier03,bouret03,fullerton06, fullerton08,bouret12}, and hence we consider similar solutions.

Various solutions have been proposed for solving the \ionnl{P}{v}{1117,1128}\ problem, including
the following:
\begin{enumerate}
\item
Sub-solar abundance. While factors of 2 variation between stars cannot
be ruled out, much larger variations are unlikely. In the present case,
other S lines rule this out as the main cause.
\item
Wrong wind ionization, possibly due to the influence of X-rays. However,
the required X-ray flux is higher than observed. 
\item
Wrong wind ionization, possibly because the bulk of the wind has been shocked
and not cooled.
\item
The wind is porous spatially, or porous in velocity space (hereafter called vorosity).
Porosity will affect both observed X-ray and UV line profiles, whereas vorosity will only affect UV line profiles.
\end{enumerate}

The influence of X-rays on the ionization structure was analyzed
 by \cite{marcolino09} in the context of the weak wind 
problem of late O dwarf stars. They found that the 
X-ray luminosity necessary to fit the \ionn{C}{iv} line strength
is approximately 3 dex above the standard value of $L/L_{bol}{\rm{=}-7.0}$.
In the same sense, \cite{krticka09,krticka12} 
explored the influence of X-rays on the phosphorous ion 
fraction $q(P^{+4})$ aiming to fit the \ionn{P}{v} lines. 
They concluded that hard X-ray emission does not have a strong 
effect on $q(P^{+4})$, whereas XUV (54 - 124 ev) emission 
yields a $q(P^{+4})$ according to observations. 
However, X-ray and XUV emission affects the
ionization structure of other ions (e.g. \ionn{C}{vi},
\ionn{N}{iv} and \ionn{O}{iii}). This influences the
wind terminal velocity value in a non-negligible way,
contradicting the observations \citep{krticka12}.

In our models the influence of X-rays on the ionization
fraction of $S^{+3}$ is less than 5\% for the 
clumped winds ($f_\infty = 0.01-0.1$), while for the 
smooth wind the $q(S^{+3})$ is $\sim$25\%. 
Manipulating the $f_X$ values for the cold
plasma (\scie{1}{6} K), we computed a model with an enhanced 
emission in the range of 160-250 \AA, just above the
ionization threshold of \ionn{S}{iv}.
We found that even though the bluer absorption component of 
\ionn{S}{iv} lines is weakened, it wasn't enough to 
fit the observed profiles. Furthermore, the XUV radiation
needed to fit the lines would be extremely high.
Moreover, we obtained reasonable \ionn{S}{iv} profiles setting \Ro\
deeper in the wind (\Ro=1.1 R$_*$). Nevertheless, this change
increases the \ionn{P}{v} abundance in such a way that the 
lines \ionnll{P}{v}{1118,1128} become too strong. Besides that, such \Ro\
value is ruled out by X-ray analysis.

Macroclumping (porosity) could reproduce the resonance lines of
\ionn{P}{v} without requiring extremely low mass-loss rates, 
since optically thick clumps will reduce the effective opacity 
of the lines \cite[e.g.][]{owocki08, oskinova07}. 

Another effect to be taken into account is the so-called ``velocity-porosity'' or
\emph{vorosity} \citep{owocki08} that evaluate the effect
of a non-monotonic velocity or a velocity field with  strong jumps or holes (in the velocity space) on line radiative transfer. 

\cite{owocki08} showed that \emph{vorosity} yields a reduction in line absorption
but it is low (10-20\%). This makes the \emph{vorosity} effects weak
when compared with \emph{porosity} ones.
However, simulation from \cite{sundqvist10} and \cite{sundqvist11}
\citep[see also][]{sundqvist14}
point out that \emph{vorosity} has an important effect on 
UV profiles that improves the agreement between
optical and UV line mass-loss rates.

Using 3D Monte-Carlo simulations \cite{surlan12,surlan13}
argue that porosity dominates over vorosity effects.
An issue with all of these studies is that they primarily utilize the \ionn{P}{v} profiles --
consequently the fit parameters are not unique.

We postulate that the problem of \ionnll{S}{iv}{1063,1073}
is analogous to the \ionn{P}{v} problem, and this gives us a
good chance to study clumping in early B stars where
\ionn{P}{v} is not the dominant ion. Furthermore, looking at
the \copernicus\ and \fuse\ catalogs of galactic B stars, 
it is clear that \ionn{S}{iv} lines are weak even though 
we expect that it is the dominant ion in the early B 
stellar winds.

More evidence for the existence of porosity and vorosity effects 
can be found in the optical depth ratio of UV resonance doublets. 
\cite{prinja10} pointed out that this ratio should be equal to the ratio 
of the oscillator strengths when the clumps are
optically thin while for optically thick clumps this ratio is equal to 1. 
The doublet \ionnll{Si}{iv}{1393,1402} is
an excellent candidate for this diagnostic since the oscillator
strength ratio is $\sim$2 and the splitting between 
the components ($\Delta v{\sim}1940$ \kms) is large (although it is 
less than 3600 \kms, twice the computed terminal velocity).
The data show that the absorption associated with the
\ionn{Si}{iv} doublet components are  significantly weaker than
 the model profiles which are relatively 
insensitive to the assumed volume filling factor (see 
second panel from figure \ref{fA5}). This could be evidence for a reduction in the effective opacity by porosity and/or vorosity.

In the case of \ionnll{S}{iv}{1062,1073}, the red component is
stronger, with a $gf$ value twice that of the blue component.
Thus, if vorosity were the dominant effect models that fit
the red component (by adjusting the abundance or filling factor)
would underestimate the strength of the blue component. 
Surprisingly this is not  the case -- if anything our models do 
a worse job at matching the bluer component with our model having 
too much absorption at high velocities. Excluding issues with 
the observational (and atomic) data, this would suggest that the
problem is with the ionization structure of the wind, not an issue
with vorosity. Further evidence comes from the shape of the profile. 
In standard calculations holes in velocity space decrease with velocity, 
leading  to enhanced absorption near the terminal velocity 
which is not seen in the observed profiles 
\citep[e.g.][]{sundqvist14}.
\cite{petrov14} suggested that H$\alpha$ could
be affected by vorosity, but only for stars below the bi-stability
jump. \epsori\ is above that limit, furthermore, \cite{sundqvist11}
showed that this effects are weak on H$\alpha$. Thus H$\alpha$
does not provide a suitable diagnostic of porosity.

\subsection{X-rays}\label{dxrays}

An important aspect of our calculations is the derivation of the spatial distribution
of the X-ray emitting plasmas in the wind. The \Ro\ values from
Table \ref{txmodel} are higher than previous estimates
for other massive stars \citep[see e.g.][]{leutenegger06} or even
with those reported  by \cite{cohen14} for \epsori\ itself.
We also found a trend of \Ro\ vs emitting ion. 
This trend does not contradict the numerical 
simulations that predict strong acceleration of some elements of 
the wind close to the surface \citep{feldmeier97,runacres02,dessart03,dessart05},
since it could be caused by optical depth effects.

The main reason for the larger \Ro\ is the high value of 
$\beta$. With a larger $\beta$ the wind velocity 
necessary to fit the line widths occurs at larger radii. 
For a model with $\beta{=}1$, $R_0$ is closer
to the surface, and variation of \Ro\ with the temperature
is weaker. This is consistent with the finding of constant 
$R_0$ for \epsori\ by \cite{cohen14}. However, we stress 
that the link between \Ro\ and $\beta$ that we found 
in this work, provides a new challenge for understanding
 shock formation in the wind and the X-ray emission
of massive stars. What does \Ro\ actually mean and
what is behind its relation with wind acceleration?
In order to understand the underlying physics
of these connections, it is necessary to expand the analysis to more systems. 

From the models it is possible to estimate the region where 
the bulk of each component of He-like lines is emitted. 
The local contribution to line flux is proportional  to 
$r^2n^2_p(r)\epsilon(r)$, where $\epsilon(r)$ is the line 
emissivity and $n_p$ is the density of the emitting 
plasma. For instance, for \ionn{O}{vii} \ionn{He}{$\alpha$}, the intercombination 
component ($i$) has its maximum contribution at $\sim$4.6 R$_*$.
For the forbidden component ($f$), the corresponding value is $\sim$28.5 R$_*$. 
These values show that the different components of He-like triplets are 
mostly produced in different regions in the wind.

One caveat with our analysis is that 
our assumption of a small set of single-temperature plasmas
to describe the shock emissions is not realistic.
In the inner wind the cooling time 
($T_{cool}\sim1/\rho$) for each shock is less than the dynamic time, and
hence modeling of the shock structure is required to accurately predict
the X-ray emission.This will be explored in the future.

\begin{figure*}
 \centering
 \includegraphics[width=0.48\linewidth]{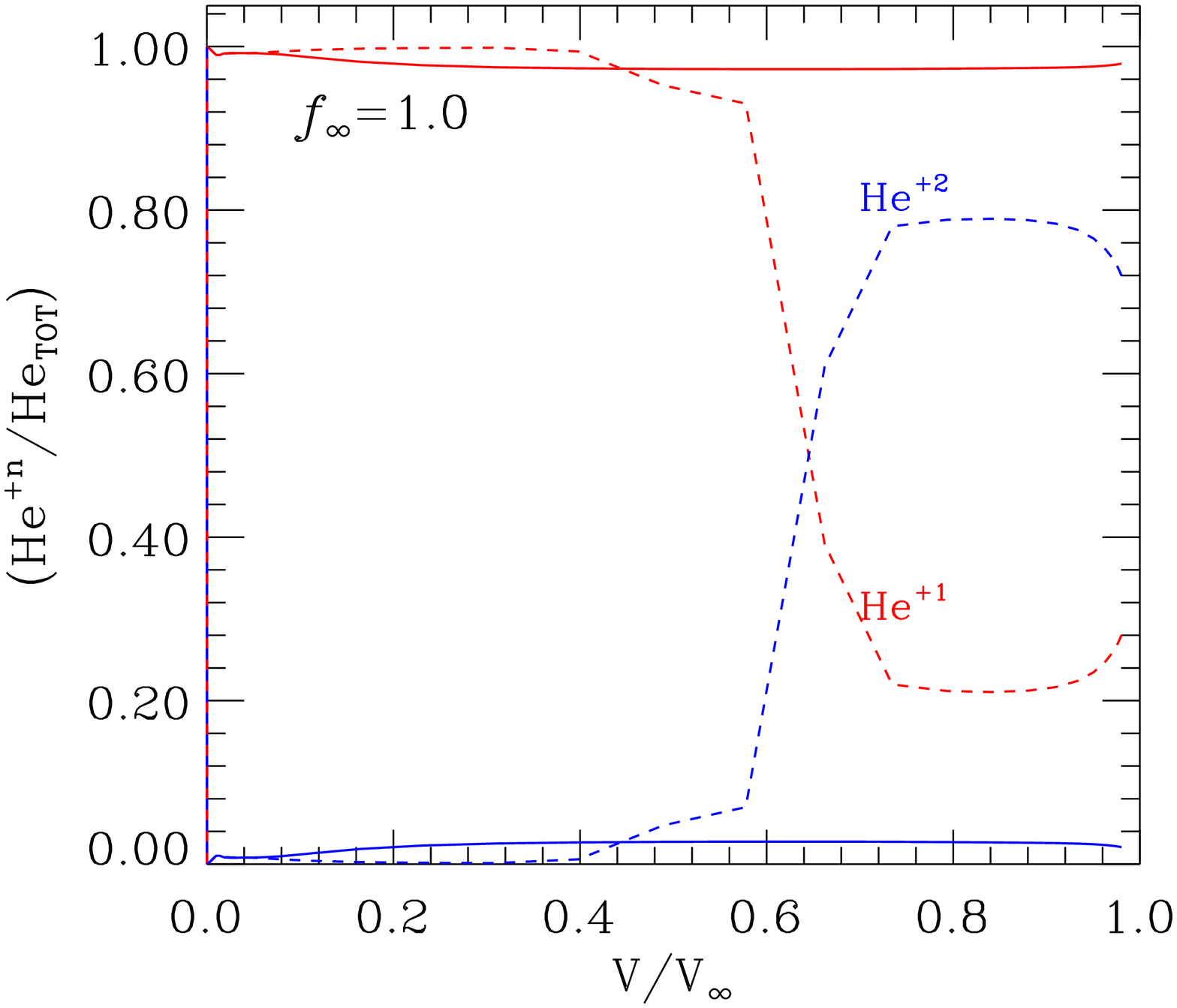}
 \includegraphics[width=0.48\linewidth]{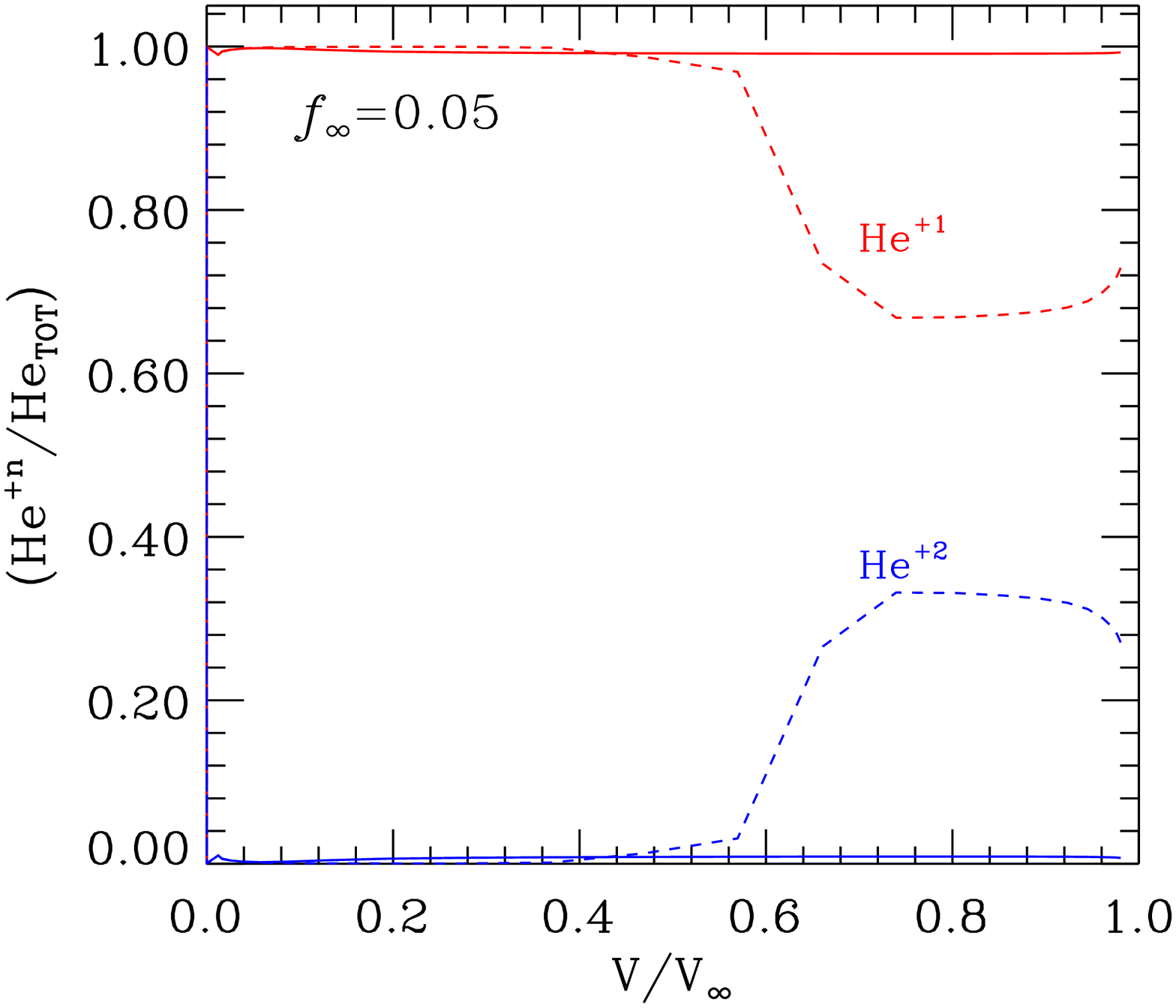}
 \caption{Effect of including X-rays on the ionization structure
          of He. The solid line corresponds to a model without
          X-rays, dashed lines correspond to models that
          include X-ray emission that fit the \chandra\ and
          \xmm\ data.
         }
          \label{fhelium}
\end{figure*}

All of our models yield asymmetric lines, with a lower 
degree of skewness for models with lower filling factor.
The observed blue-shifts of approximately ${\sim}210{\pm}123$ \kms\ 
for lines with wavelength $\lambda{\gtrsim}10$ \AA\ provide  
clear evidence of wind absorption.
While a clumped wind
is necessary to fit the line profiles we cannot
distinguish between the  ``A'' or ``B'' models with $f_\infty$=0.1 
and 0.05 respectively. Both models yield adequate X-ray profiles 
with reasonable mass-loss rates without requiring the inclusion of porosity or resonance scattering effects. 
This confirms the mass-loss rate derived by \cite{cohen14} that 
did not use porosity effects to fit X-ray line profiles. 
Even though, our results and the \citeauthor{cohen14} ones
cannot confirm the absence of porosity. Much higher signal-to-noise X-ray observations are needed if tighter
constraints are to be obtained. Moreover, due to the strong 
variability of H$\alpha$, the influence of non-spherical winds 
and temporal variability should be considered.

It is generally assumed that X-rays don't strongly affect the 
X-ray opacity of the ``cool'' wind. In the case of \epsori\ we found 
that this is  not  necessarily true. In \epsori\ the  X-rays can ionize 
He$^+$ in the outer wind (Fig.~\ref{fhelium}), reducing the X-ray 
opacity (see the behaviour at outer regions of the red 
line in Fig.~\ref{ftaux}). 
The effect is largest in the unclumped 
model. Normally, X-rays don't strongly affect the 
dominant ionization stage as X-ray flux is generally 
small compared with the UV ionizing flux near the relevant
thresholds -- this is not
the case for \epsori.

The effect of X-rays on He ionization in wind  has two consequences: 
1) For modeling B or later stars, it is important to take into account
the interaction of X-rays with the wind consistently in
order to estimate the wind parameters, and not separately
as commonly done for O stars. 2) The value of the
mass-loss rates computed using X-ray line profiles may be biased
toward low values if this effect is not taken into account.

Finally, our preferred models, A and B, show a very strong
trend of higher X-ray filling factors for low temperature
emission components (see table \ref{txmodel}). This model
fitting result accounts for both wind absorption and the
different emissivities of different plasma temperature 
components, and thus reflects the instantaneous plasma
temperature distribution.
These results are broadly consistent with a shock-heating
analysis that shows a very strong preference for weaker shocks
over strong ones, with almost no shock heating above
10$^7$ K.

\section{SUMMARY AND CONCLUSIONS}\label{conclusions}

We have presented a multi-wavelength analysis of the 
early B supergiant \epsori. This analysis utilizes a modified version of 
\cmfgen\  that allows for the radiative interactions between the 
X-ray emitting plasma and the cool wind.

We get excellent fits to X-ray, UV and optical data.
The derived photospheric parameters for \epsori\ are quite similar
to those  previously derived
(T$_{eff}$=27000$\pm$500 K and $\log g$=3.0$\pm$0.05). The 
observed UV fluxes together with the known $m_V$ yields  
\ebv=0.091$\pm$0.01. Using the last \hipparcos\ distance 
(606 pc), the \epsori\ luminosity is $\log (L_*/L\odot){\sim}5.93$. 
This luminosity is the highest reported value for \epsori. 
Using the former \hipparcos\ distance 
(412 pc), the \epsori\ luminosity is $\log (L_*/L\odot){\sim}5.60$.
   
The metal abundances derived from the different passbands
(X-ray, optical and UV) are consistent within errors.
We derive [N/O]=0.15 and [N/C]=0.33 for \epsori. Optically
derived CNO abundances show a factor of two dispersion about the
mean, suggesting there are still some inadequacies in the atomic
models, and/or the treatment of turbulence. 

The UV \ionnll{S}{iv}{1068,1073} profiles
show the same problem as reported for \ionnll{P}{v}{1117,1128} in 
O stars \cite[e.g.,][]{fullerton06}. To fit this doublet (actually three lines but two are blended) 
it is necessary to use an extremely clumped wind and/or a very low sulfur abundance.
The latter is incompatible with other UV and optical sulfur lines.
Close examination of the profiles suggests an issue with the calculated 
ionization structure of the outer wind (possibly related to the presence of a hot plasma), 
although porosity and vorosity effects may also be important.

A clumped wind, with a volume filling factor of order 0.1 or less, 
produces X-ray profiles consistent with observation, favoring 
\Mdot$\leq$\massrate{4.9}{-7}, a value approximately 4 times
lower than the one theoretically predicted by \cite{vink00,vink01}.
However the signal-to-noise ratio of the X-ray data does not
allow us to determine an accurate $f_X$ value, or to examine the effects of
porosity. We also  found evidence that the UV profiles \ionn{Si}{iv} and \ionn{S}{iv}
could be affected by porosity or problems with the ionization structure.

As found for $\zeta$~Pup \citep{zsargo08}, we found evidence of resonance scattering    
from the interclump medium -- the traditional clumped model cannot reproduce 
the observed \ionn{O}{vi} and \ionn{N}{v} resonance profiles.

The $\beta{=}2$ value reported in this work is higher than found in earlier work, 
and than that expected from standard wind theory. However the observed H$\alpha$ 
is variable, and thus time variable and non-spherically symmetric models are 
needed to analyze the profiles.

\section*{Acknowledgments}   
   
Support for this work was provided by the National Aeronautics 
and Space Administration through Chandra Award Number ARO-11002A  
issued by the Chandra X-ray Observatory Center, which is operated 
by the Smithsonian Astrophysical Observatory for and on behalf of
the National Aeronautics Space Administration under contract NAS8-03060.
This work was also supported by NASA Chandra grants:
AR2-13001A \& AR0-11002B. D. John Hillier also acknowledges partial
support from  STScI theory grant HST-AR-12640.01.
Maurice A. Leutenegger also acknowledges the support from
Chandra, grants: G02-13002A and AR2-130001B.
David Cohen also acknowledges the support from Chandra, 
grants: TM3-14001B, AR0-11002B and AR2-13001A.
Janos Zsargo acknowledges CONACyT grant CB-2011-01 No. 168632.
We also acknowledge Francisco Najarro for his
highly valuable comments and suggestions on this manuscript.
We are also grateful to Randall Smith for providing us the source 
code of APEC and to the Chandra X-ray Center for the use of ATOMDB.
We also thank the anonymous referee for his valuable comments 
that helped us to improve this manuscript.



\bibliography{bibliography.bib}

\appendix
\renewcommand\thefigure{A.\arabic{figure}}

\section*{Appendix A}

In this appendix we present best fit model for each filling 
factor used in this work $f_\infty$=1.0 (red), 0.1 (blue),
0.05 (green) and 0.01 (orange). Figures A.1 
to A.3 show the data (black) and the models in the optical band.
Figure A.4 shows the \copernicus\ data (black) and the 
models in FUV. Figure A.5 shows the \iue\ data (black) and the 
models in UV. The shown models have a rotation  
$v\sin i$=70 \kms and $v_{macro}$=70 \kms.  

\begin{figure*}
 \centering
 \includegraphics[width=\linewidth]{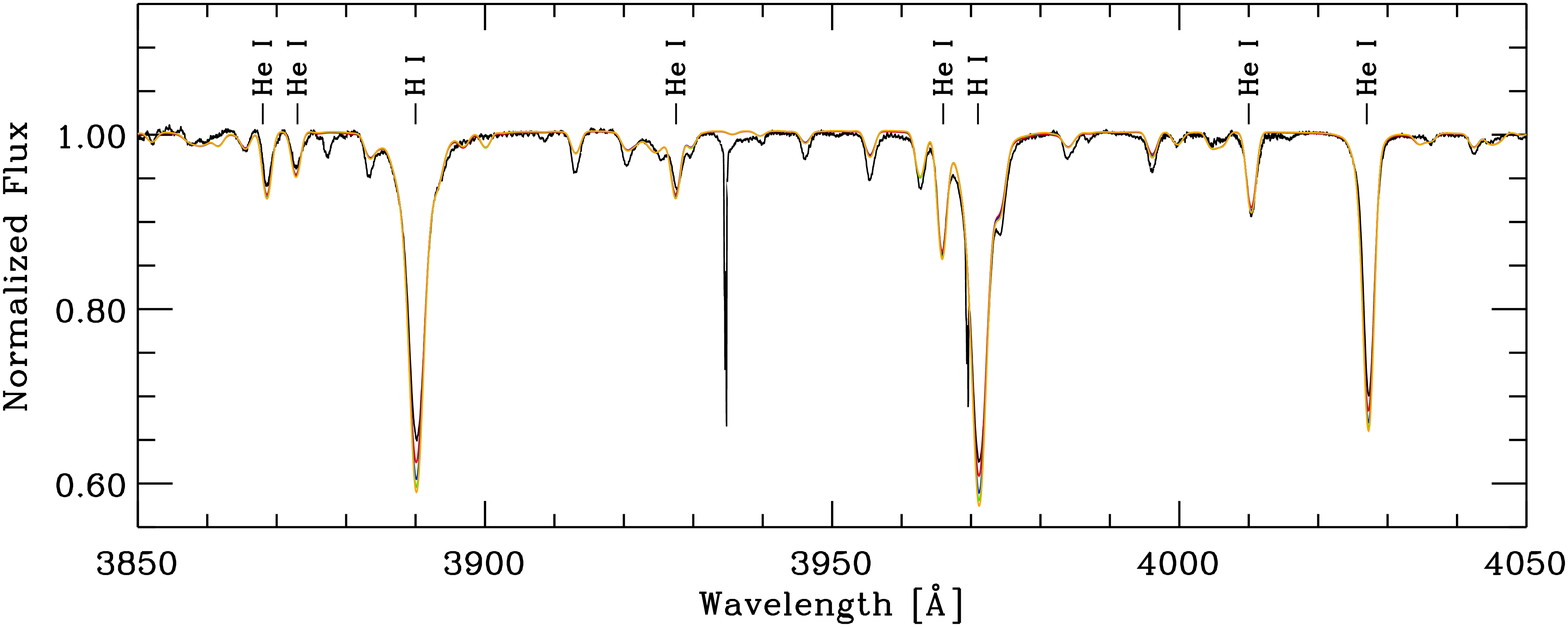} \\
 \includegraphics[width=\linewidth]{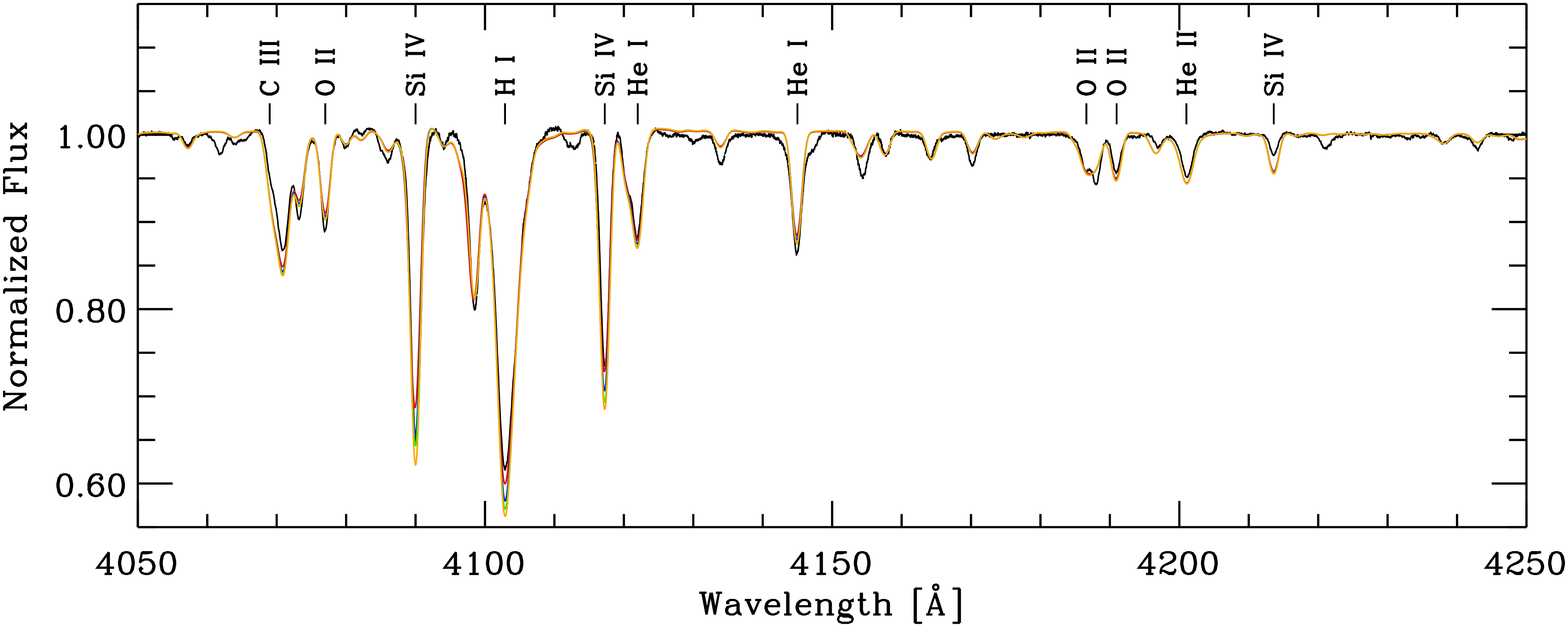} \\
 \includegraphics[width=\linewidth]{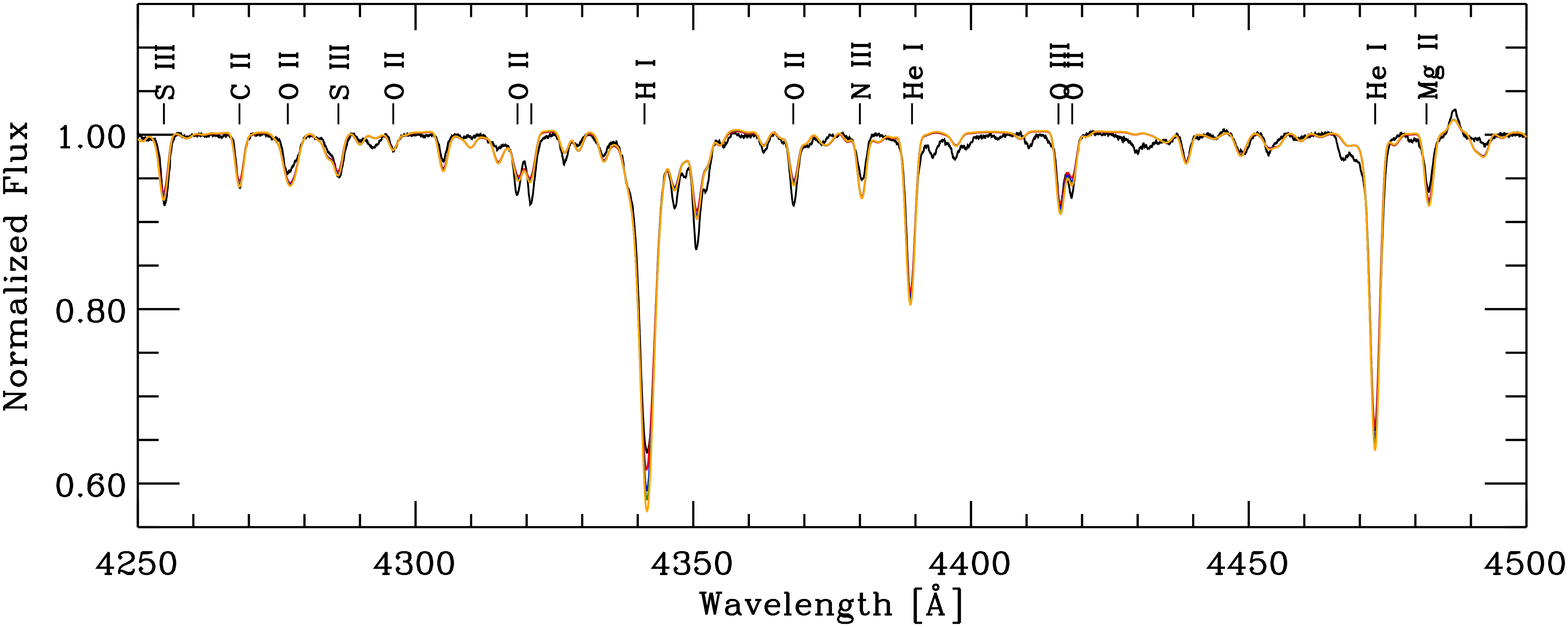} 
 \caption{Best model fits for \epsori\ for each $f_\infty$ value: 
          1.0 (red), 0.1 (blue), 0.05 (green) and 0.01 (orange). Optical
          data are shown by the black line.}\label{fA1}
\end{figure*}

\begin{figure*}
 \centering
 \includegraphics[width=\linewidth]{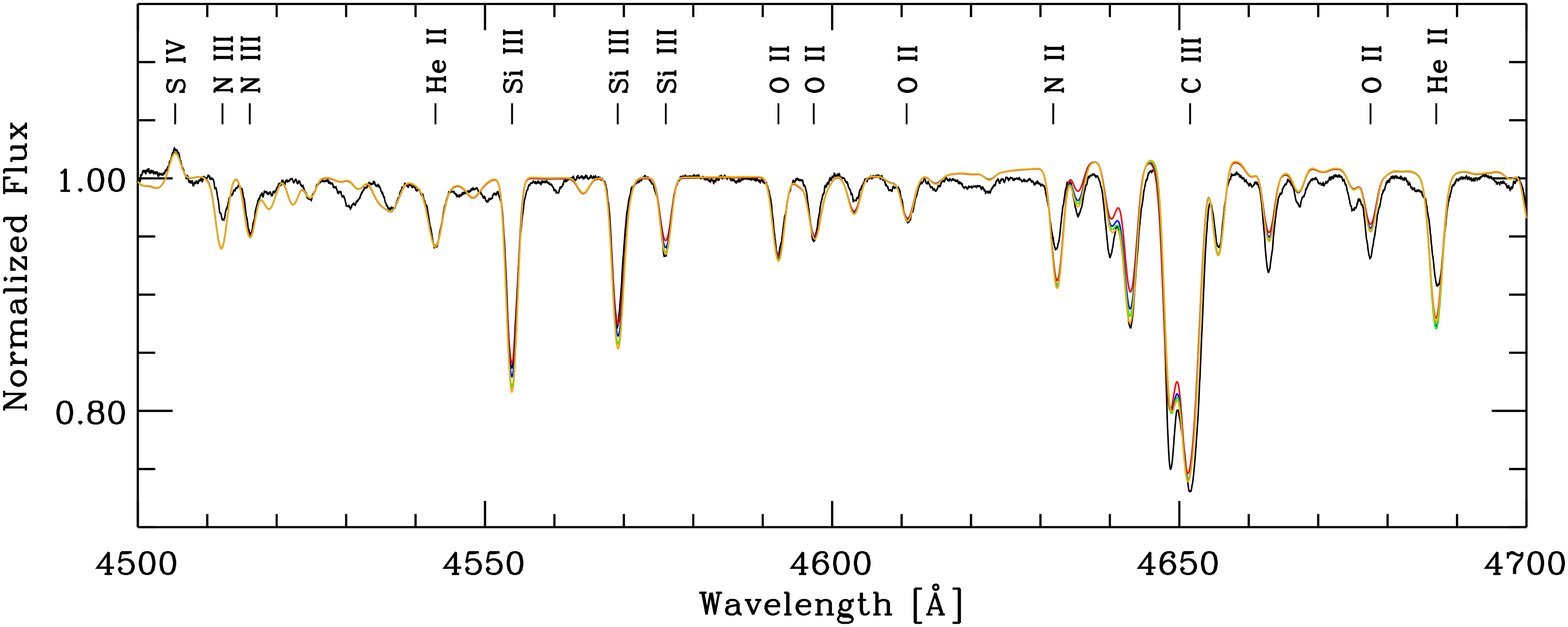} \\  
 \includegraphics[width=\linewidth]{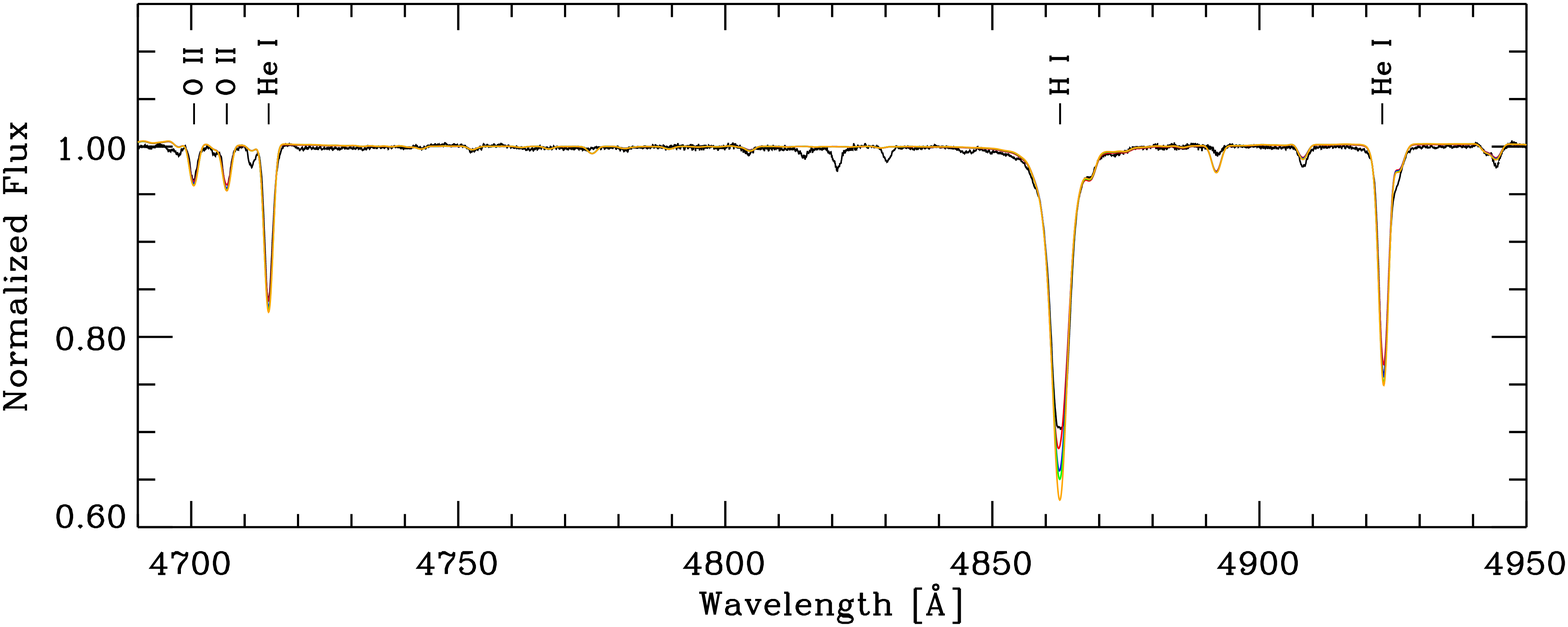} \\
 \includegraphics[width=\linewidth]{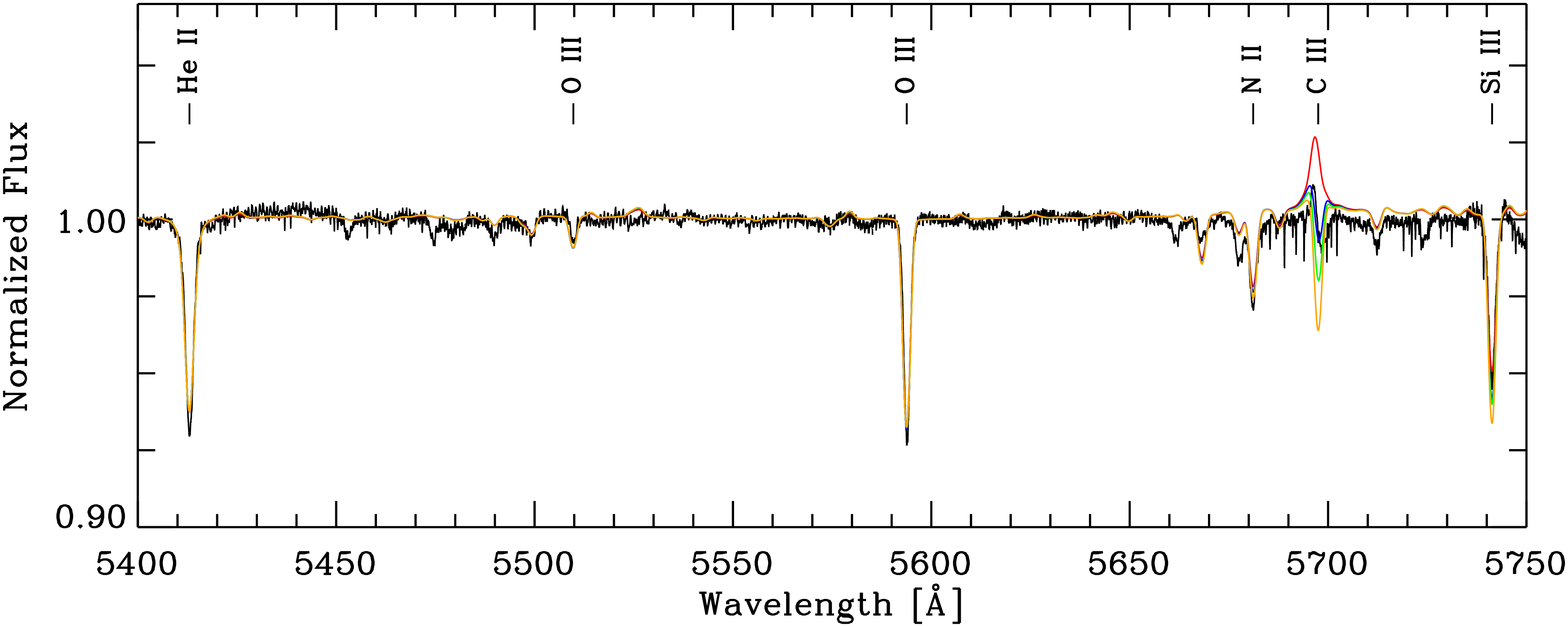}
 \caption{Best model fits for \epsori\ for each $f_\infty$ value: 
          1.0 (red), 0.1 (blue), 0.05 (green) and 0.01 (orange). Optical
          data are shown by the black line.}\label{fA2}
\end{figure*}

\begin{figure*}
 \centering
 \includegraphics[width=\linewidth]{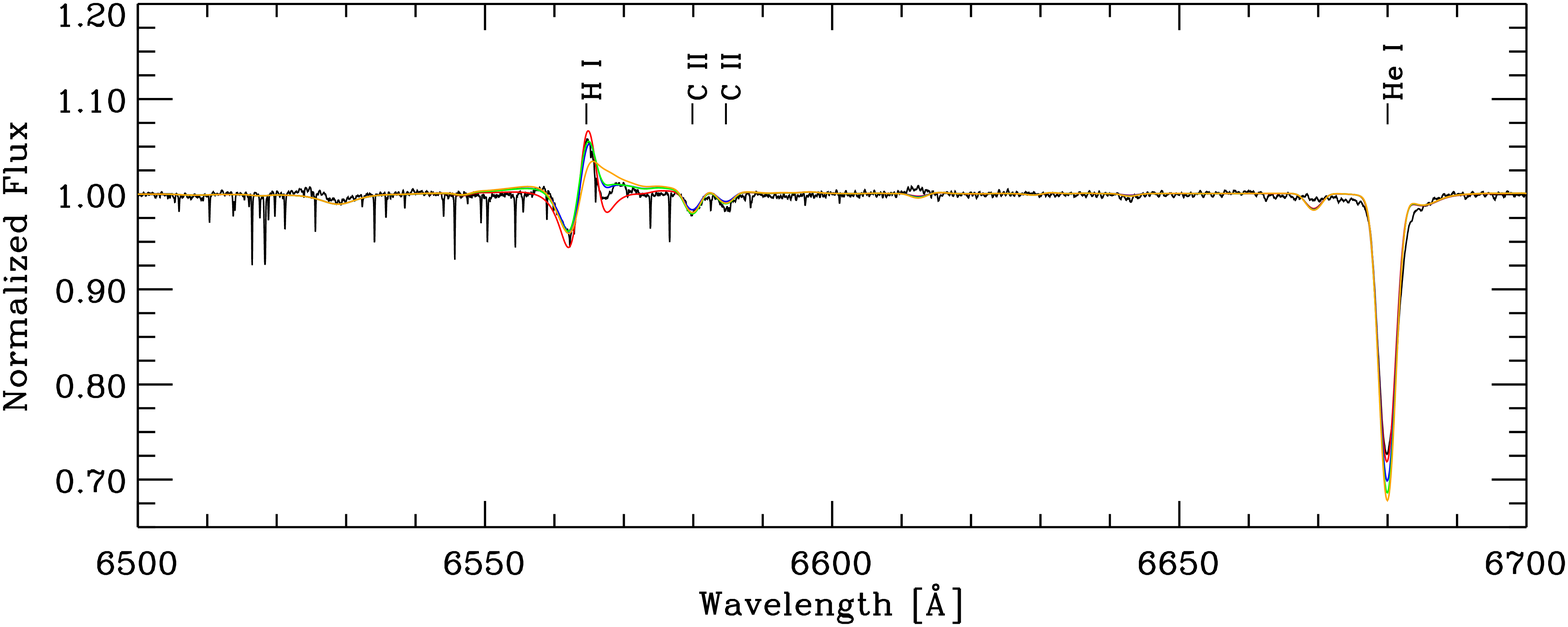} 
 \caption{Best model fits for \epsori\ for each $f_\infty$ value: 
          1.0 (red), 0.1 (blue), 0.05 (green) and 0.01 (orange). Optical
          data are shown by the black line.}\label{fA3}
\end{figure*}

\begin{figure*}
 \centering
 \includegraphics[width=\linewidth]{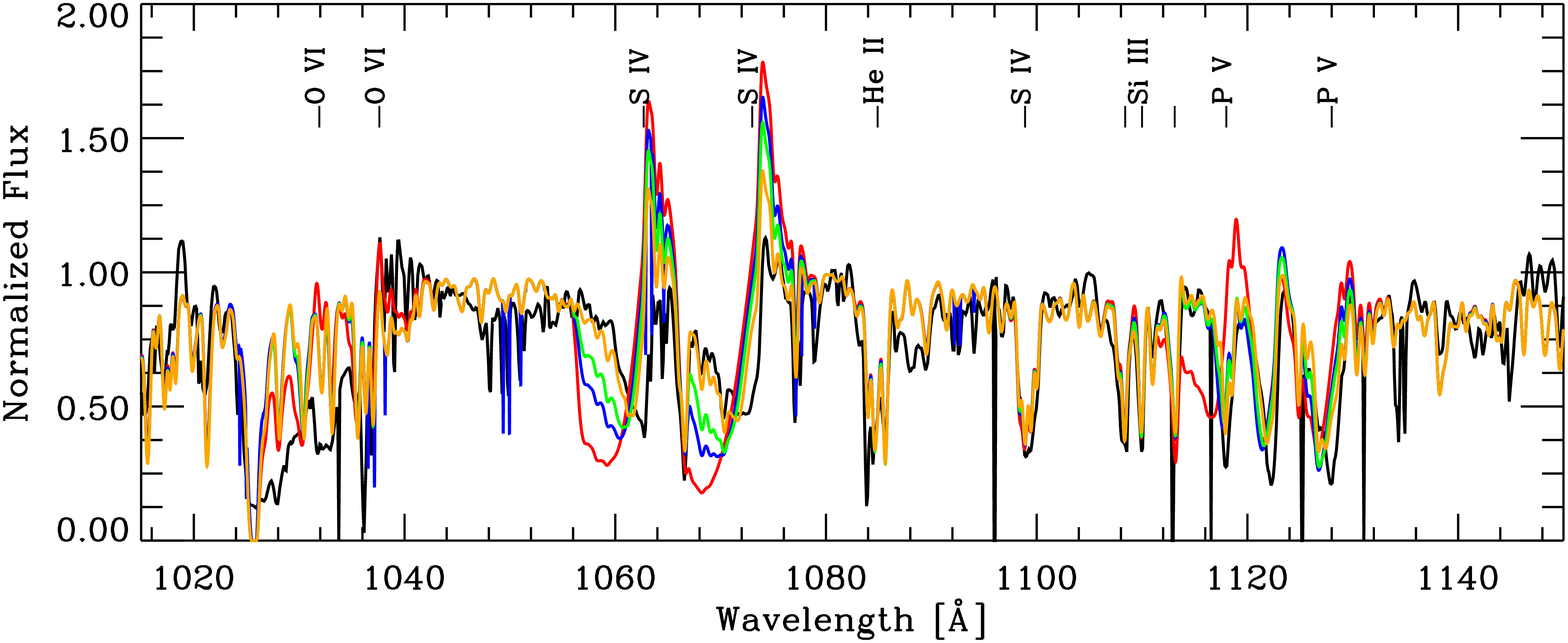} 
 \caption{Best model fits for \epsori\ for each $f_\infty$ value: 
          1.0 (red), 0.1 (blue), 0.05 (green) and 0.01 (orange). The 
          black line represents the \copernicus\ data. The influence
          of the inter clump medium on \ionnll{O}{vi}{1032,1038} has not been included.
          }\label{fA4}
\end{figure*}

\begin{figure*}
 \centering
 \includegraphics[width=\linewidth]{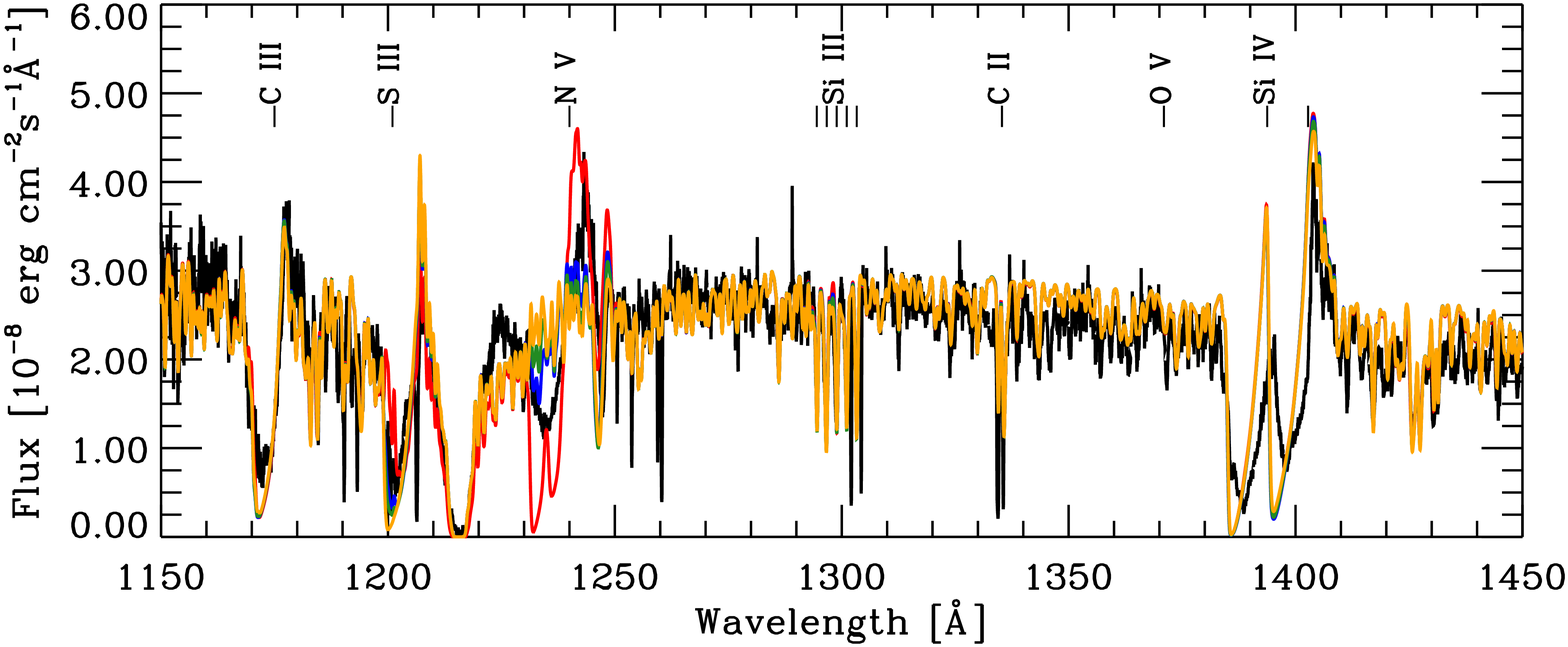} \\
 \includegraphics[width=\linewidth]{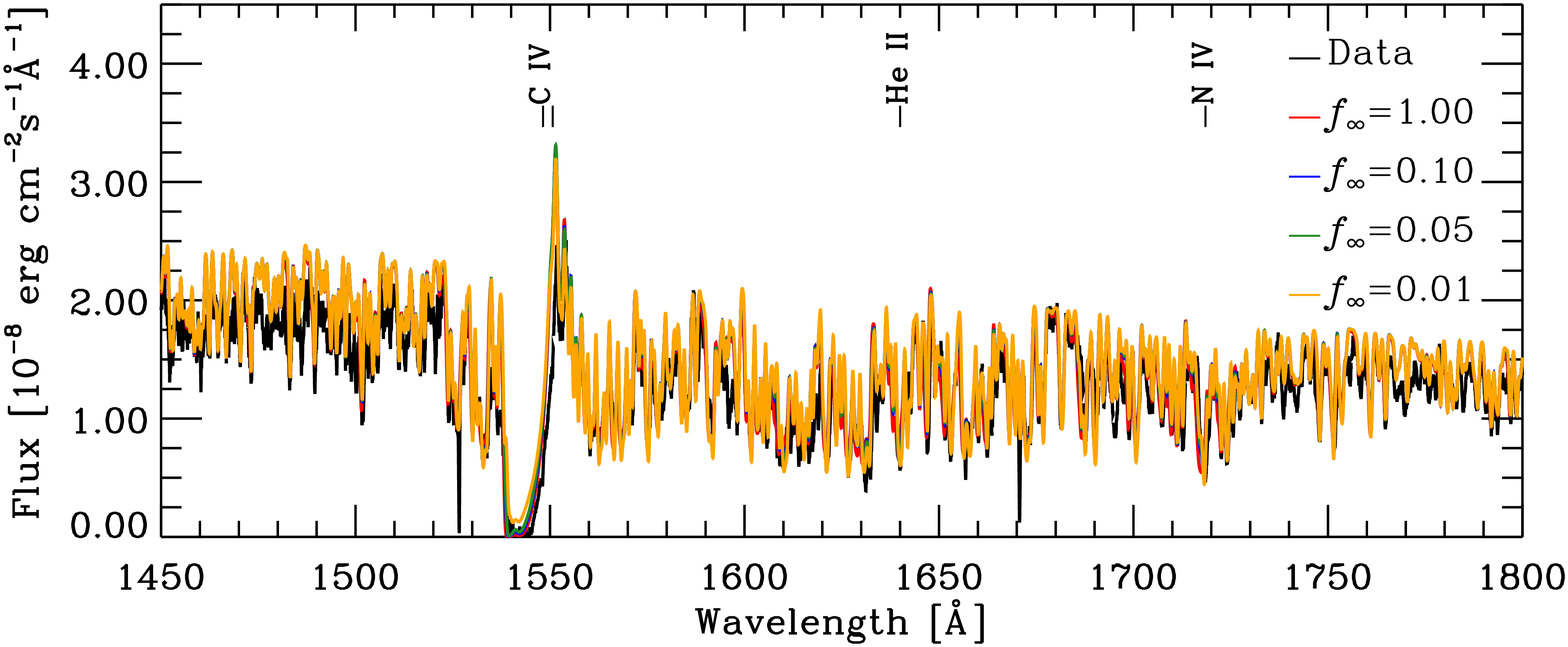}  \\
 \includegraphics[width=\linewidth]{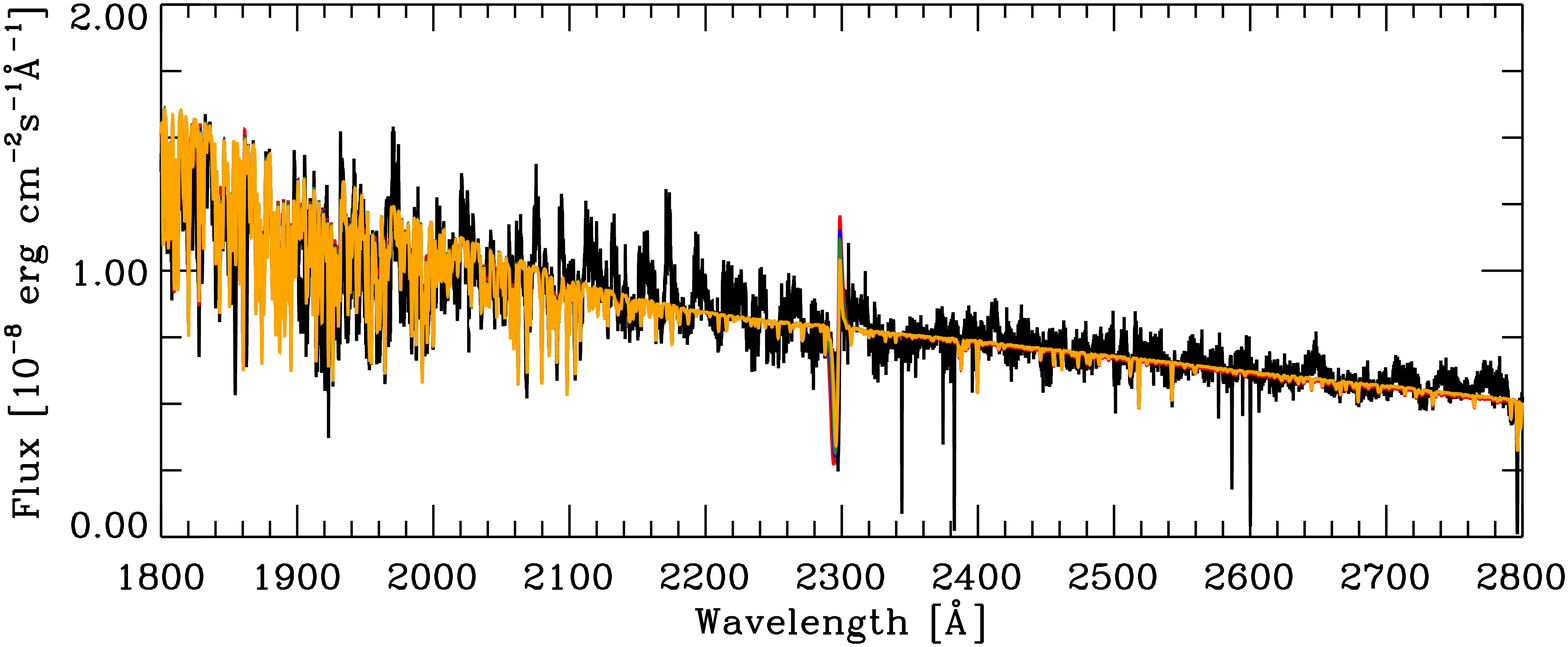}
 \caption{Best model fits for \epsori\ for each $f_\infty$ value: 
          1.0 (red), 0.1 (blue), 0.05 (green) and 0.01 (orange). The 
          black line represents the \iue\ data. The influence
          of the inter clump medium on \ionnll{N}{v}{1238,1242} has not been included.
          }\label{fA5}
\end{figure*}

\bsp

\label{lastpage}

\end{document}